\documentclass[aps,groupedaddress, superscriptaddress,showpacs,amsmath,amssymb,twocolumn,pra]{revtex4-1}

\usepackage[dvips]{graphicx}
\usepackage{wrapfig,subfigure}
\usepackage{amssymb} 
\usepackage{xcolor}
\usepackage[normalem]{ulem}
\usepackage{amsmath,bm}
\usepackage{epstopdf}
\usepackage{mathtools} 
\usepackage{extarrows} 

\usepackage{amsfonts}
\usepackage{braket}

\begin{document}

\title{Continuous quantum error correction for evolution under time-dependent Hamiltonians}

\author{J. Atalaya*}
\affiliation{Department of Chemistry, University of California, Berkeley,  CA 94720, USA}
\affiliation{Berkeley Quantum Information and Computation Center, University of California, Berkeley,  CA 94720, USA}
\author{S. Zhang*}
\affiliation{Berkeley Quantum Information and Computation Center, University of California, Berkeley,  CA 94720, USA}
\affiliation{Department of Physics, University of California, Berkeley,  CA 94720, USA}

\author{M. Y. Niu}
\affiliation{Google Research, 340 Main Street, Venice Beach, California 90291, USA}
\author{A. Babakhani}
\affiliation{Department of Physics, University of California, Berkeley,  CA 94720, USA}
\author{H. C. H. Chan}
\affiliation{Department of Physics, University of California, Berkeley,  CA 94720, USA}
\author{J. M. Epstein}
\affiliation{Department of Physics, University of California, Berkeley,  CA 94720, USA}
\author{K. B. Whaley}
\affiliation{Department of Chemistry, University of California, Berkeley,  CA 94720, USA}
\affiliation{Berkeley Quantum Information and Computation Center, University of California, Berkeley,  CA 94720, USA}

\date{\today}

\begin{abstract}
We develop a protocol for continuous operation of a quantum error correcting code  for protection of coherent evolution due to an encoded Hamiltonian against environmental errors, using the three qubit bit flip code and bit flip errors as a canonical example. To detect errors in real time, we filter the output signals from continuous measurement of the error syndrome operators and use a double thresholding protocol for error diagnosis, while correction of errors is done as in the conventional operation. We optimize our continuous operation protocol for {evolution under quantum memory and under quantum annealing}, {by maximizing the fidelity between the target and actual logical states at a specified final time.} {In the case of quantum memory} we show that our continuous operation protocol yields a logical  error rate that is slightly larger than the one obtained from using the optimal Wonham filter for error diagnosis. The advantage of our protocol is that it can be simpler to implement. For quantum annealing, we show that our continuous quantum error correction protocol can significantly reduce the final logical state infidelity when {the} continuous measurements are sufficiently strong {relative to the strength of the time-dependent Hamiltonian,} {and that it can also significantly reduces the run time relative to that of a classical encoding.} These results suggest that a continuous implementation is suitable for quantum error correction in the presence of encoded time-dependent Hamiltonians, opening the possibility of many applications in quantum simulation and quantum annealing. 
\end{abstract}

\maketitle

\section{Introduction}
Quantum error correction (QEC) is an essential component of quantum information processing.  The need to either avoid or correct errors on quantum states due to imperfect quantum operations {or} decohering interactions with the environment places stringent requirements on realization of the promise of quantum computation and quantum simulations.  Various tools have been developed to mitigate the effect of such errors, including encoding into decoherence free subspaces or subsystems \cite{lidar1998decoherence,lidar2003decoherence}, addition of penalty Hamiltonians \cite{jordan2006error,bookatz2015error,marvian2017error}, dynamical decoupling methods \cite{viola1999dynamical,pasini2010optimized,lidar2014review} and other applications of pulse sequences \cite{Gorman2012pulses}, {as well as} the use of quantum error correcting codes (QECC) that delocalize the errors over multiple physical qubits, combined with error recovery operations \cite{gottesman2002introduction,devitt2013quantum,lidar2013quantum,terhal2015quantum}. The latter provides a powerful approach to systematically correct errors that can also be made fault tolerant \cite{shor1996fault}.  

{The canonical operation mode for} quantum error correction codes~\cite{Shor1995,Steane1996,Gottesman1996,ChuangBook} employ projective measurements and discrete recovery operations to provide reduction of errors that are treated as discrete events occurring at a specified rate.  The formalism of QEC has been developed to provide firm guarantees of protection in terms of reduced scaling of the logical error rate for an encoded state. However, in practice, few measurements can be described as projective, and are instead better described as finite strength weak measurements that are characterized by a gradual collapse of the measured system wavefunction~\cite{KrausBook,Carmichael1993,Korotkov2016,Wiseman2010,Molmer1992,Diosi1988,Katz2006,
PalaciosLaloy2010,Devoret2013, KMurch2013,DiCarlo2013,Hacohen-Gourgy2016,Huard2018}. A continuous quantum error correction code, i.e., a CQEC, is based on the continuous quantum measurement of the error syndrome operators of the conventional QEC code. Previous theoretical work on such continuous quantum error correction has been devoted primarily to analysis of the continuous operation performance of stabilizer~\cite{Zurek1998,Ahn2002,Ahn2003,Ahn2004,Sarovar2004,Sarovar2005,Brun2007,Geremia2008,Mabuchi2009, Mabuchi_new_j_phys2009, Brun2016,Dressel2019,Cardona2019} and subsystem~\cite{Atalaya2017, Atalaya2020} QEC codes for quantum memory, where the Hamiltonian of the encoding physical qubits is disregarded in the analysis. In contrast, in this work we focus on protecting the coherent evolution of an encoded qubit system evolving under a time-dependent Hamiltonian, against environmental decoherence. This problem is particularly important for the development of quantum error correction for a broad range of  quantum information applications employing continuously varying Hamiltonians.  These include quantum annealing  {and} adiabatic quantum computation~\cite{Lidar2018}, and quantum simulation~\cite{Nori2014}. 

A major challenge for application of either discrete or continuous QEC to protect coherent evolution of an encoded qubit system is that perfect identification and  correction of errors (in the example studied here, these will be bit-flip errors) does not imply absence of logical errors~\cite{KYoung2013}. We can understand this difficulty by thinking of the action of errors on the  Hamiltonian instead of on the quantum state---{a perspective} somewhat similar to the Heisenberg picture. In this picture, an error causes the Hamiltonian to effectively change from $H(t)$ to $EH(t)E$, where $E$ is the operator associated to the error {that occurred and is} assumed to be a {single-qubit} Pauli operator. Subsequent coherent evolution is due to the new Hamiltonian $EH(t)E$, until the moment when the error {that occurred} is detected and corrected. During this period of error diagnosis and correction, logical errors will accrue if the original Hamiltonian does not commute with the error operators, i.e., {if} $H(t)\neq EH(t)E$. Since Hamiltonians that commute with all error operators are difficult to implement~\cite{KYoung2013}, this problem has constituted a major stumbling block for the development of quantum error correction for quantum annealing and for analog quantum simulation in general.  This is precisely the situation that we address in this work.

We consider here the continuous operation of  a quantum code that is designed to protect the coherent evolution of the encoded qubit system.  As a specific example we take the three-qubit bit flip code~\cite{ChuangBook},  which is a stabilizer code~\cite{Gottesman1996} with two commuting stabilizer operators that constitute the measurement operators. 
Section~\ref{sec:3qubit_continuous} describes the continuous operation of this code.  In Section~\ref{sec:CQEC_2threshold} we propose and analyze an error detection protocol based on time-averaging (filtering) of the bare readout signals from simultaneous continuous monitoring of the error syndrome operators, together with a double error thresholding scheme that is applied to the filtered readout signals in order to explicitly diagnose errors. Unlike previous schemes~\cite{Ahn2002,Sarovar2004}, partial errors are not acted on---the error diagnosis is acted on only when occurrence of a complete, i.e., discrete, error has been diagnosed with high probability.  Filtering is necessary in the protocol to reduce (but not eliminate) the amount of noise in the filtered readout signals, while double error thresholding is essential to reduce the probability of mis-identification of single bit-flip errors that affect several readout signals at the same time~\cite{Atalaya2020}.

In Section~\ref{sec:effectiveQME} we develop an effective open quantum system model to describe the Hamiltonian-driven evolution of the encoded qubit system in the presence of both bit-flip errors and CQEC. In Section~\ref{sec:Optimization} we use this model to optimize the performance of the proposed continuous QEC protocol for {operation under} quantum annealing. In this case, the performance of our protocol depends on the relative strength of three parameters; namely, the error rate $\gamma$, the Hamiltonian strength parameter $\Omega_0$, and the  strength $\Gamma_{\rm m}$ {of the} continuous measurements. We find that our CQEC protocol yields a significant reduction of the final logical state infidelity when the  measurements are sufficiently strong relative to the Hamiltonian. We show that further improvements are possible by using modified error correction operators. To demonstrate the capability of our proposed CQEC approach, we present detailed results for one  logical qubit and then show that the strategy can be readily generalized to quantum simulation with many encoded qubits.  We show explicitly  that a high level of protection can be obtained for quantum annealing of two logical qubits. In addition, we analyse the performance of the proposed continuous QEC protocol for quantum memory operation, in the absence of a Hamiltonian. We minimize the logical error rate and find that its optimal value is slightly larger than the logical error rate obtained from using the linear variant~\cite{Dressel2019} of the optimal Wonham filter~\cite{Mabuchi2009,Mabuchi_new_j_phys2009} for error diagnosis. The advantage of our continuous QEC protocol is  that it can be simpler to implement. We also show that the resulting optimized double thresholding error diagnosis scheme is very effectively combined with discrete recovery operations to obtain the reduced scaling of the logical error rate that is necessary for a valid quantum error correcting code. Finally, in Section~\ref{sec:discussion}, we provide a summary and discussion, with additional analysis of the time-to-solution metric for estimating the cost of applying CQEC to quantum annealing problems. We conclude with a prognosis for future applications and useful directions for further work.

\section{Continuous operation of the three-qubit bit flip code.}
\label{sec:3qubit_continuous}

In contrast to the discrete operation of the three-qubit bit flip code~\cite{ChuangBook}, in the continuous operation, the error syndrome operators (stabilizer generators), 
\begin{align}
\label{eq:stabilizers}
S_1=Z_{12}=Z_1Z_2\;\;\; {\rm and} \;\;\; S_2=Z_{23}=Z_2Z_3,
\end{align}
are continuously measured at the same time. In Eq.~\eqref{eq:stabilizers}, $Z_1$ represents the Pauli $z$ operator that acts on the first physical qubit; that is, $Z_1\,|q_1\,q_2\,q_3\rangle=(-1)^{q_1}|q_1\,q_2\,q_3\rangle$, where the set of states $|q_1\,q_2\,q_3\rangle$ with $q_1$, $q_2$, $q_3=\{0,1\}$ defines the computational basis. Similar definitions hold for the Pauli $z$ operators $Z_2$ and $Z_3$. The corresponding normalized readout signals are given by ($k=1,2$)
\begin{align}
\label{eq:Ik}
I_{k}(t) = {\rm Tr}[S_k\,\rho(t)] + \sqrt{\tau_k}\,\xi_k(t), 
\end{align}
where $\rho(t)$ is the $8\times8$ density matrix of the three physical qubits and $\tau_k$ is the so-called ``measurement time'' to distinguish between the $\pm1$ eigenvalues of the stabilizer generator $S_k$ with a signal-to-noise ratio (SNR) of 1~\cite{Korotkov2001}. Note that the detector readout signals $I_k(t)$ are given by the sum of the ``signal part'' ${\rm Tr}[S_k\,\rho(t)]$ and the noise part $\xi_k(t)$, which has a vanishing mean. In the Markovian approximation, the noises $\xi_k(t)$ are assumed Gaussian and white with a two-time correlation function: 
\begin{align}
\label{eq:noises-corr-func}
\langle \xi_k(t)\xi_{k'}(t')\rangle = \delta_{kk'}\delta(t-t'), 
\end{align}
where $\langle\cdot\rangle$ denotes average over an ensemble of noise realizations. 

The evolution of the three-qubit quantum state $\rho(t)$ in the absence of environmental decoherence is described by (in It\^o interpretation {\cite{Wiseman2010}})
\begin{align}
\label{eq:rho-eom-no-errors}
&\dot{\rho}(t) = -i[H(t),\rho] + \nonumber \\
&\sum_{k=1,2}\Big[\frac{\Gamma_{k}}{2}({S}_k\rho S_k - \rho) + \frac{\xi_k}{\sqrt{\tau_k}}\big(\frac{S_k\rho + \rho S_k}{2} - \rho\,{\rm Tr}[S_k \rho]\big)\Big].
\end{align}
The first line of Eq.~\eqref{eq:rho-eom-no-errors} describes the coherent evolution of the three physical qubits due to a time dependent Hamiltonian.  In this work we shall focus on the quantum annealing Hamiltonian ($\hbar=1)$ 
\begin{align}
\label{eq:HL}
H(t) =-{\Omega_0}{}\left[a(t)X_{\rm L} + b(t)Z_{\rm L}\right],
\end{align}
where the frequency parameter $\Omega_0$ sets the energy scale of the above Hamiltonian, and the coefficients $a(t)$ and $b(t)$ are  functions of time with magnitudes smaller than 1. The operators $X_{\rm L}$ and $Z_{\rm L}$ denote the logical $X$ and $Z$ operators, given by 
\begin{align}
\label{eq:XL-ZL}
X_{\rm L} = X_1X_2X_3 \;\;\; {\rm and} \;\;\; Z_{\rm L} = \frac{Z_1+Z_2+Z_3}{3},  
\end{align}
where $X_q$  represents the Pauli $x$ operator that acts on the $q$th physical qubit. Note that the system Hamiltonian~\eqref{eq:HL} and the stabilizer generators~\eqref{eq:stabilizers} exhibit a block-diagonal matrix representation in the computational basis. \newline 
The second line of Eq.~\eqref{eq:rho-eom-no-errors} describes the measurement-induced quantum back-action on the three-qubit quantum state {that is} due to simultaneous continuous measurement of the stabilizer generators $Z_{12}$ and $Z_{23}$. Each measurement channel is characterized by the measurement time parameter $\tau_k$  and the measurement-induced ensemble dephasing rate $\Gamma_k$, which are related via the quantum efficiency $\eta_k$ as follows $\tau_k=1/(2\Gamma_k\eta_k)$~\cite{Korotkov2001}. For ideal detectors, the quantum efficiency is unity, while for nonideal detectors the quantum efficiency is less than one. For simplicity of notation, we shall assume below that both detectors have identical parameters: 
\begin{align}
\label{eq:symm-cont-measurements}
\Gamma_k=\Gamma_{\rm m},\;\;\; \tau_k= \tau_{\rm m},\;\;\; {\rm and}\;\;\; \eta_k = \eta\;\;\;\;(k=1,2).
\end{align}
(This assumption can be readily removed and the analysis continued with different parameters for each detector.)

Encoding with the three-qubit bit flip code effectively divides the full eight-dimensional Hilbert space of the three physical qubits into four two-dimensional subspaces, where the stabilizer generators $Z_{12}$ and $Z_{23}$ have definite $\pm1$ values. As usual, the two-dimensional subspace where both stabilizer generators have values $+1$ is referred to as the code space, denoted as $\mathcal{Q}_0$, while the two-dimensional subspaces where $(Z_{12},Z_{23})$ have values $(-1,+1), (-1,-1)$ and $(+1,-1)$ are referred to as the error subspaces, denoted as $\mathcal{Q}_1$, $\mathcal{Q}_2$ and $\mathcal{Q}_3$, respectively. The code space is spanned by the zero and one logical states, which are  expressed in the computational basis as 
\begin{align}
|0_{\rm L}\rangle = |0\,0\,0\rangle\;\;\;{\rm and}\;\;\;|1_{\rm L}\rangle = |1\,1\,1\rangle,
\end{align}
respectively. In the absence of errors, the (target) logical wavefunction 
\begin{align}
\label{eq:psiL}
|\psi_{\rm L}(t)\rangle = \alpha_{\rm L}(t)|0_{\rm L}\rangle  + \beta_{\rm L}(t)|1_{\rm L}\rangle
\end{align}
evolves according to the following Schr\"odinger equation for the probability amplitudes  of the zero ($\alpha_{\rm L}$) and one ($\beta_{\rm L}$) logical states:
\begin{align}
\label{eq:psiL-eom}
\begin{bmatrix}\dot\alpha_{\rm L}(t)  \\ \dot\beta_{\rm L}(t) \end{bmatrix} = -ih_{\rm L}(t)\begin{bmatrix}\alpha_{\rm L}(t)  \\ \beta_{\rm L}(t) \end{bmatrix}.  
\end{align}
In the above equation, $h_{\rm L}(t)$ represents the Hamiltonian of the logical qubit and {is} given by the $2\times2$ diagonal sub-matrix of $H(t)$ that corresponds to the code space,    
\begin{align}
\label{eq:hL}
h_{\rm L}(t) = -\Omega_0\left[a(t)\sigma_x + b(t)\sigma_z\right],
\end{align}
where $\sigma_x$ and $\sigma_z$ denote the conventional Pauli $x$ and $z$ matrices, and the coefficients $a(t)$ and $b(t)$ are the coefficients given in Eq.~\eqref{eq:HL}. (In this work, {we shall use the notation $|\psi_{\rm L}(t)\rangle$ to denote} the column matrix $[\alpha_{\rm L}(t)\;\beta_{\rm L}(t)]^{\rm T}$.) We emphasize that evolution of the target logical wavefunction~\eqref{eq:psiL} is not affected by measurement, because the system Hamiltonian~\eqref{eq:HL} and the stabilizer generators~\eqref{eq:stabilizers} commute with each other; i.e., there is no quantum Zeno effect ({unlike the non-commuting situation, e.g.,~\cite{Slichter2015}}). \newline
The error subspace $\mathcal{Q}_1$ is spanned by the computational states $|1\,0\,0\rangle = X_1\,|0_{\rm L}\rangle$ and $|0\,1\,1\rangle = X_1\,|1_{\rm L}\rangle$; the error subspace $\mathcal{Q}_2$ is spanned by the computational states $|0\,1\,0\rangle = X_2\,|0_{\rm L}\rangle$ and $|1\,0\,1\rangle = X_2\,|1_{\rm L}\rangle$; and the error subspace $\mathcal{Q}_3$ is spanned by the computational states $|0\,0\,1\rangle = X_3\,|0_{\rm L}\rangle$ and $|1\,1\,0\rangle = X_3\,|1_{\rm L}\rangle$. In addition, the $2\times2$ diagonal sub-matrices of $H(t)$ that correspond to these error subspaces are  {all identical} and equal to 
\begin{align}
\label{eq:hsp}
h_{\rm spurious}(t) = -\Omega_0\left[a(t)\sigma_x + \frac{1}{3}b(t)\sigma_z\right].   
\end{align}
Note the factor of $1/3$ in the above equation. This {derives} from  the action of the system Hamiltonian $H(t)$, Eq.~\eqref{eq:HL}, on a state with support in {one of the}  error subspaces. For instance, for the system state, $|\psi(t)\rangle = \alpha \,X_1\,|0_{\rm L}\rangle + \beta\, X_1\,|1_{\rm L}\rangle$, which is in the error subspace $\mathcal{Q}_1$, we obtain $H(t)|\psi(t)\rangle = -\Omega_0a(t)\big[\alpha\,X_1\,|1_{\rm L}\rangle + \beta\,X_1\,|0_{\rm L}\rangle\big] -(1/3) \big[\alpha\,X_1\,|0_{\rm L}\rangle - \beta\,X_1\,|1_{\rm L}\rangle\big]\Omega_0b(t)$. In contrast, this factor of $1/3$ does not appear when the system Hamiltonian $H(t)$ acts on (code space) logical states, {Eq.~\eqref{eq:psiL}}. We {can therefore} say that when the system state is in the error subspaces, coherent evolution in those subspaces is due to the spurious Hamiltonian~\eqref{eq:hsp}, instead of the intended logical Hamiltonian~\eqref{eq:hL}. 

In the presence of bit-flip errors, the (mixed) three-qubit state $\rho(t)$ evolves according to the evolution equation that results from adding to the right-hand side of Eq.~\eqref{eq:rho-eom-no-errors} the following decoherence terms 
\begin{align}
\label{eq:rho-decoh-eom}
\dot{\rho}_{\rm decoh}(t) = \sum_{q=1,2,3}\gamma_q\left[X_q\, \rho\,X_q - \rho \right],
\end{align}
where $\gamma_q$ denotes the bit-flip error rate of the $q$th physical qubit. Thus in the presence of bit-flip errors, the full three-qubit state evolves as 
\begin{align}
\label{eq:rho-eom-no-errors-v2}
&\dot{\rho}(t) = -i[H(t),\rho] + \nonumber \\
&\sum_{k=1,2}\!\left[\frac{\Gamma_{k}}{2}({S}_k\rho S_k - \rho) + \frac{\xi_k}{\sqrt{\tau_k}}\big(\frac{S_k\rho + \rho S_k}{2} - \rho\,{\rm Tr}[S_k \rho]\big)\right]
\!+\nonumber\\ 
&\sum_{q=1,2,3}\gamma_q\,[X_q\rho X_q - \rho].
\end{align}

Our analysis of logical errors presented below is based on the jump/no-jump method~\cite{ChuangBook} for bit-flip errors. In this method,  gradual decoherence due to the terms~\eqref{eq:rho-decoh-eom} is described as the average effect of  bit-flip errors $X_1$, $X_2$ or $X_3$ that occur at random times, as follows. At the infinitesimal time interval $(t,t+\delta t)$, a bit-flip error $X_q$ occurs with probability $\delta t \gamma_q$. If this error occurs, the system state ``jumps'' from $\rho(t)$ to $\rho(t+\delta t) = X_q\,\rho(t)\,X_q$; otherwise, the system state continuously evolves according to Eq.~\eqref{eq:rho-eom-no-errors}, without environmental decoherence. On averaging over many instances of the bit-flip errors, the jump/no-jump approach reduces to the open quantum system model~\eqref{eq:rho-eom-no-errors-v2}, where errors {continuously} change the mixed system state $\rho(t)$. 

The encoded logical state is obtained from the mixed three-qubit state $\rho(t)$ as follows 
\begin{align}
\label{eq:rhoL-def}
\varrho_{\rm L}(t) =\left\langle \frac{1}{p_{\textrm{code-space}}(t)}\begin{bmatrix}\rho_{000,000}(t) &  \rho_{000,111}(t) \\
\rho_{111,000}(t) &  \rho_{111,111}(t)\end{bmatrix}\right\rangle,  
\end{align}
where $p_{\textrm{code-space}}(t)=\rho_{000,000}(t) + \rho_{111,111}(t)$ is the probability of {the system being in} the code space. 

Our goal is to maximize the fidelity between the target logical wavefunction~\eqref{eq:psiL} and the true (mixed) logical state~\eqref{eq:rhoL-def}, at some final time, {where the} evolution includes the decoherence effect of bit-flip errors {as well as the effect of the spurious coherent evolution in the error subspaces due to an added Hamiltonian. To counteract the latter two effects, we introduce the double threshold CQEC protocol} described in the following section.  

\section{The double threshold CQEC protocol} 
\label{sec:CQEC_2threshold}

In the three-qubit bit flip code, the conventional error correction operations are (we will modify these operations to improve the continuous operation performance in Section~\ref{sec:correction_logical_errors}): 
\begin{align}
C_{\rm op}=X_1,\; X_2\;{\rm or}\; X_3.
\label{eq:Cop}
\end{align}
These operations are applied on the physical qubits when the error syndrome (defined as the values of the stabilizer generators $Z_{12}$ and $Z_{23}$, in this order) is equal to $(-1,+1)$, $(-1,-1)$ or $(+1,-1)$, respectively. To apply these error correction operations in the continuous operation, we have to estimate the error syndrome from the noisy readout signals $I_k(t)$  given in Eq.~\eqref{eq:Ik}. To do this, we filter the latter to obtain smoother signals $\mathcal{I}_k(t)$ that obey the following filter equation:
\begin{align}
\label{eq:filter-eq}
\dot{\mathcal{I}}_k(t) = -\frac{\mathcal{I}_k(t)}{\tau} + \frac{{I}_k(t)}{\tau},
\end{align}
where $\tau$ plays the role of an averaging time parameter. The initial condition for Eq.~\eqref{eq:filter-eq} is discussed below. In practice, the filtered readout signals $\mathcal{I}_k(t)$ can be obtained, e.g., by passing the bare readout signals $I_k(t)$ through a resistor-capacitor circuit (RC lowpass filter~\cite{HorowitzBook}). Note that the SNRs of the filtered readout signals can be increased by choosing {a larger value of} $\tau$. For instance, in the absence of bit-flip errors, the filtered readout signals read as 
\begin{align}
\label{eq:Ik-filtered-noerrors}
\mathcal{I}_k(t) = \int_0^t dt'\, \frac{e^{-\frac{t-t'}{\tau}}}{\tau}\,I_k(t') 
\end{align}
in the stationary regime ($t\gg \tau$) and their SNRs are equal to $2\tau/\tau_{\rm m}$. The averaging time parameter $\tau$, however, should not be chosen arbitrarily large; there is an optimal value that is obtained below.

\begin{figure}[t!]
\centering
\includegraphics[width=0.99\linewidth, trim =0cm 0.25cm 0cm 0cm,clip=true]{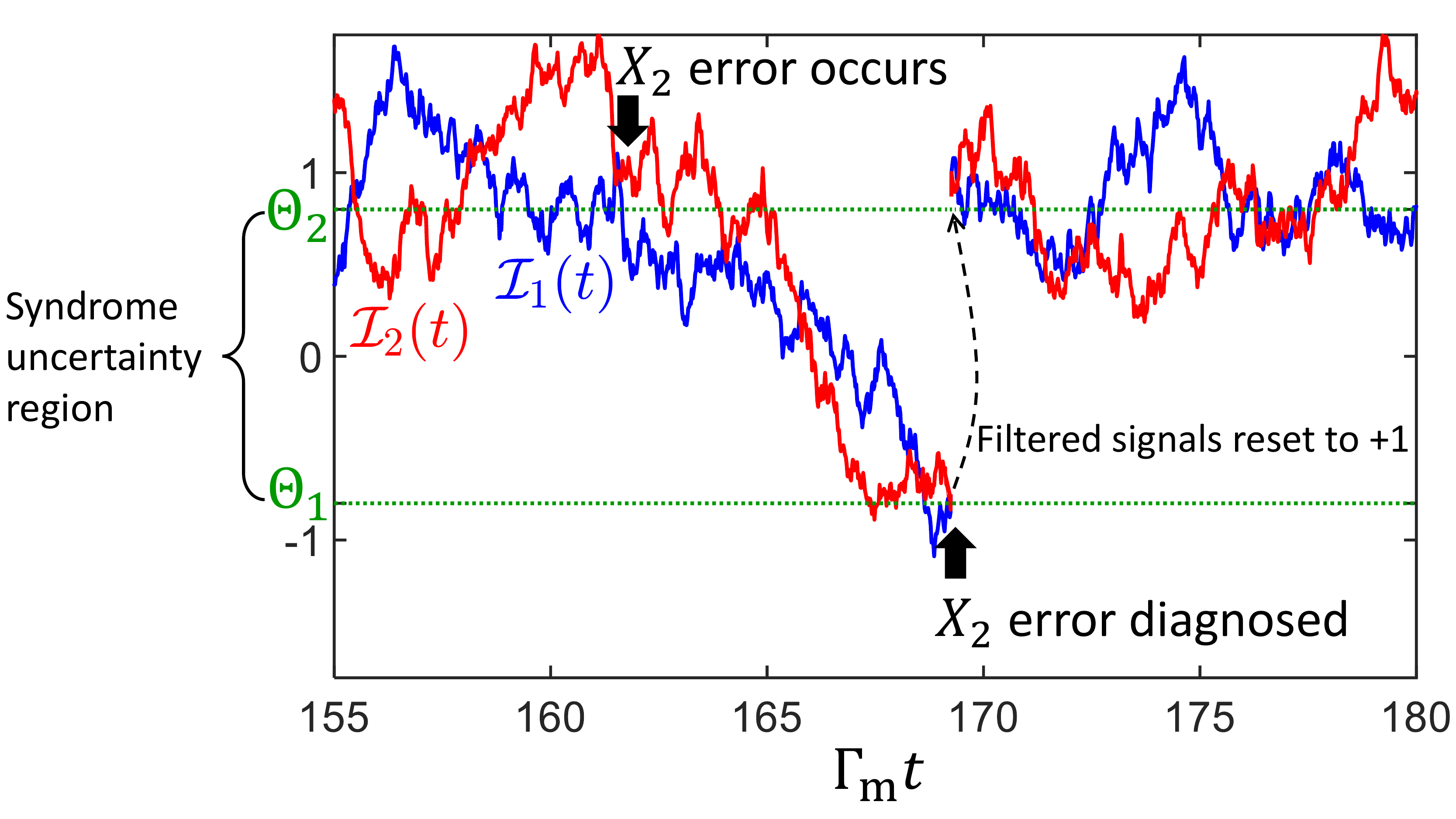}
\caption{{Example of} filtered readout signals $\mathcal{I}_1(t)$ and $\mathcal{I}_2(t)$ when a bit-flip error $X_2$ occurs. This error is detected by the CQEC protocol (see main text) at the moment when both filtered readout signals {have} exited the ``syndrome uncertainty region'' below the lower error threshold $\Theta_1$. The filtered readout signals $\mathcal{I}_k(t)$ are discontinuous since the CQEC protocol reset them to the value $+1$ at the moment when the occurred error is diagnosed.}
\label{fig:filtered-signals-X2-err}
\end{figure}

To diagnose the error syndrome, we use a double thresholding scheme that is applied to the filtered readout signals $\mathcal{I}_1(t)$ and $\mathcal{I}_2(t)$. We introduce two error threshold parameters $\Theta_1$ and $\Theta_2$ ($\Theta_1<\Theta_2$) that define the interval $[\Theta_1,\Theta_2]$, which is referred to as the ``syndrome uncertainty region'', see Fig.~\ref{fig:filtered-signals-X2-err}. If at least one of the filtered readout signals {lies within} this interval, we say that we are not certain about {the value of} the error syndrome, and do nothing. More precisely, the double thresholding scheme works as follows.  If $\mathcal{I}_1(t)$ and $\mathcal{I}_2(t)$ are both larger than $\Theta_2$, the diagnosed error syndrome is $(+1,+1)$ and no error correction operation is applied, since the system quantum state is most likely in the code space. If $\mathcal{I}_1(t)<\Theta_1$ and $\mathcal{I}_2(t) > \Theta_2$, the diagnosed error syndrome is  $(-1,+1)$ and the error correction operation to be applied is $C_{\rm op}=X_1$, since the system quantum state is most likely in the error subspace $\mathcal{Q}_1$. If $\mathcal{I}_1(t)$ and $\mathcal{I}_2(t)$ are both smaller than $\Theta_1$, the diagnosed error syndrome is  $(-1,-1)$ and the error correction operation to be applied is $C_{\rm op}=X_2$,  since the system quantum state is most likely in the error subspace $\mathcal{Q}_2$. If $\mathcal{I}_1(t)>\Theta_2$ and $\mathcal{I}_2(t)< \Theta_1$, the diagnosed error syndrome is  $(+1,-1)$ and the error correction operation to be applied is  $C_{\rm op}=X_3$,  since the system quantum state is most likely in the error subspace $\mathcal{Q}_3$. 

The error correction operations $C_{\rm op}$ must now be applied immediately after an error is detected. Note that this contrasts with the situation in operation of a quantum memory, where correction of errors can be delayed to the end of the continuous operation of the code~\cite{Mabuchi2009,Mabuchi_new_j_phys2009,Dressel2019,Atalaya2020}. In the present analysis, we shall assume that the error correction operations are applied instantaneously on the physical qubits, changing the three-qubit state from $\rho(t)$ to $C_{\rm op}\rho(t) C_{\rm op}$ when the error correction operation $C_{\rm op}$ is applied. 

Finally, the filtered readout signals $\mathcal{I}_k(t)$ are reset to the initial condition $+1$ at the moment when an error is  diagnosed {(see Fig.~\ref{fig:filtered-signals-X2-err}). Their subsequent values are dictated by the filter equation~\eqref{eq:filter-eq} until the next error is diagnosed, and so on.}

Figure~\ref{fig:filtered-signals-X2-err} depicts an example showing how the filtered readout signals $\mathcal{I}_1(t)$ and $\mathcal{I}_2(t)$ are affected by the occurrence of a bit-flip error $X_2$ at the moment $t_{\rm err}=162\Gamma_{\rm m}^{-1}$. Before this error occurs, the system state is in the code space, so the filtered readout signals fluctuate around 1. After the occurrence of the error $X_2$, the ``signal part'' of the filtered readout signals becomes (for $t\geq t_{\rm err}$)
\begin{equation}
\label{eq:Ik_mean}
\langle \mathcal{I}_1(t)\rangle=\langle \mathcal{I}_2(t)\rangle = -1 + 2\,e^{-(t-t_{\rm err})/\tau}.
\end{equation}
Equation~\eqref{eq:Ik_mean} is the solution of Eq.~\eqref{eq:filter-eq} with $I_k(t)$ replaced by $-1$, which is the ``signal part'' of the bare readout signal Eq.~\eqref{eq:Ik} after the error $X_2$ occurs. Even though both filtered readout signals have the same ``signal part'', we see in Fig.~\ref{fig:filtered-signals-X2-err} that these signals follow different paths due to noise. This indicates that if we had used a single error threshold to detect errors, the error $X_2$ would have been most likely misdiagnosed, because the filtered readout signals do not cross the given error threshold at the same time, see Fig.~\ref{fig:filtered-signals-X2-err}. In contrast, our double thresholding scheme performs well unless relatively large fluctuations occur in the filtered readout signals. For instance, in the example considered above, the
error $X_2$ would be diagnosed as  $X_1$ if a relatively large positive fluctuation (of magnitude of the order of $\Theta_2-\Theta_1$) had made the filtered readout signal $\mathcal{I}_2(t)$ be above the upper error threshold $\Theta_2$ at the moment when the other filtered readout signal $\mathcal{I}_1(t)$ is below the lower error threshold $\Theta_1$. We will show below that the probability to misdiagnose errors in our double thresholding scheme can be made exponentially small by both increasing the length of the ``syndrome uncertainty region'' and increasing the averaging time parameter $\tau$, see Fig.~\ref{fig:prob-misdiagnosis}. Generally speaking, detecting errors that affect several error syndrome signals $I_k(t)$ at the same time are the most difficult to detect under continuous monitoring (e.g., error $X_2$ in the three-qubit bit flip code), and the performance of the latter critically depends on suppressing misdiagnosis of such errors~\cite{Atalaya2020}. 

\section{Effective open-system model for the logical qubit}
\label{sec:effectiveQME}
\subsection{Quantum master equation for the logical qubit}
\label{subsec:QME_logical}
In this subsection we develop an approximate evolution equation for the mixed logical state $\varrho_{\rm L}(t)$  that describes the combined action of {both} bit-flip errors and the above CQEC protocol, {and} the action of an applied time-dependent Hamiltonian. We are particularly interested in the limit of sufficiently small bit-flip error rates $\gamma_q$, where single bit-flip errors are the most probable, followed by two bit-flip errors, and so on. 
{In this regime there} are three different scenarios that can give rise to logical errors during the time evolution---a single misdiagnosed bit-flip error, spurious coherent evolution in an error subspace prior to a  correctly diagnosed bit-flip error, and two bit-flip errors that are misdiagnosed as one. We analyze {each of} these in turn below.

{For the following analysis} it is convenient to introduce a timestep $\Delta t$ such that
\begin{align}
\label{eq:Deltat-ineq}
t_{\rm det}\ll \Delta t,\;\; \Delta t\ll\gamma_q^{-1},\;\; \Delta t\ll\frac{\hbar}{|H_{\rm L}(t)|},\;\; {\rm and}\;\; \Delta t\ll t_{\rm op},
\end{align}
where $t_{\rm det}$ denotes the characteristic time to detect a bit-flip error by our CQEC protocol, and $t_{\rm op}$ is the operation time of the continuous implementation. Because of the second inequality of Eq.~\eqref{eq:Deltat-ineq}, we assume below that at most two bit-flip errors occur {within} each timestep $\Delta t$.   {We shall eventually send $\Delta t$ to zero, to obtain an effective evolution equation for the encoded density matrix $\varrho_{\rm L}(t)$.}

We consider first the scenario where a single bit-flip error that occurs in the time interval $(t,t+\Delta t)$ is misdiagnosed by the CQEC protocol. In this case, a wrong error correction operation is applied to one of the physical qubits: {this incorrect operation} transfers the system state to another error subspace, instead of {back} to the code space. For instance, if the {actual} error is $X_2$ {but} the diagnosed error syndrome is $(-1,+1)$ instead of $(-1,-1)$, the error correction operation {that will be applied} is $C_{\rm op}=X_1$ {instead of $C_{\rm op}=X_2$.} {This will  incorrectly} transfer the system state from error subspace $\mathcal{Q}_2$ to error subspace $\mathcal{Q}_3$, {resulting in a logical X error,} since $X_1X_2=X_3\,X_{\rm L}$ and $X_{\rm L}$ is the logical $X$ operator. The system state will be returned to the code space by the next iteration of the CQEC protocol if this iteration successfully diagnoses the new error syndrome. We {shall} assume that the probability to misdiagnose a bit-flip error is small {enough} that a series of two consecutive misdiagnoses is unlikely, {and the next iteration does {indeed} return the system state to the code space}. After completion of the next (successful) iteration of the CQEC protocol, the system state at the moment $t+\Delta t$ is equal to $X_{\rm L}\,\rho(t)\,X_{\rm L}$, which implies that the $2\times2$ logical density matrix at that moment is 
\begin{align}
\label{eq:quantum-proc-2}
\varrho_{\rm L}^{\textrm{scn-1}}(t+\Delta t) = \sigma_x\,\varrho_{\rm L}(t)\,\sigma_x. 
\end{align}
The probability of this scenario is given by
\begin{align}
\label{eq:quantum-proc-2-prob}
p_{\textrm{scn-1}}=\Delta t\,\Big(\gamma_1\, p_{\rm misdiag}^{(X_1)} + \gamma_2\, p_{\rm misdiag}^{(X_2)}+\gamma_3\,p_{\rm misdiag}^{(X_3)}\Big),
\end{align}
where $p_{\rm misdiag}^{(X_q)}$ denotes the probability to misdiagnose the bit-flip error $X_q$. We show in appendix \ref{sec:appendix_prob} that this probability depends exponentially on the parameters of the CQEC protocol, as is illustrated in Fig.~\ref{fig:prob-misdiagnosis}. This scenario {results in a contribution $\Delta \varrho_{\rm L}^{(1)}$ to the actual logical state $\varrho_{\rm L}(t+\Delta t)$ at the moment $t+\Delta t$ (Eq.~\eqref{eq:rhoL-deriv}), with}
\begin{align}
\label{eq:contribution-1}
\Delta \varrho_{\rm L}^{(1)} = p_{\textrm{scn-1}}\varrho_{\rm L}^{\textrm{scn-1}}(t+\Delta t).
\end{align}
Note that in the argument leading to Eq.~\eqref{eq:quantum-proc-2}, we have disregarded the  coherent evolution of the system state in the error subspaces because this leads to correction terms of the order of $(t_{\rm det}\, \Omega_0)^2 \ll(\Delta t\Omega_0)^2$. These can be neglected since we shall take the limit $\Delta t \rightarrow 0$ below and thus only need to keep terms up to first order in $\Delta t$ in Eq.~\eqref{eq:contribution-1}. 

The second scenario corresponds to the case of a single bit-flip error that is correctly diagnosed by the CQEC protocol. {The probability for this scenario is 
\begin{align}
\label{eq:quantum-proc-3-logical-prob}
p_{\textrm{scn-2}}^{(q)}= {\Delta t\gamma_q}(1 - p_{\rm misdiag}^{(X_q)}). 
\end{align}
}
In contrast to the {first} scenario, logical errors are {now due only} to spurious coherent evolution in the corresponding error subspace during the time that it takes to diagnose and correct the occurred error. Let us assume that the bit-flip error $X_q$ occurs at the instant $t'\in [t,t+\Delta t]$. 
{We shall} denote the time to detect such an error as $t_{\rm det}^{(q)}$, where the upper index $q$ indicates that {in general} the error detection time {may} depend on the bit-flip error type, $X_q=X_1$, $X_2$ or $X_3$. The system density matrix at the moment $t+\Delta t$ is 
\begin{equation}
\label{eq:quantum-proc-3}
\rho^{\textrm{scn-2}}(t+\Delta t) = {\frac{\gamma_q}{\mathcal{N}}}\int_{t}^{t+\Delta t - t_{\rm det}^{(q)}}\mathcal{V}_q(t')\,\rho(t)\,\mathcal{V}_q^{\dagger}(t')\,dt',
\end{equation}
where $\mathcal{N} = (\Delta t - t_{\rm det}^{(q)}) \gamma_q$ is a normalization factor such that ${\rm Tr}[\rho^{\textrm{scn-2}}(t+\Delta t)]=1$. The integral in Eq.~\eqref{eq:quantum-proc-3} evaluates the average over the error instant $t'$, 
\begin{align}\label{eq:V-def}
\mathcal{V}_q(t') =\, \mathcal{U}(t'+t_{\rm det}^{(q)},t+\Delta t)\,X_q\,\mathcal{U}(t',t'+t_{\rm det}^{(q)})\,X_q\,\mathcal{U}(t,t'),
\end{align}
and $\mathcal{U}(t_1,t_2)$ with $t_1\leq t_2$ denotes the unitary evolution operator associated to the  system Hamiltonian~\eqref{eq:HL}. If we read the right-hand side of Eq.~\eqref{eq:V-def} from right to left, the first $X_q$ operator accounts for the error that occurred, {and} the second $X_q$ operator accounts for the application of the error correction operation, which is $C_{\rm op}=X_q$ since the occurred error is correctly diagnosed. We now seek to approximate $\rho^{\textrm{scn-2}}(t+\Delta t)$ to first order in $\Delta t$. Because the integral in Eq.~\eqref{eq:quantum-proc-3} is over a time interval of duration approximately equal to $\Delta t$, we may write $\rho^{\textrm{scn-2}}(t+\Delta t) \approx  \mathcal{V}_q(t)\, \rho(t) \mathcal{V}_q^{\dagger}(t)$, where the integrand of Eq.~\eqref{eq:quantum-proc-3} has been evaluated at $t'=t$.
In addition, {the operator} $\mathcal{V}_q(t)$ may be replaced by its zero-order approximation in $\Delta t$: 

\begin{align}
\label{eq:Vq_zeroth}
\mathcal{V}_q(t) \approx e^{iH(t)t_{\rm det}^{(q)}}\,C_{\rm op}\,e^{-iH(t)t_{\rm det}^{(q)}}\,X_q \\ \nonumber
= e^{iH(t)t_{\rm det}^{(q)}}\,X_q\,e^{-iH(t)t_{\rm det}^{(q)}}\,X_q \\ \nonumber
= e^{iH(t)t_{\rm det}^{(q)}}\,e^{-iX_q H(t)X_qt_{\rm det}^{(q)}}.
\end{align}
Note that the $8\times8$ matrices $X_q H(t) X_q$ and $H(t)$ exhibit a similar block-diagonal matrix representation in the computational basis,
since both commute with the stabilizer generators. This block-diagonal structure consists of $2\times2$ diagonal submatrices for each subspace $\mathcal{Q}_\ell$. In particular, the  $2\times2$ diagonal submatrices of $X_qH(t)X_q$ and $H(t)$ 
corresponding to the code space are given by the spurious Hamiltonian $h_{\rm spurious}(t)$ and the logical Hamiltonian $h_{\rm L}(t)$ defined in Eqs.~\eqref{eq:hL} and \eqref{eq:hsp}, respectively. This implies that the $2\times2$ diagonal submatrix of $\mathcal{V}_q(t)$ that corresponds to the code space can be approximated as 
\begin{align}
\label{eq:Vq-def}
V_q {(t)} = \exp\left[{i t_{\rm det}^{(q)}\,h_{\rm L}(t)}\right]\,\exp\left[-i t_{\rm det}^{(q)}\,h_{\rm spurious}(t)\right]. 
\end{align}
Up to first order in $\Delta t$, the logical state at the moment $t+\Delta t$ is {then} given by 
\begin{align}
\label{eq:quantum-proc-3-logical}
[\varrho_{\rm L}^{\textrm{scn-2}}]_{q}(t+\Delta t) =  V_q{(t)}\,\varrho_{\rm L}(t)\,V_q^{\dagger}{(t)}.
\end{align}

Equation~\eqref{eq:Vq-def} provides an effective parameterization of the effective action of the logical error operation $V_q{(t)}$ due to spurious coherent evolution in an error subspace during detection of a single bit-flip error, in terms of the error detection time $t_{\rm det}^{(q)}$. We can estimate this time from the ``signal part'' of the filtered readout signals $\mathcal{I}_k(t)$,  {i.e.,} disregarding the noise. In this noiseless approximation, the error-detection time is the same for all bit-flip errors; i.e., $t_{\rm det}^{(q)}=t_{\rm det}$, so we {may} consider  {a} particular case.  {Let us consider} the bit-flip error $X_{2}$. If we apply the CQEC protocol to the ``signal part'' of the filtered readout signals, the error $X_2$ will be diagnosed when $\langle \mathcal{I}_k(t_{\rm err}+t_{\rm det})\rangle=\Theta_1$ {for $k= 1,2$.} From this {condition} and Eq.~\eqref{eq:Ik_mean}, we obtain   the error-detection time 
\begin{align}
\label{eq:t_det}
t_{\rm det}^{(q)} =t_{\rm det}= \tau\ln\left[\frac{2}{1+\Theta_1} \right], \, q=1,2,3. 
\end{align}
More generally, the presence of noise in the filtered readout signals {will make} the error-detection times random. For simplicity, {and to obtain analytic estimates, we shall} assume in this work that they are deterministic and given by Eq.~\eqref{eq:t_det}. 

The contribution of this scenario to the logical state $\varrho_{\rm L}(t+\Delta t)$ at the moment $t+\Delta t$ is (see Eq.~\eqref{eq:rhoL-deriv})
\begin{align}
\label{eq:contribution-2}
\Delta \varrho_{\rm L}^{(2)} = \sum_{q=1,2,3} p_{\textrm{scn-2}}^{(q)}\,  [\varrho_{\rm L}^{\textrm{scn-2}}]_{q}(t+\Delta t).
\end{align}
\begin{figure}[t!]
\centering
\includegraphics[width=\linewidth, trim =0.75cm 3.25cm 2.5cm 1cm,clip=true]
{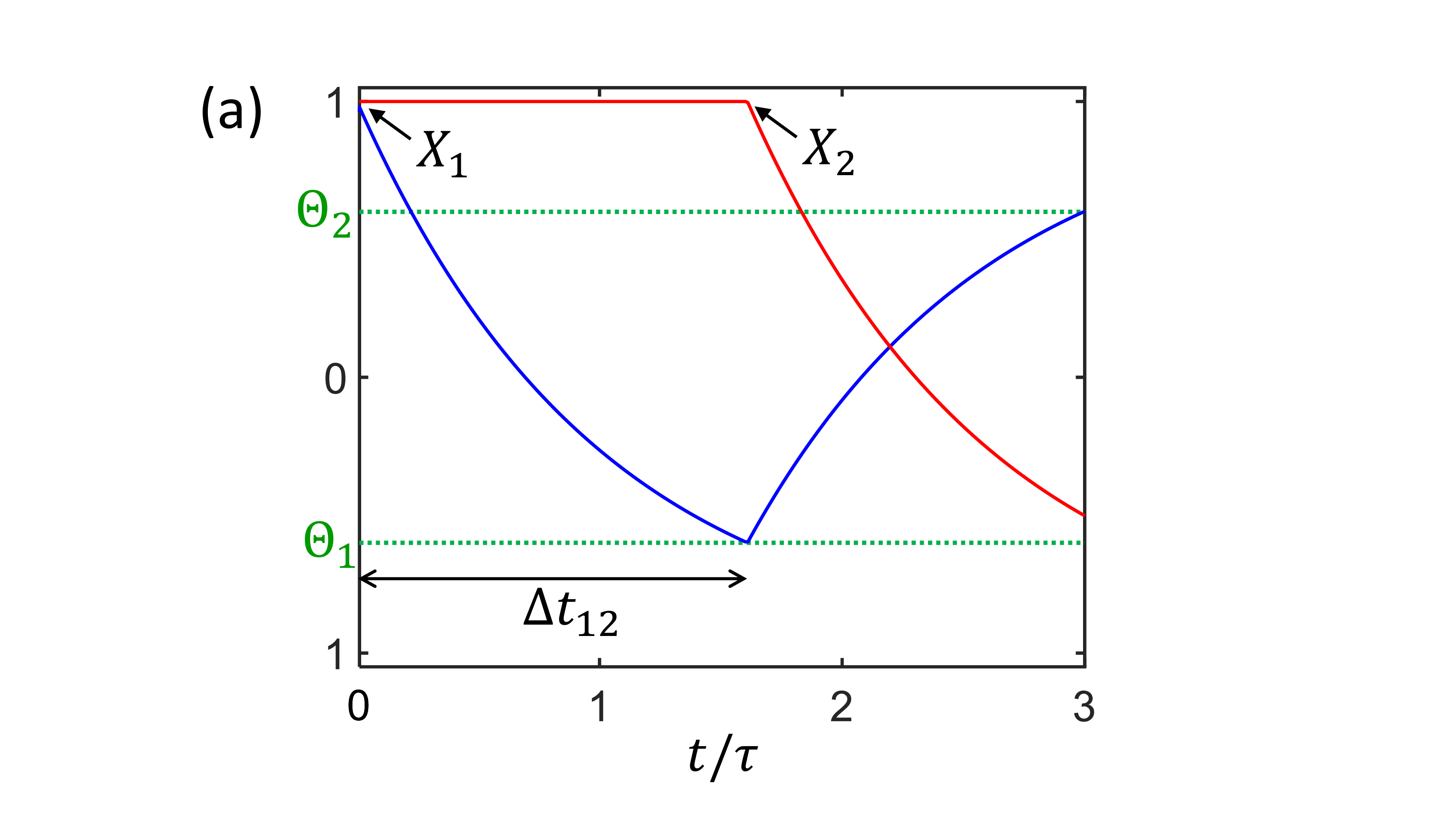}
\includegraphics[width=\linewidth, trim =0.75cm 3.25cm 2.5cm 2cm,clip=true]{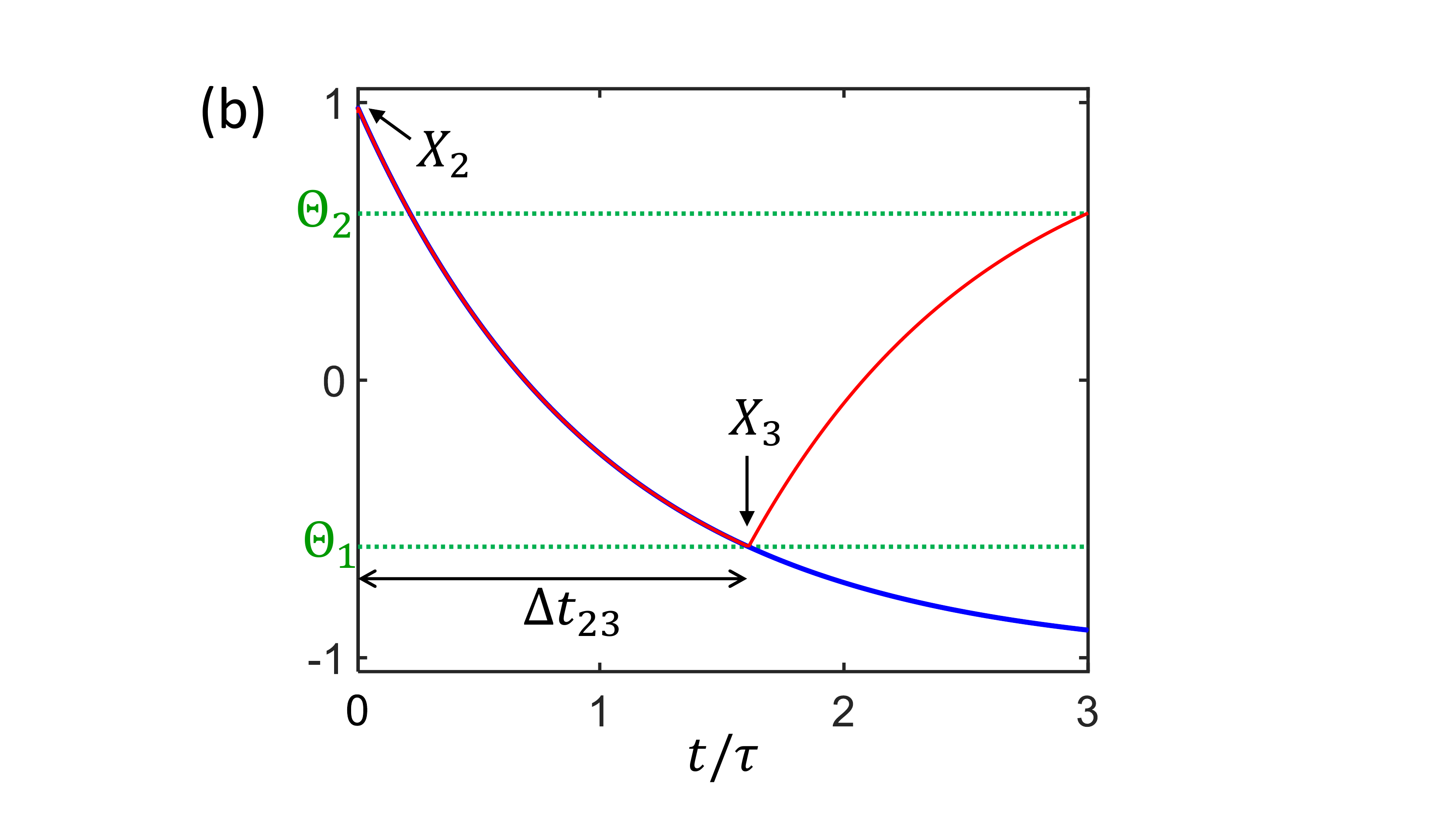}
\includegraphics[width=\linewidth, trim =0.75cm 1cm 2.5cm 2cm,clip=true]{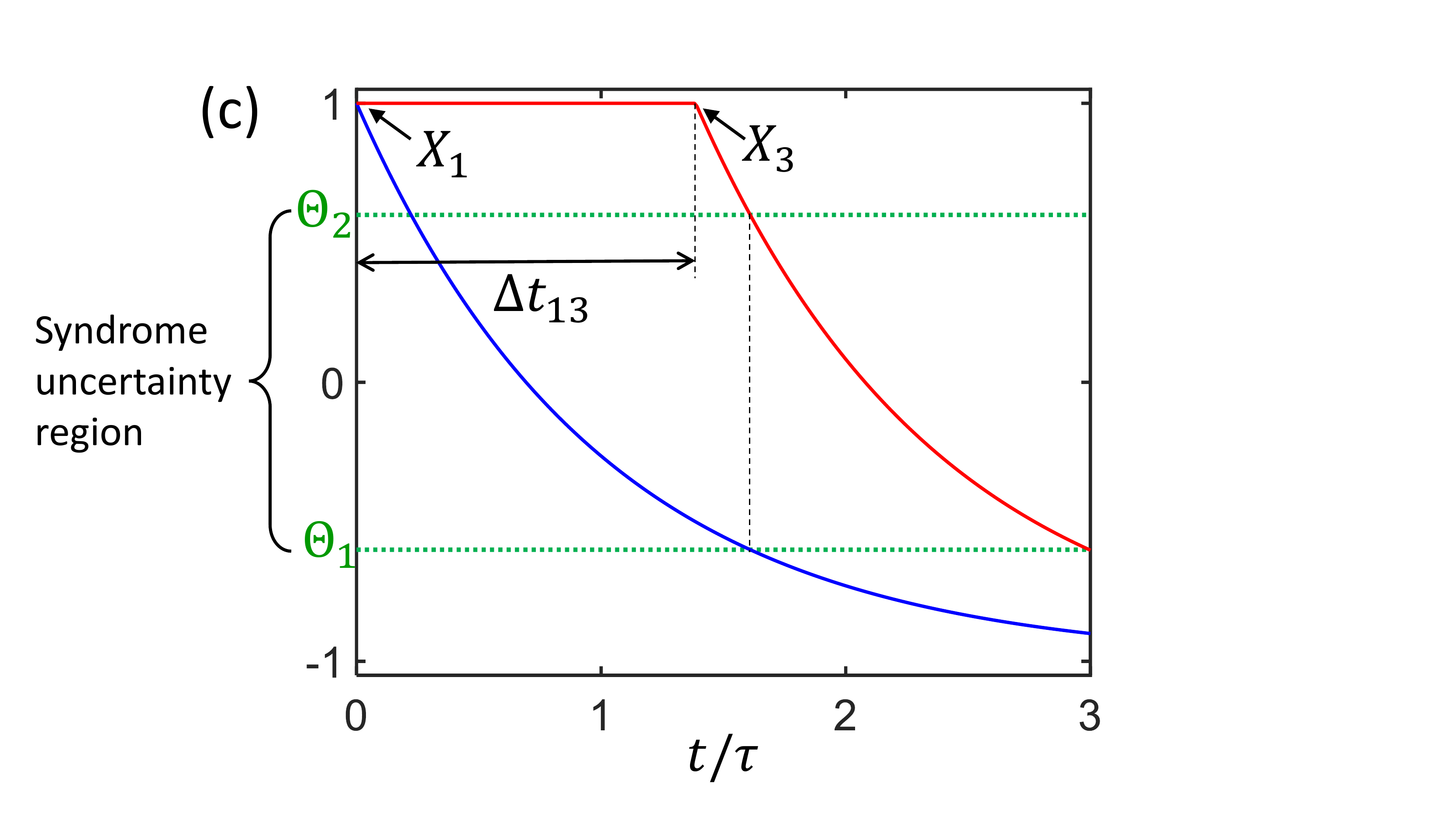}
\caption{Time windows $\Delta t_{qq'}$ for two consecutive bit-flip errors $X_q$ (first) and $X_{q'}$ (second) to be diagnosed as a single (false) error $X_{q''}$, in the noiseless approximation. 
{This is scenario three of the main text.} The first error ($X_q$) occurs at the moment $t=0$. The solid lines depict the ``signal part'' of the filtered readout signals $\mathcal{I}_1(t)$ [blue lines] and $\mathcal{I}_2(t)$ [red lines]. In panels (a) and (b), the window times $\Delta t_{12}$ and $\Delta t_{23}$ are both equal to the time that the blue solid line [i.e., $\langle \mathcal{I}_1(t)\rangle$, given in Eq.~\eqref{eq:Ik_mean} of the main text] takes to cross the lower error threshold $\Theta_1$. In panel (c), the window time $\Delta t_{13}$ is given by the time that the blue solid line spends inside the ``syndrome uncertainty region''. Explicit formulas for the windows times $\Delta t_{qq'}$ are given in Eq.~\eqref{eq:window-times} of the main text. If the second error occurs within the time window $t=\Delta t_{qq'}$, the depicted two-error combinations are diagnosed in the noiseless approximation as $X_{q''}=X_3$ [panel (a)], $X_{1}$ [panel (b)] and $X_2$ [panel (c)].}
\label{fig:fig1}
\end{figure}

The third scenario is the case of two errors that occur sufficiently close in time that they are not individually diagnosed by the CQEC protocol; instead, the protocol  diagnoses {a different} (false) error. {Now it} is clear that if two consecutive errors occur sufficiently far apart in time, both errors will be correctly diagnosed. On the other hand, if these errors occur sufficiently close in time, the CQEC protocol can fail, since our protocol determines the error syndrome from the filtered readout signals $\mathcal{I}_k(t)$, which are slow and take some time (proportional to the averaging time parameter $\tau$) to exit the ``syndrome uncertainty region'', as evident in Fig.~\ref{fig:filtered-signals-X2-err}. Let us denote  $\Delta t_{{q}{q'}}$ as the time window {in which} two consecutive errors, first $X_q$ and then $X_{q'}$, are misdiagnosed as the false error $X_{q''}$ ($q\neq q'\neq q''$). Neglecting spurious coherent evolution in the error subspaces, application of the wrong error correction operation $C_{\rm op}=X_{q''}$ effectively induces a logical $X$ operation on the system state $\rho(t)$ since 
$C_{\rm op}X_{q'}X_{q} = X_{q''}X_{q'}X_{q}=X_{\rm L}$, and then the logical density matrix changes from 
$\varrho_{\rm L}(t)$ to 
\begin{align}
\label{eq:quantum-proc-4-logical}
\varrho^{\textrm{scn-3}}_{\rm L}(t+\Delta t) = \sigma_x\, \varrho_{\rm L}(t)\, \sigma_{x}
\end{align}
at the moment $t+\Delta t$ {(see also Eq.~\eqref{eq:quantum-proc-2})}. The probability for this scenario is given by 
\begin{align}
\label{eq:quantum-proc-4-logical-prob}
p_{\textrm{scn-3}} = 2\left(\gamma_1\gamma_2 \Delta t_{12} + \gamma_2\gamma_3\Delta t_{23} + \gamma_1\gamma_3\Delta t_{13}\right)\Delta t,
\end{align}
where the time windows $\Delta t_{12}$, $\Delta t_{23}$  and $\Delta t_{13}$ can be easily evaluated in the noiseless approximation, {by an analogous procedure to that above for $t_{\rm det}^{(q)}$ and using Fig.~\eqref{fig:fig1}. This yields} 
\begin{align}
\label{eq:window-times}
\Delta t_{12}=\Delta t_{23} = \tau \ln\!\left[\frac{2}{1+\Theta_1}\right]\;{\rm and}\; \Delta t_{13}=\tau \ln\!\left[\frac{1+\Theta_2}{1+\Theta_1}\right].
\end{align}
The factor of 2 in Eq.~\eqref{eq:quantum-proc-4-logical-prob} is due to the fact that the time window $\Delta t_{qq'}$ is the same as $\Delta t_{q'q}$, which is the corresponding time window for the case where the error $X_{q'}$ occurs before the error $X_q$. The contribution of this scenario to the logical state $\varrho_{\rm L}(t+\Delta t)$ at the moment $t+\Delta t$ is (see Eq.~\eqref{eq:rhoL-deriv})
\begin{align}
\label{eq:contribution-3}
\Delta \varrho_{\rm L}^{(3)} = p_{\textrm{scn-3}}\, \varrho_{\rm L}^{\textrm{scn-3}}(t+\Delta t).
\end{align}

Finally, if none of the above  three scenarios occur, the logical state at the moment $t+\Delta t$ is given by the time evolved state under the logical Hamiltonian $h_{\rm L}(t)$ of Eq.~\eqref{eq:hL} and is equal to 
\begin{align}
\label{eq:quantum-proc-1}
\varrho_{\rm L}^{\textrm{scn-0	}}(t+\Delta t) = \varrho_{\rm L}(t)-i[h_{\rm L}(t),\varrho_{\rm L}(t)]\Delta t, 
\end{align}
where 
we have disregarded terms of order $(\Delta t)^2$. 

The logical state at the moment $t + \Delta t$ that takes into account {all of} the above four  scenarios {is then given by}
\begin{align}
\label{eq:rhoL-deriv}
&\varrho_{\rm L}(t+\Delta t) =\nonumber \\
&\hspace{0.25cm}\Big(1 - p_{\textrm{scn-1}} - \sum_{q=1,2,3} p_{\textrm{scn-2}}^{(q)} - p_{\textrm{scn-3}}\Big) \varrho_{\rm L}^{\textrm{scn-0}}(t+\Delta t) + \nonumber \\
&\hspace{0.25cm} p_{\textrm{scn-1}}\,\varrho_{\rm L}^{\textrm{scn-1}}(t+\Delta t) + \sum_{q=1,2,3}\Big\{ p_{\textrm{scn-2}}^{(q)}\,[\varrho_{\rm L}^{\textrm{scn-2}}]_{q}(t+\Delta t)\Big\} + \nonumber \\
&\hspace{0.25cm}p_{\textrm{scn-3}}\,\varrho_{\rm L}^{\textrm{scn-3}}(t+\Delta t).
\end{align}
Inserting the approximations Eqs.~\eqref{eq:quantum-proc-2}--\eqref{eq:quantum-proc-2-prob},~\eqref{eq:quantum-proc-3-logical}--\eqref{eq:quantum-proc-3-logical-prob},~\eqref{eq:quantum-proc-4-logical}--\eqref{eq:quantum-proc-4-logical-prob}  and~\eqref{eq:quantum-proc-1} into Eq.~\eqref{eq:rhoL-deriv} and {then} taking the limit $\Delta t\to 0$, we obtain the following effective evolution equation for the logical state $\varrho_{\rm L}(t)$: 
\begin{align}
\label{eq:rhoL-eom}
\dot{\varrho}_{\rm L} =&\, -i[h_{\rm L}(t),\varrho_{\rm L}] + \Gamma_{\rm L}\,\big[\sigma_x\,\varrho_{\rm L}\, \sigma_x - \varrho_{\rm L}\big] + \nonumber \\
&\,\sum_{q=1,2,3}\gamma_q \Big(1-p_{{\rm misdiag}}^{(X_q)}\Big)\big[ V_{q}\,\varrho_{\rm L}\, V_q^{\dagger}  - \varrho_{\rm L}\big]. 
\end{align} 
Here  
\begin{align}
\label{eq:GL}
\Gamma_{\rm L}=&\;\gamma_1\, p_{\rm misdiag}^{(X_1)} + \gamma_2\, p_{\rm misdiag}^{(X_2)}+\gamma_3\,p_{\rm misdiag}^{(X_3)} + \nonumber \\
&\;2\,\big(\gamma_1\gamma_2 \Delta t_{12} + \gamma_2\gamma_3\Delta t_{23} + \gamma_1\gamma_3\Delta t_{13}\big) 
\end{align}
is now the logical $X$ error rate for quantum memory operations~\cite{Atalaya2017,Dressel2019}. 
The initial condition for Eq.~\eqref{eq:rhoL-eom} reads as 
\begin{equation}
\label{eq:rhoL-t0}
\varrho_{\rm L}(0) = |\psi_{\rm L}(0)\rangle \langle \psi_{\rm L}(0)|= \begin{bmatrix}|\alpha^2_{\rm L}(0)| & \alpha_{\rm L}(0)\beta^*_{\rm L}(0)\\\alpha^*_{\rm L}(0)\beta_{\rm L}(0) & |\beta^2_{\rm L}(0)| 
\end{bmatrix}.
\end{equation}

Equation~\eqref{eq:rhoL-eom} is the main result of this subsection. To the best of our knowledge, the last term at the right-hand side of Eq. \eqref{eq:rhoL-eom} has not been previously discussed in the context of QEC for quantum simulation or quantum annealing. This term quantifies {the} logical errors due to spurious coherent evolution in the error subspaces. 

\begin{figure}[t!]
\centering
\includegraphics[width=\linewidth, trim =0cm 0cm 1cm 0cm,clip=true]{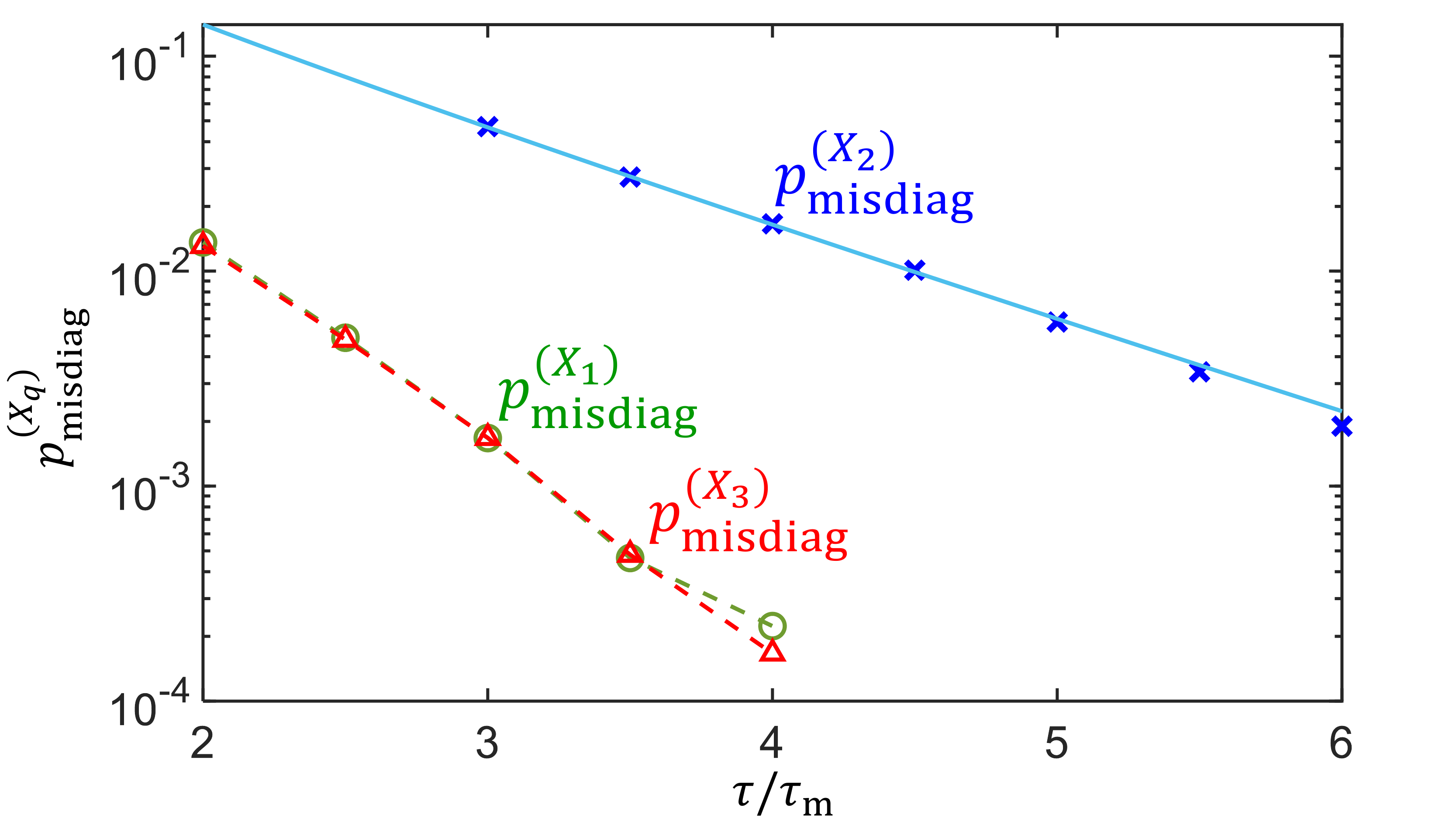}
\caption{Probability $p_{\rm misdiag}^{(X_q)}$ that the CQEC protocol misdiagnoses the bit-flip error $X_q$ as function of the measurement averaging time parameter $\tau$. The error thresholds are fixed at $\Theta_1=-0.54$ and $\Theta_2=0.8$. Symbols indicate numerical results and the solid line depicts the analytical formula~\eqref{eq:p-mis-2} of the main text with $c=1.607$. The dashed lines are guides to the eye. } 
\label{fig:prob-misdiagnosis}
\end{figure}

We now estimate the {probabilities} $p_{\rm misdiag}^{(X_q)}$ that the CQEC protocol misdiagnoses the bit-flip {errors} $X_q$. Note that the bit-flip errors $X_1$ and $X_3$ are equivalent in the three-qubit bit flip code. Thus we expect that $p_{\rm misdiag}^{(X_1)}=p_{\rm misdiag}^{(X_3)}$, 
which is numerically verified in Fig.~\ref{fig:prob-misdiagnosis}. 
Figure~\ref{fig:prob-misdiagnosis} also shows that for the range of values of interest of the measurement averaging time parameter $\tau\gtrsim 2\tau_{\rm m}$ (see Fig.~\ref{fig:opt-results-tau}), the probability to misdiagnose the $X_1$ or $X_3$ errors is much smaller than the probability to misdiagnose the $X_2$ error. Thus, we may not only assume that $p_{\rm misdiag}^{(X_1)}=p_{\rm misdiag}^{(X_3)}$, but we can also neglect these terms in Eqs.~\eqref{eq:rhoL-eom}--\eqref{eq:GL}, i.e., we can set
\begin{align}
\label{eq:p-mis-13}
p_{\rm misdiag}^{(X_1)}=p_{\rm misdiag}^{(X_3)}=0.
\end{align}
In addition, the probability $p_{\rm misdiag}^{(X_2)}$ to misdiagnose the error $X_2$ can be approximated as 
\begin{align}
p_{\rm misdiag}^{(X_2)} = \, c\,\frac{e^{-(\Theta_2-\Theta_1)^2\tau/2\tau_{\rm m}}}{(\Theta_2-\Theta_1)\sqrt{\tau/\tau_{\rm m}}},   
\label{eq:p-mis-2}
\end{align}
where the coefficient $c= 1.607$ is obtained from {the fit shown} in Fig.~\ref{fig:prob-misdiagnosis}. The exponential dependence of the probability $p_{\rm misdiag}^{(X_2)}$ on the parameters of the CQEC protocol is {derived} in appendix \ref{sec:appendix_prob}.

\begin{figure}[t!]
\centering
\includegraphics[width=\linewidth, trim =0cm 0cm 2.5cm 0cm,clip=true]{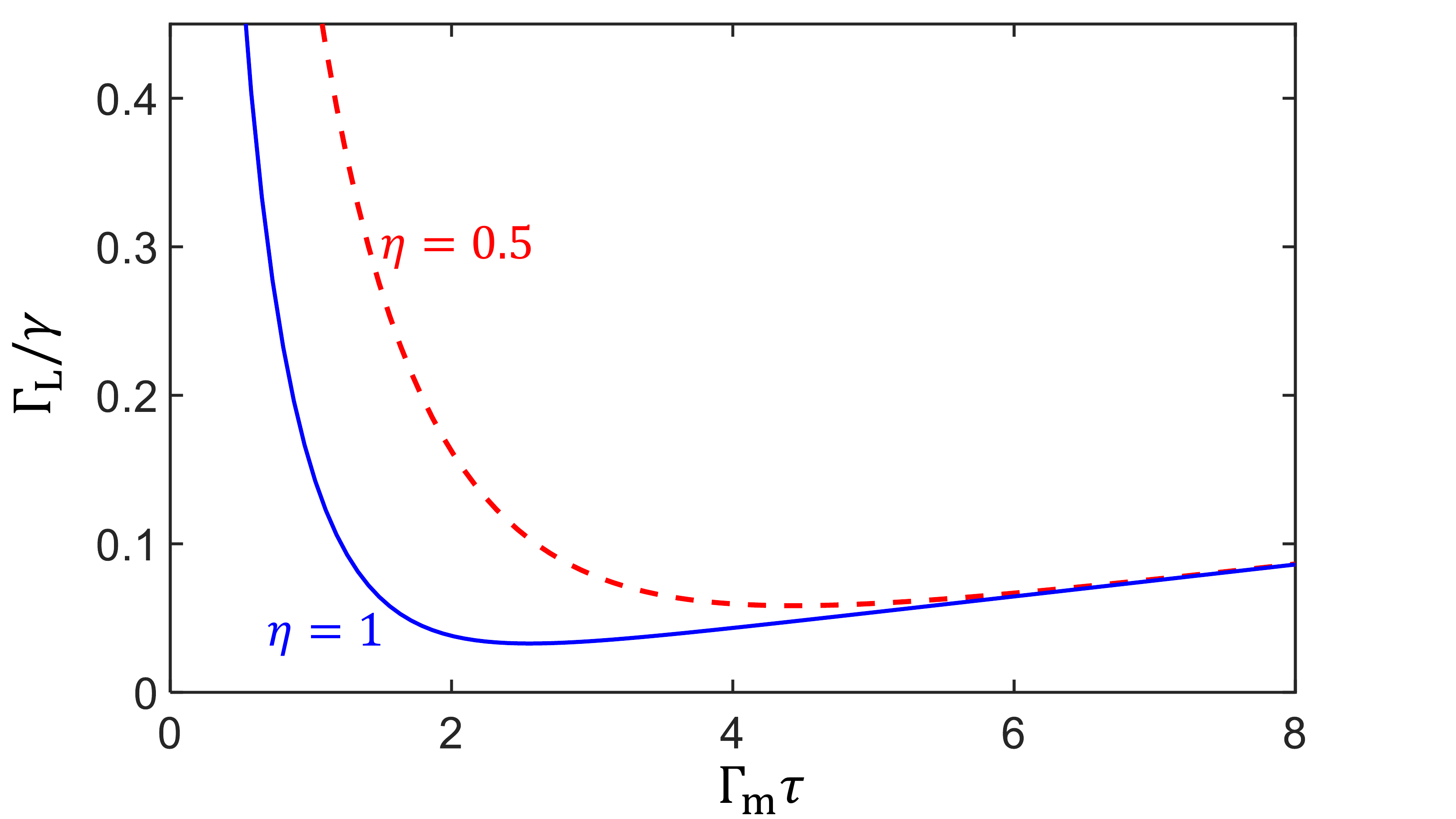}
\caption{Logical error rate $\Gamma_{\rm L}$ as function of the measurement averaging time parameter $\tau$. The curves depict formula Eq.~\eqref{eq:GL-v2} of the main text, evaluated for bit-flip error rates $\gamma_q=\gamma=1.25\times10^{-3}\Gamma_{\rm m}$ ($q=1,2,3$),  error threshold parameters $\Theta_1=-0.54$ and $\Theta_2=0.8$, and quantum efficiencies $\eta=0.5$ (dashed curve) and $\eta=1$ (solid curve). $\Gamma_{\rm m}$ denotes measurement strength from continuous measurement.} 
\label{fig:GL-vs-tau}
\end{figure}

Using {these estimates for $p_{\rm misdiag}^{(X_q)}$}, the logical $X$ error rate formula Eq.~\eqref{eq:GL} can be {rewritten in terms of all relevant parameters} as 
\begin{align}
\label{eq:GL-v2}
\Gamma_{\rm L} =&\;1.607\,\gamma_2\frac{e^{-(\Theta_2-\Theta_1)^2\tau/2\tau_{\rm m}}}{(\Theta_2-\Theta_1)\sqrt{\tau/\tau_{\rm m}}}  + 2(\gamma_{1}\gamma_{2} + \gamma_2\gamma_3)\tau\times\nonumber\\
&\;\ln\left[\frac{2}{1+\Theta_1}\right] + 2\gamma_1\gamma_3\tau\ln\left[\frac{1+\Theta_2}{1+\Theta_1} \right].
\end{align}
{Note that $\Gamma_{\rm L}$ implicitly depends on the efficiency of the measurement, $\eta$, via the explicit dependence on measurement time $\tau_{\rm m} =1 /(2\Gamma_{\rm m}\eta$).  In a given experimental setup, the parameters $\tau, \Theta_1, \Theta_2$ would constitute a minimal set of tunable parameters.} 

Figure~\ref{fig:GL-vs-tau} shows the non-monotonic dependence of $\Gamma_{\rm L}$ on the time-averaging parameter $\tau$, for fixed values of the error threshold parameters $\Theta_1=-0.54$ and $\Theta_2=0.8$, and equal bit-flip error rates $\gamma_q=\gamma=1.25\times10^{-3}\Gamma_{\rm m}$. Note that, in the limit of relatively small $\tau$, the logical $X$ error rate increases exponentially because the SNR of the filtered readout signals decreases, leading to more frequent false diagnoses of $X_2$ errors. In this limit, the first term of Eq.~\eqref{eq:GL-v2} is dominant. In the opposite limit of relatively large $\tau$, the logical $X$ error rate increases linearly in $\tau$, due to misdiagnosis of two errors that occur sufficiently close in time. {We see that measurement inefficiency $\eta\leq1$ affects the logical error rate only for small averaging times $\tau$ and has no effect at large $\tau$.  This reflects the fact that while the mis-diagnosis of single qubit errors that dominates $\Gamma_{\rm L}$ at small $\tau$ depends on measurement efficiency via $\tau_{\rm m}$ (measurement time parameter), the mis-diagnosis of two errors occurring close in time was evaluated in the noiseless approximation and does not depend on $\eta$.}

The numerical calculations presented at the end of this subsection show that the effective open-system model for the logical qubit [Eq.~\eqref{eq:rhoL-eom}] together with the estimates Eqs.~\eqref{eq:GL-v2},~\eqref{eq:Vq-def}, \eqref{eq:p-mis-13} and \eqref{eq:p-mis-2} for the parameters $\Gamma_{\rm L}$ (logical $X$ error rate), $V_q(t)$ (logical error operation parameterized in terms of error-detection times $t_{\rm det}^{(q)}$, see Eq.~\eqref{eq:t_det}) and $p_{\rm misdiag}^{(X_q)}$ (probability to misdiagnose bit-flip error $X_q$) provide a good description for the true evolution of the logical state $\varrho_{\rm L}(t)$ that is encoded into the full system state $\rho(t)$, which evolves according to Eq.~\eqref{eq:rho-eom-no-errors-v2}.

\subsection{Final logical state fidelity} 

The figure of merit that we aim to maximize {under evolution due to a time-dependent Hamiltonian} is the final fidelity $\mathcal{F}$ between the target, Eq.~\eqref{eq:psiL}, and the true, Eq.~\eqref{eq:rhoL-def} logical states, defined as 
\begin{align}
\label{eq:F}
\mathcal{F} = \left\langle\psi_{\rm L}(t_{\rm op})|\varrho_{\rm L}(t_{\rm op})|\psi_{\rm L}(t_{\rm op}) \right\rangle.    
\end{align}
Note that the target state $\psi_{\rm L}(t_{\rm op})$ is exactly equal to the ground state of the problem Hamiltonian only in the adiabatic limit as $t_{\rm op} \rightarrow \infty$.

Using the effective evolution equation~\eqref{eq:rhoL-eom} for the logical state $\varrho_{\rm L}(t)$, we {can derive} the following analytical expression for the final logical state infidelity 
\begin{align}
\label{eq:F-sol}
&1-\mathcal{F} =\;\Gamma_{\rm L}\int_0^{t_{\rm op}} \left(1-\big|\langle\psi_{\rm L}(t)|\sigma_x|\psi_{\rm L}(t)\rangle\big|^2\right) dt+ \nonumber \\
&\sum_{q=1,2,3}\gamma_q\big(1-p^{(q)}_{\rm mis}\big)\int_0^{t_{\rm op}} \left(1-\big|\langle\psi_{\rm L}(t)|V_q|\psi_{\rm L}(t)\rangle\big|^2\right) dt, 
\end{align}
which is expressed in terms of the {coherent} evolution of the target logical state $|\psi_{\rm L}(t)\rangle$. The first term {on} the right-hand side of Eq.~\eqref{eq:F-sol} is the usual term in quantum memory, i.e., $\Gamma_{\rm L} t_{\rm op}$, generalized here to the case of a finite  and time-dependent logical Hamiltonian~\eqref{eq:hL}. {Note that the time integral accounts for the accumulated loss of fidelity due to logical {$X$} errors on the time-evolving logical state.}  The second term is due to the spurious coherent evolution in the error subspaces.  Note that this term is positive, i.e., contributes a finite infidelity, because the operator $V_q$, given in Eq.~\eqref{eq:Vq-def}, is unitary. Equation~\eqref{eq:F-sol} is the main result of this subsection. 

{To obtain this result in Eq.~\eqref{eq:F-sol}, we have applied} the jump/no-jump method in Eq.~\eqref{eq:rhoL-eom} to estimate $\varrho_{\rm L}(t_{\rm op})$ as follows: 
\begin{align}
\label{eq:rhoL-sol}
&\varrho_{\rm L}(t_{\rm op}) = \left[1-\Gamma_{\rm L}t_{\rm op}-\tilde\gamma_{\rm tot} t_{\rm op}\right]|\psi_{\rm L}(t_{\rm op})\rangle\langle \psi_{\rm L}(t_{\rm op})|\,+ \nonumber\\
&\;\Gamma_{\rm L}\int_0^{t_{\rm op}}dt'\, U_{\rm L}(t',t_{\rm op})\,\sigma_x\,|\psi_{\rm L}(t')\rangle \langle\psi_{\rm L}(t')|\,\sigma_x\, U_{\rm L}^\dagger(t',t_{\rm op}) \,+ \nonumber \\
&\sum_{q}\tilde\gamma_q\!\int_0^{t_{\rm op}}\!dt'\, U_{\rm L}(t',t_{\rm op})\,V_q\,|\psi_{\rm L}(t')\rangle \langle\psi_{\rm L}(t')|\,V_q^{\dagger}\, U_{\rm L}^\dagger(t',t_{\rm op}), 
\end{align}
where $\tilde\gamma_q = \gamma_q\big(1-p_{\rm mis}^{(q)}\big)$, $\tilde\gamma_{\rm tot}=\tilde\gamma_1+\tilde\gamma_2+\tilde\gamma_3$ and $U_{\rm L}(t_1,t_2)$ is the unitary evolution operator associated to the {error free} Schr\"odinger evolution equation~\eqref{eq:psiL-eom}. When the jump/no-jump approach is applied to Eq.~\eqref{eq:rhoL-eom}, we see that logical errors come in two forms.  First, the usual logical $X$ errors that change the logical wavefunction from $|\psi_{\rm L}(t)\rangle$ to $\sigma_x|\psi_{\rm L}(t)\rangle$ (second term in Eq.~\eqref{eq:rhoL-sol}). These occur at the logical $X$ error rate $\Gamma_{\rm L}$ given in Eq.~\eqref{eq:GL-v2}. Second, logical errors that are characterized by the logical error operation $V_q$ given in Eq.~\eqref{eq:Vq-def} (third term in Eq.~\eqref{eq:rhoL-sol}). This new type of logical errors is specifically due to spurious coherent evolution in the error subspaces. Such errors change the logical wavefunction from $|\psi_{\rm L}(t)\rangle$ to $V_q|\psi_{\rm L}(t)\rangle$ and occur at the rate $\tilde\gamma_q$. In addition, we also have the coherent no-jump evolution that is described by the unitary evolution operator $U_{\rm L}(t_1,t_2)$ (first term in Eq.~\eqref{eq:rhoL-sol}). Note that in Eq.~\eqref{eq:rhoL-sol} we have disregarded cases where there are more than one logical error occurrences during the continuous operation duration $t_{\rm op}$.  This approximation is valid in the limit of small bit-flip error rates $\gamma_q$ that we assume here.

\section{Optimization and benchmarking}
\label{sec:Optimization}
\subsection{Optimization of the CQEC protocol} 

In this subsection we derive the optimal parameters ($\Theta_1^{\rm opt}$, $\Theta_2^{\rm opt}$ and $\tau^{\rm opt}$) of the CQEC protocol that maximize the final logical state fidelity~\eqref{eq:F}. The optimization will be specific to a particular choice of Hamiltonian evolution, i.e., to the choice of $h_{\rm L}(t)$, since the temporal dependence of $|\psi_{\rm L}(t)\rangle$ is determined by this. We shall consider here both quantum memory, $h_{\rm L}(t) =0$, and quantum annealing with the specific choice of a linear schedule. In the latter case the logical Hamiltonian $h_{\rm L}(t)$ is given by Eq.~\eqref{eq:hL} with the coefficients $a(t)$ and $b(t)$ equal to
\begin{align}
\label{eq:a-b}
a(t) = 1-\frac{t}{t_{\rm op}}\;\;\;{\rm and}\;\;\;b(t) = \frac{t}{t_{\rm op}}.
\end{align}
Following convention, we shall refer to the Hamiltonian component multiplied by $b(t)$ as the problem Hamiltonian, i.e., the final Hamiltonian at $t=t_{\rm op}$. In the context of quantum annealing we shall assume that the adiabatic limit holds, $t_{\rm op}\Omega_0\gg1$, so that we may approximate the target logical wavefunction as the instantaneous ground state, which reads as
\begin{align}
\label{eq:psiL-sol}
|\psi_{\rm L}(t)\rangle = \cos\left(\frac{\theta(t)}{2}\right)|0_{\rm L}\rangle +  \sin\left(\frac{\theta(t)}{2}\right)|1_{\rm L}\rangle,
\end{align}
where $\theta(t) = \arctan\big(a(t)/b(t)\big)$. 

Inserting Eq.~\eqref{eq:psiL-sol} into Eq.~\eqref{eq:F-sol}, we obtain for the final logical state infidelity
\begin{align}
\label{eq:F-QA}
1-\mathcal{F} = \frac{\Gamma_{\rm L}t_{\rm op}}{2} + \!\sum_{q=1,2,3}\!\frac{3\pi-8}{54}\!\left(1-p_{\rm mis}^{(q)}\right)\!\left[\Omega_0 t_{\rm det}^{(q)}\right]^2\!\gamma_qt_{\rm op},  
\end{align}
which is the cost function that we use in the optimization procedure. We emphasize that the result~\eqref{eq:F-QA} applies to the special case of quantum annealing with a linear schedule, and note {also} that we have included terms up to second order in $\Omega_0 t_{\rm det}^{(q)}$. The linear dependence of the infidelity Eq.~\eqref{eq:F-QA} on the operation time $t_{\rm op}$ is valid for small infidelities, less than 0.1 for the parameters used in this work. 
More generally, the final infidelity $1-\mathcal{F}$ for arbitrary {annealing schedule parameters} $a(t)$ and $b(t)$ can also be easily obtained, as long as these coefficients also satisfy the adiabatic condition $|\dot a(t)|, |\dot b(t)|\ll\Omega_0$.  This can be accomplished by writing the first integrand of Eq.~\eqref{eq:F-sol} as $1 - |\langle\psi_{\rm L}(t)|\sigma_x|\psi_{\rm L}(t)\rangle|^2=\cos^2\big(\theta(t)\big)$ and the second integrand as $1-\big|\langle\psi_{\rm L}(t)|V_q|\psi_{\rm L}(t)\rangle\big|^2= \sin^2\big(\theta(t)-\tilde\theta(t)\big)\sin^2\big(\tilde\Omega(t) t_{\rm det}^{(q)}\big)$, where $\tilde\theta(t) = \arctan\big(3a(t)/b(t)\big)$ and $\tilde\Omega(t) = \Omega_0\sqrt{a^2(t) + b^2(t)/9}$ is half the instantaneous energy gap of the spurious Hamiltonian~\eqref{eq:hsp}. {To obtain a final numerical value for the infidelity, the integrals of Eq.~\eqref{eq:F-sol} would have to be  evaluated numerically for evolution under a specific annealing Hamiltonian.} 

\begin{figure}[t!]
\centering
\includegraphics[width=\linewidth, trim =0.5cm 0cm 1cm 0cm,clip=true]{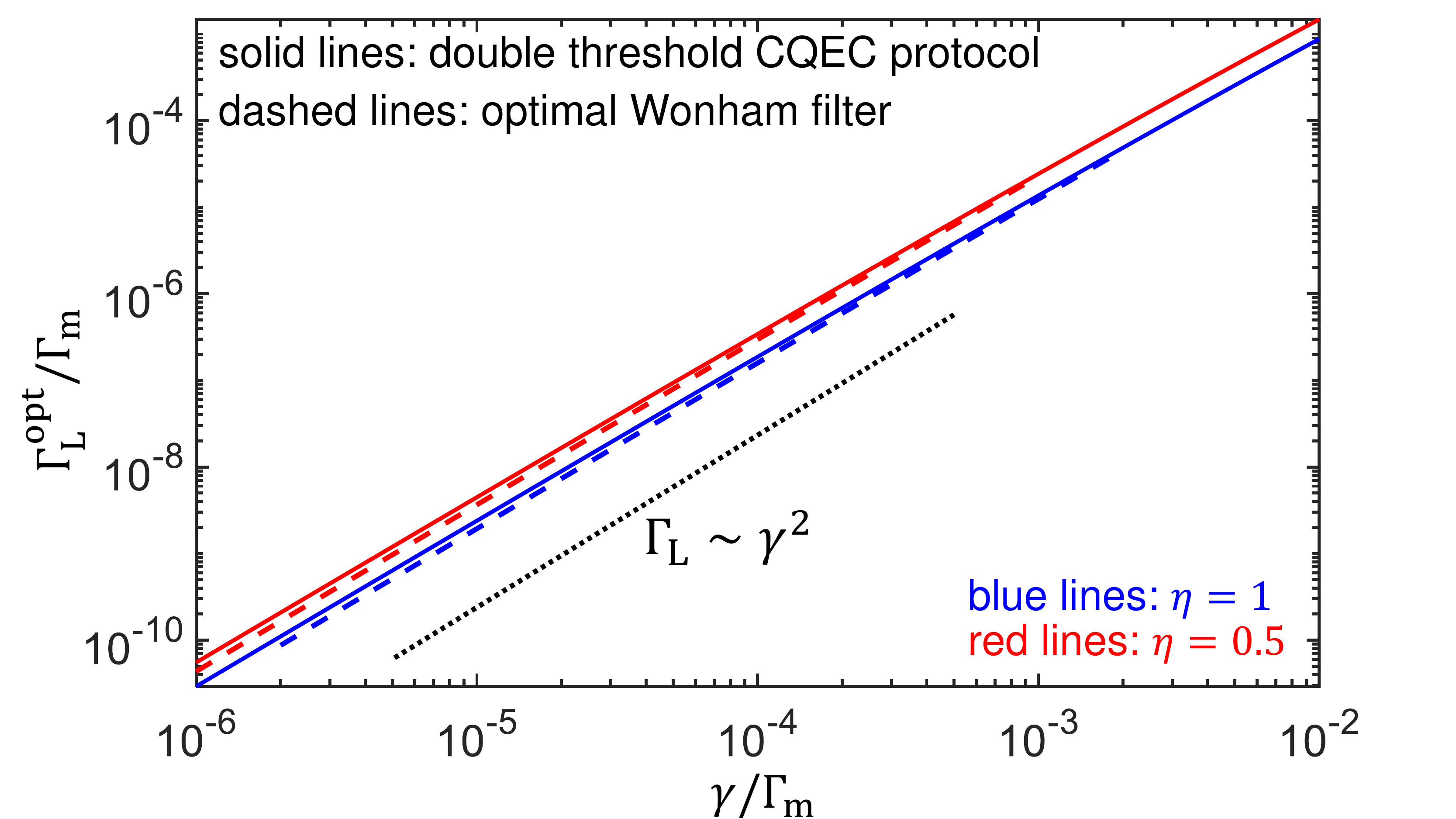}
\caption{Optimized logical $X$ error rate $\Gamma^{\rm opt}_{\rm L}$ for quantum memory operation. Solid red and blue lines depict $\Gamma_{\rm L}^{\rm opt}$, obtained using our double threshold CQEC protocol (Eq.~\eqref{eq:GL-v2}), and the dashed red and blue lines depict the logical $X$ error rate, $\Gamma_{\rm L}^{\rm Wonham} = 3\gamma^2\tau_{\rm m}\ln (2/\gamma\tau_{\rm m})$, for the {linear variant of the} optimal Wonham filter~\cite{Dressel2019}. $\Gamma_{\rm m}$ and $\tau_{\rm m}=(2\Gamma_{\rm m}\eta)^{-1}$ denote the measurement strength and ``measurement time'' from continuous measurement, respectively. Red and blue lines depict results for quantum efficiencies $\eta=0.5$ and 1, respectively. The dotted line depicts the quadratic scaling $\Gamma_{\rm L}\sim \gamma^2$.} 
\label{fig:GL-vs-gamma}
\end{figure}

Since when $\Omega_0=0$ the system operates as a quantum memory, we first present results for the logical error rate before presenting results for quantum annealing ($\Omega_0\neq0$). For simplicity, we discuss {here} the case of equal bit-flip error rates 
\begin{align}
\gamma_1=\gamma_2=\gamma_3=\gamma.
\end{align}

In quantum memory {operation}, the final logical state infidelity $1-\mathcal{F}$ is given by the first term of Eq.~\eqref{eq:F-sol} since the second term exactly vanishes. Assuming that the initial logical state is $|\psi_{\rm L}(0)\rangle = |0_{\rm L}\rangle$ or $|1_{\rm L}\rangle$, we find that $1-\mathcal{F}$ reduces to $\Gamma_{\rm L} t_{\rm op}$, because the target logical evolution is trivial in the quantum memory case ($|\psi_{\rm L}(t)\rangle$ is constant). In addition, we may assume that the operation duration $t_{\rm op}$ is fixed. 
{Then} minimization of the final infidelity in quantum memory is equivalent to optimization of the logical $X$ error rate $\Gamma_{\rm L}$  in Eq.~\eqref{eq:GL-v2}. 

Figure~\ref{fig:GL-vs-gamma} depicts the optimized value of the logical $X$ error rate $\Gamma_{\rm L}^{\rm opt}$ {for quantum memory}. We find that {this logical error rate} scales approximately quadratically with the error rate $\gamma$: 
\begin{align}
\label{eq:GL-fitting-QM}
\Gamma_{\rm L}^{\rm opt} \approx&\; 27.3208\left(\gamma\Gamma^{-1}_{\rm m}\right)^{1.897}\Gamma_{\rm m}\;\;\;\;\;\;(\eta=0.5),\nonumber\\
\approx&\; 15.7580\left(\gamma\Gamma^{-1}_{\rm m}\right)^{1.904}\Gamma_{\rm m}\;\;\;\;\;\;(\eta=1). 
\end{align}
The numerical factors and exponents in the above equation are obtained from fitting for $\gamma\in[10^{-6}\Gamma_{\rm m},10^{-4}\Gamma_{\rm m}]$. The approximate quadratic scaling of $\Gamma_{\rm L}^{\rm opt}$ {with} $\gamma$ indicates that the double threshold CQEC protocol 
{is both effective and accurate} in diagnosing single bit-flip errors. Figure~\ref{fig:GL-vs-gamma} also shows the logical $X$ error rate for the linear variant of the optimal Wonham filter, $\Gamma_{\rm L}^{\rm Wonham} = 3\gamma^2\tau_{\rm m}\ln(2/\gamma\tau_{\rm m})$, that was obtained in Ref.~\cite{Dressel2019}. We point out that our optimized logical error rate $\Gamma_{\rm L}^{\rm opt}$ is very close to that of the linear variant of the optimal Wonham filter.  

In addition, we find that the discrete and continuous operations can exhibit similar performance if the cycle time $\Delta t_{\rm cycle}$ from the discrete operation is related to the strength $\Gamma_{\rm m}$ of the continuous measurements as follows:
\begin{align}
\label{eq:cont-discrete-performances}
t_{\rm cycle} \approx&\; 9.1069\gamma^{-1}\left(\gamma\Gamma^{-1}_{\rm m}\right)^{0.897} \;\;\;\;(\eta=0.5),\nonumber\\
\approx &\; 5.2527 \gamma^{-1}\left(\gamma\Gamma^{-1}_{\rm m}\right)^{0.904} \;\;\;\;(\eta=1).
\end{align}
The above results are obtained from the relation $\Gamma^{\rm opt}_{\rm L}=\Gamma_{\rm L}^{\rm disc}$, where $\Gamma_{\rm L}^{\rm disc} = 3\gamma^2 \Delta t_{\rm cycle}$ is the logical $X$ error rate for the discrete operation.

We now discuss the results of optimizing the {double threshold error detection parameters} in the {specific case} of quantum annealing. To quantify the effectiveness of the CQEC protocol in correcting logical errors, we introduce here the {ratio of the infidelity for an unencoded calculation, to the infidelity for an encoded calculation using the optimized double-thresholding parameters. This ratio, $R(\gamma,\Omega_0)$, is defined for a given error rate and annealing Hamiltonian, which we shall denote here only by its strength $\Omega_0$. Specifically,}
\begin{align}
\label{eq:reduction-factor}
R(\gamma,\Omega_0) = \frac{1-\mathcal{F}_{\rm unenc}}{1 - \mathcal{F}_{\rm opt}},
\end{align}
where $\mathcal{F}_{\rm opt}$ is the  value of the final logical state fidelity Eq.~\eqref{eq:F-QA}, {optimized with respect to $\tau, \Theta_1, \Theta_2$ (see below)}, and 
\begin{align}
\label{eq:infidelity-unencoded}
\mathcal{F}_{\textrm{unenc}} = \langle \psi_{\rm L}(t_{\rm op})|\rho_{\rm unenc}(t_{\rm op}) |\psi_{\rm L}(t_{\rm op})\rangle
\end{align}
is the fidelity between the final target logical state $|\psi_{\rm L}(t_{\rm op})\rangle$ and the final state $\rho_{\rm unenc}(t_{\rm op})$ of an unencoded qubit subject to bit-flip errors with rate $\gamma$ and coherent evolution due to a Hamiltonian $h_{\rm L}(t)$. {We refer to $R(\gamma,\Omega_0) $ as the ``reduction factor" of the final logical state infidelity, since by construction it shows by how much the infidelity is reduced by encoding together with optimization of the error detection.}  

It is easy to see, using the jump/no-jump method, that the unencoded final infidelity $1-\mathcal{F}_{\rm unenc}$ can be estimated by the first term of Eq.~\eqref{eq:F-sol} with $\Gamma_{\rm L}$ replaced by $\gamma$. For the quantum annealing problem with a linear schedule, we obtain 
\begin{align}
\label{eq:infidelity-unencoded-result}
1-\mathcal{F}_{\textrm{unenc}} = \frac{\gamma t_{\rm op}}{2}, 
\end{align}
which is linear in the operation time $t_{\rm op}$ for sufficiently small physical error rate $\gamma$.

\begin{figure}[t!]
\centering
\includegraphics[width=\linewidth, trim =0.25cm 0.15cm 1.25cm 0.25cm,clip=true]
{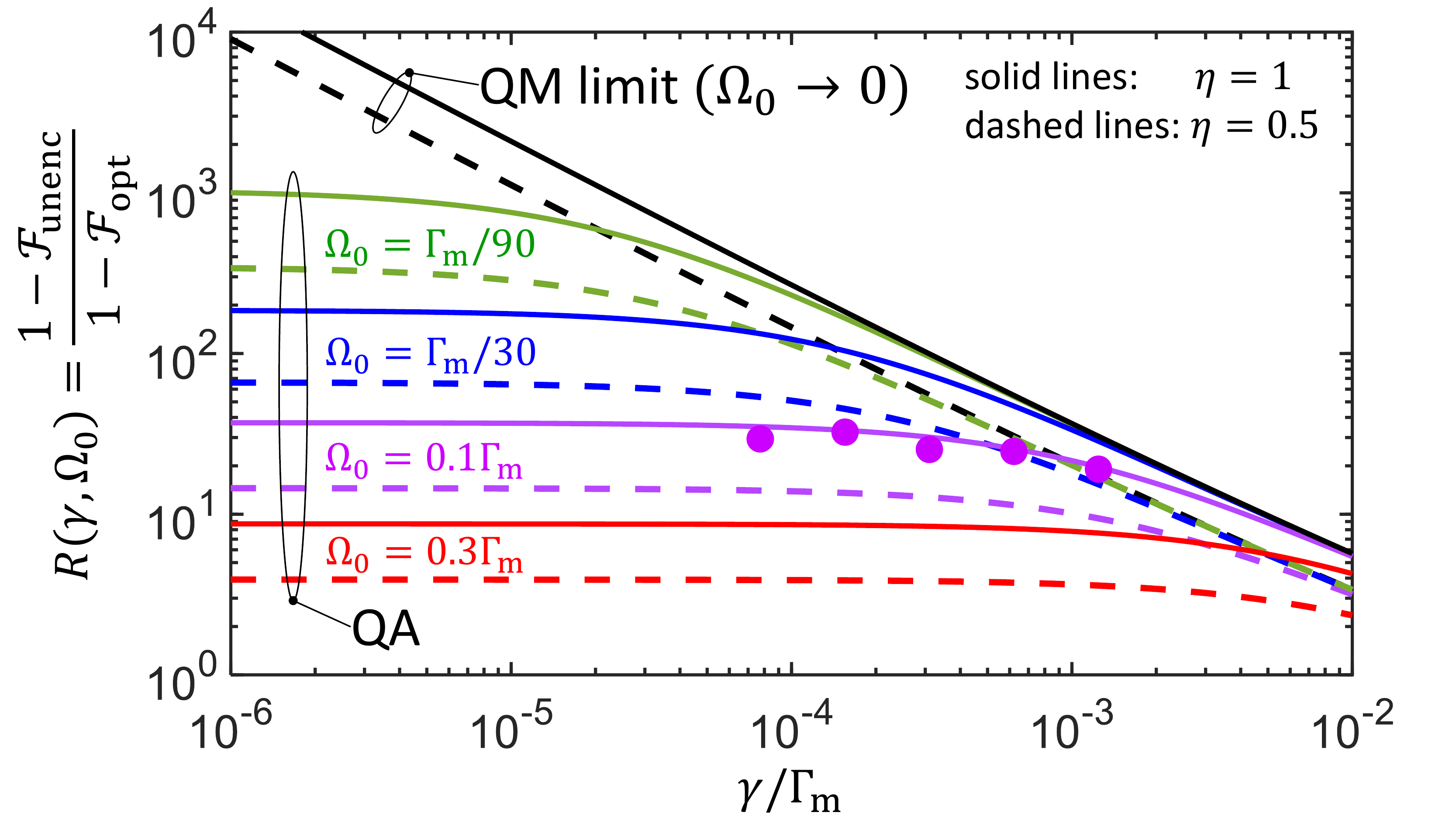}
\caption{{Reduction factor $R(\gamma,\Omega_0)$ of the final logical state infidelity due to continuous QEC, Eq.~\eqref{eq:reduction-factor}, for operation time $t_{\rm op} = 500 \Gamma_{\rm m}$, shown as a function of the bit-flip error rate $\gamma$ for a range of strengths $\Omega_0$ of the annealing Hamiltonian. {Lines present analytic estimates, symbols depict numerical results averaged over an ensemble of 60,000-80,000 realizations of the conditional quantum master equation Eq.~\eqref{eq:rho-eom-no-errors-v2} together with our CQEC protocol.} Black lines depict results for the quantum memory (QM) limit, $\Omega_0\to0$, see Eq.~\eqref{eq:hL} of the main text. The other lines depict the results for quantum annealing (QA) for $\Omega_0=\Gamma_{\rm m}/90$ (green lines), $\Gamma_{\rm m}/30$ (blue lines), $0.1\Gamma_{\rm m}$ (purple lines) and $0.3\Gamma_{\rm m}$ (red lines). {The plateau values $R_{\rm plateau}$ are evident on the left hand side of the plot.} Solid and dashed lines depict the results for quantum efficiencies $\eta=1$ and $0.5$, respectively. $\Gamma_{\rm m}$ denotes the strength of the continuous measurement.}}
\label{fig:opt-results}
\end{figure}

Figure~\ref{fig:opt-results} shows the dependence of the reduction factor~\eqref{eq:reduction-factor} on the physical qubit error rate $\gamma$. We see that {$R(\gamma,\Omega_0)$ increases as $\gamma$ decreases, saturating at} the value $R_{\rm plateau}$ in the limit of small bit-flip error rate $\gamma$. {This} plateau value  increases with decreasing $\Omega_0$ as follows 
\begin{align}
\label{eq:Rmax-fitting}
R_{\rm plateau} \approx&\; \frac{0.7174}{\left[\Omega_0\Gamma_{\rm m}^{-1}\ln\left(1.694\,\Omega_0\Gamma_{\rm m}^{-1}\right) \right]^2} \;\;\;\;(\eta=0.5)\nonumber \\
 \approx&\; \frac{2.512}{\left[\Omega_0\Gamma_{\rm m}^{-1}\ln\left(1.289\,\Omega_0\Gamma_{\rm m}^{-1}\right) \right]^2} \;\;\;\;(\eta=1).
\end{align}
The numerical factors and exponents of Eq.~\eqref{eq:Rmax-fitting} are obtained from fitting  for $\Omega_0=\Gamma_{\rm m}/2430$, $\Gamma_{\rm m}/810$, $\Gamma_{\rm m}/270$, $\Gamma_{\rm m}/90$, $\Gamma_{\rm m}/30$ and $\Gamma_{\rm m}/10$. Note that, in the quantum annealing operation considered here, the operation duration $t_{\rm op}$ and the frequency parameter $\Omega_0$ have to satisfy the adiabatic condition,  $\Omega_0t_{\rm op}\gg1$, which allowed us to use the instantaneous ground state~\eqref{eq:psiL-sol} as the target logical state. Assuming that this condition is satisfied, the reduction factor~\eqref{eq:reduction-factor} of the final logical state infidelity due to CQEC is independent of $t_{\rm op}$. 

\begin{figure}[t!]
\centering
\includegraphics[width=\linewidth, trim =2.25cm 0.5cm 2.5cm 0.5cm,clip=true]{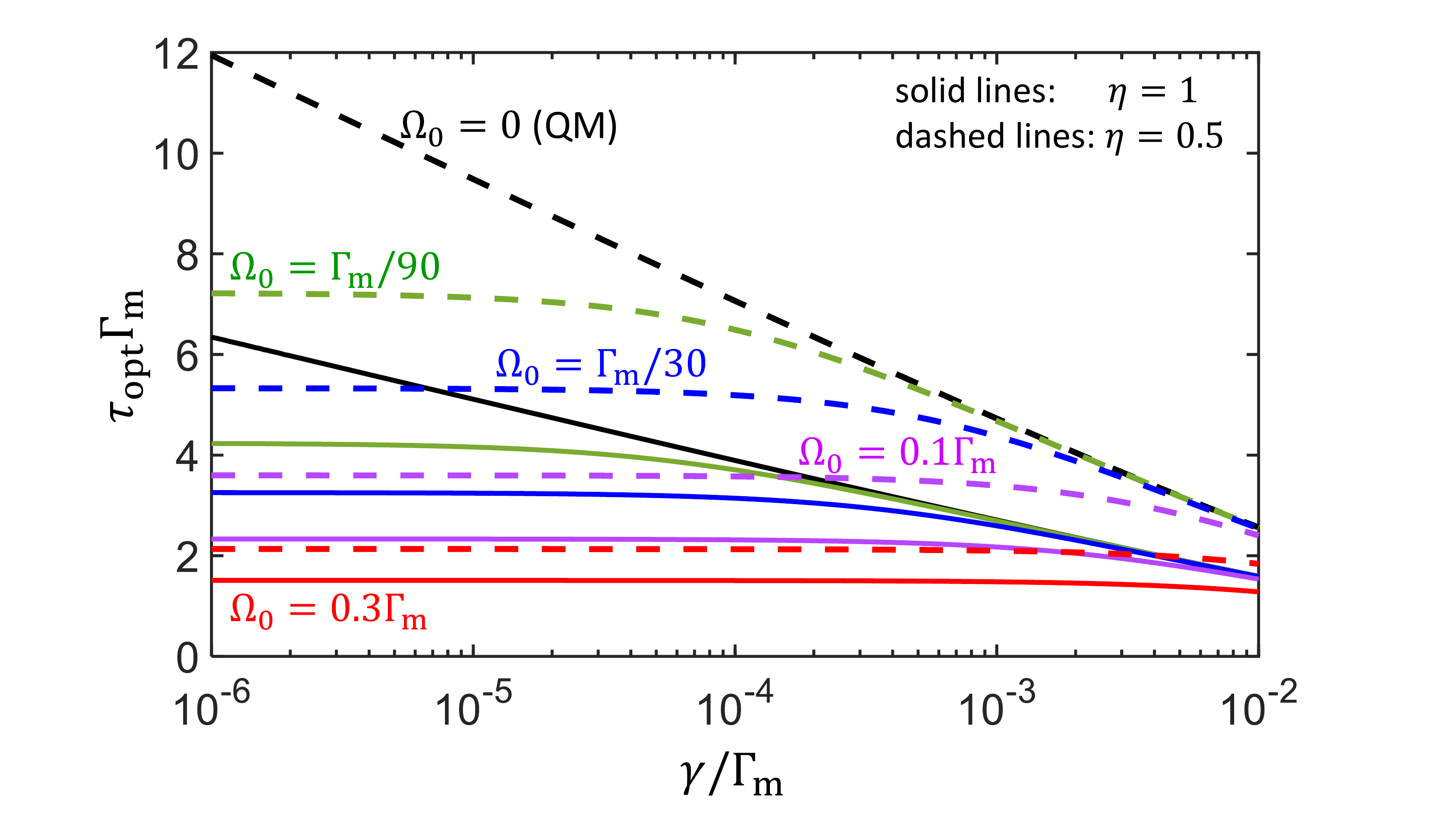}
\caption{Optimal measurement averaging time parameter $\tau_{\rm opt}$. Black solid and dashed lines depict the results for quantum memory (QM), where the frequency parameter $\Omega_0$ vanishes, see Eq.~\eqref{eq:hL} of the main text. The other lines depict the results for quantum annealing (QA) with $\Omega_0=\Gamma_{\rm m}/90$ (green lines), $\Gamma_{\rm m}/30$ (blue lines), $0.1\Gamma_{\rm m}$ (purple lines) and $0.3\Gamma_{\rm m}$ (red lines). Solid and dashed lines depict the results for quantum efficiencies $\eta=1$ and $0.5$, respectively. $\Gamma_{\rm m}$ denotes the strength of the continuous measurement.} 
\label{fig:opt-results-tau}
\end{figure}

{Finally, we summarize the optimized parameters $\tau_{\rm opt}, \Theta_1^{\rm opt}, \Theta_2^{\rm opt}$ employed in Fig.~\ref{fig:opt-results}.}  
Figure~\ref{fig:opt-results-tau} depicts the results for the optimal measurement averaging time $\tau_{\rm opt}$ that minimizes the logical $X$ error rate $\Gamma_{\rm L}$ in the  case of quantum memory (black lines) and the final logical state infidelity~\eqref{eq:F-QA} in the case of quantum annealing for $\Omega_0=\Gamma_{\rm m}/90$ (green lines), $\Gamma_{\rm m}/30$ (blue lines), $0.1\Gamma_{\rm m}$ (purple lines) and $0.3\Gamma_{\rm m}$ (red lines). We see that the optimal averaging time parameter $\tau_{\rm opt}$ generally increases when the measurement quantum efficiency $\eta$ decreases, due to the additional noise at the output of the readout signals, see Eq.~\eqref{eq:Ik}~\cite{Korotkov2001}.  In the particular case of quantum memory ($\Omega_0=0$), we obtain 
\begin{align}
\label{eq:tau-opt-QM}
\tau_{\rm opt}^{\rm QM}\approx&\; -1.027\,\Gamma_{\rm m}^{-1}\log\left(9.6955\,\gamma\Gamma^{-1}_{\rm m}\right)\;\;\;\;\,(\eta=0.5), \nonumber \\
\approx&\; -0.5192\,\Gamma_{\rm m}^{-1}\log\left(5.2891\,\gamma\Gamma^{-1}_{\rm m}\right)\;\;\;(\eta=1).
\end{align}
The above results are obtained from fitting $\tau_{\rm opt}^{\rm QM}$
for the range of error rates indicated in Fig.~\ref{fig:opt-results-tau}. In the case of quantum annealing, for a fixed  and finite $\Omega_0$, the optimal measurement averaging time $\tau_{\rm opt}$ tends to increase as the error rate $\gamma$ is decreased until it reaches a plateau level $\tau_{\rm opt}^{\rm plateau}$ that depends on $\Omega_0$ as follows 
\begin{align}
\label{eq:tau-opt-saturation}
\tau_{\rm opt}^{\rm plateau} \approx&\, -1.759\Gamma_{\rm m}^{-1} \log \left(1.3880\,\Omega_0\Gamma_{\rm m}^{-1} \right)\;\;\;\;(\eta=0.5), \nonumber \\
\approx &\,  -0.9079\Gamma_{\rm m}^{-1} \log \left(1.2408\,\Omega_0\Gamma_{\rm m}^{-1} \right)\;\;\,(\eta=1).
\end{align}
The relations~\eqref{eq:tau-opt-saturation} are obtained from fitting for $\Omega_0=\Gamma_{\rm m}/810$, $\Gamma_{\rm m}/270$, $\Gamma_{\rm m}/90$, $\Gamma_{\rm m}/30$ and $\Gamma_{\rm m}/10$. 

{The optimal values for the error threshold parameters ($\Theta_1^{\rm opt}$ and $\Theta_2^{\rm opt}$) are found to exhibit only} a weak dependence on $\gamma$ and $\Omega_0$ parameters. For practical purposes these {are given here for both the quantum memory and quantum annealing cases by}
\begin{align}
\Theta_1^{\rm opt} \approx -0.54 \;\;\;{\rm and}\;\;\;\Theta_2^{\rm opt} = 0.8. 
\end{align}
We point out that in our optimization procedure we have imposed two constraints: $-1\leq\Theta_{1}\leq0$ and $0\leq\Theta_{2}\leq0.8$. The reason for the constraint on $\Theta_2$ is that our analytical estimates for the logical error rate $\Gamma_{\rm L}$, see Eq.~\eqref{eq:GL-v2}, and the error detection time $t_{\rm det}$, see Eq.~\eqref{eq:t_det}, are not accurate when $\Theta_1$ approaches 1. {The optimization} finds that the optimal position of the upper error threshold should be as close to 1 as it is allowed. If we instead use the constraint $0\leq \Theta_2\leq1$ (with the same previous constraint on $\Theta_1$), the {optimization} finds that $\Theta_1^{\rm opt}\approx -0.4$ and $\Theta_2^{\rm opt}=1.0$. This indicates that the optimal position of the lower error threshold is robustly {located} around $-0.5$. 

\subsection{Overall performance of the double threshold CQEC protocol}To quantify the effectiveness of the double threshold CQEC protocol in correcting logical errors during the entire continuous operation, we introduce the time-dependent reduction factor $R_t$ of the logical state infidelity. {This} is defined analogously to Eq.~\eqref{eq:reduction-factor}; 
\begin{align}
\label{eq:Rt}
R_t = \frac{1-\mathcal{F}_{\rm unenc.}(t)}{1-\mathcal{F}_t},
\end{align}
where $\mathcal{F}_{\rm unenc.}(t)$ is now the time-dependent unencoded fidelity, defined as in Eq.~\eqref{eq:infidelity-unencoded} with the operation time $t_{\rm op}$ replaced by $t\in [0,t_{\rm op}]$, and $\mathcal{F}_t$ is the time-dependent logical state fidelity 
\begin{align}
\label{eq:Ft}
\mathcal{F}_t = \langle \psi_{\rm L}(t) | \varrho_{\rm L}(t) | \psi_{\rm L}(t)\rangle.
\end{align}
\begin{figure}[t!]
\centering
\includegraphics[width=\linewidth, trim =0.75cm 0.25cm 2cm 0cm,clip=true]{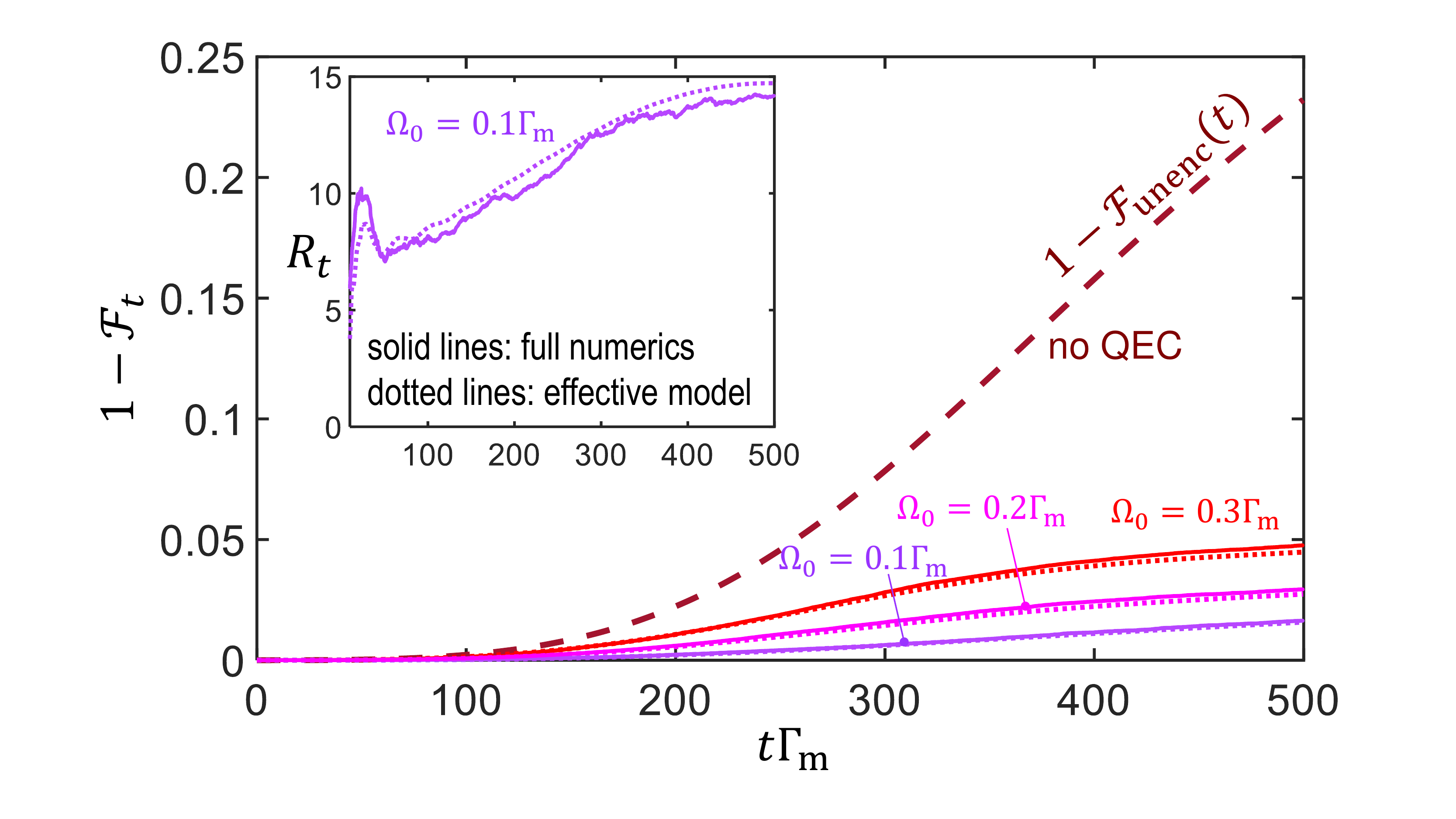}
\caption{Accuracy of the effective evolution equation~\eqref{eq:rhoL-eom} for the logical state infidelity during quantum annealing. The frequency parameter $\Omega_0$ determines the strength of the logical Hamiltonian~\eqref{eq:hL}, where the coefficients $a(t)$ and $b(t)$ correspond to quantum annealing with linear schedule, see Eq.~\eqref{eq:a-b}. Solid lines: full numerical calculations averaged over an ensemble of 20,000 realizations. Dotted lines: effective model of Eq.~\eqref{eq:rhoL-eom}. Dashed line: unencoded qubit infidelity [see Eq.~\eqref{eq:infidelity-unencoded}] for $\Omega_0=0.1\Gamma_{\rm m}$. Parameters: $\Theta_1=-0.54$, $\Theta_2=0.8$, $\tau=2.5\Gamma_{\rm m}^{-1}$, $\eta=1$,  $t_{\rm op}=500\Gamma_{\rm m}^{-1}$ and $\gamma=1.25\times10^{-3}\Gamma_{\rm m}$. }
\label{fig:accuracy-effective-model}
\end{figure}

Figure~\ref{fig:accuracy-effective-model} shows the time dependence of the logical state infidelity, $1-\mathcal{F}_t=1-\langle \psi_{\rm L}(t) | \varrho_{\rm L}(t) | \psi_{\rm L}(t)\rangle$, obtained using two approaches: ``full numerics'' and "effective model". In the first approach, the logical state $\varrho_{\rm L}(t)$ is obtained by projecting out the code space components from the full system density matrix $\rho(t)$,  where the latter evolves 
according to the evolution equation~\eqref{eq:rho-eom-no-errors-v2}, together with the action of the instantaneous error-correction operations $C_{\rm op}$ [Eq.~\eqref{eq:Cop}] that are applied to the physical qubits whenever an error is diagnosed by the double threshold CQEC protocol.
The ensemble average of Eq.~\eqref{eq:rhoL-def} is {generated} over an ensemble of 20,000 realizations, using the techniques described in appendix \ref{sec: appendix_numerics}. The results of this approach are depicted in Fig.~\ref{fig:accuracy-effective-model} by the solid lines,  for Hamiltonian strength parameters $\Omega_0=0.1\Gamma_{\rm m}$, $0.2\Gamma_{\rm m}$ and $0.3\Gamma_{\rm m}$. The second approach {is that of our effective model derived in the previous section.  Here the logical state infidelity is obtained from the numerical solution of the effective open-system model given} by Eq.~\eqref{eq:rhoL-eom}. The results of this approach are depicted in Fig.~\ref{fig:accuracy-effective-model} by the dotted lines. {The good agreement between the solid and dotted lines in} Fig.~\ref{fig:accuracy-effective-model} demonstrates that the effective open-system model accurately describes the evolution of the logical qubit during the entire continuous operation. This validates our analysis above  for the optimized performance of the double threshold CQEC protocol. The inset of Fig.~\ref{fig:accuracy-effective-model} shows the reduction factor $R_t$ {for} the logical state infidelity during the entire duration of the continuous operation for $\Omega_0=0.1\Gamma_{\rm m}$.  Here also, good agreement is found between the full numerics and the effective model approaches. Although {in this specific example} the reduction factors of the logical state infidelity are modest (varying from 5 to 15), larger reduction factors can be {readily} achieved with stronger continuous measurements. {This can be seen explicitly in Fig.~\ref{fig:opt-results}, where the increase in $R$ is evident for $\Gamma_{\rm m}$ larger than $10\Omega_0$}. 

\begin{figure}[t!]
\centering
\includegraphics[width=\linewidth, trim =2cm 1.0cm 2.25cm 0.25cm,clip=true]{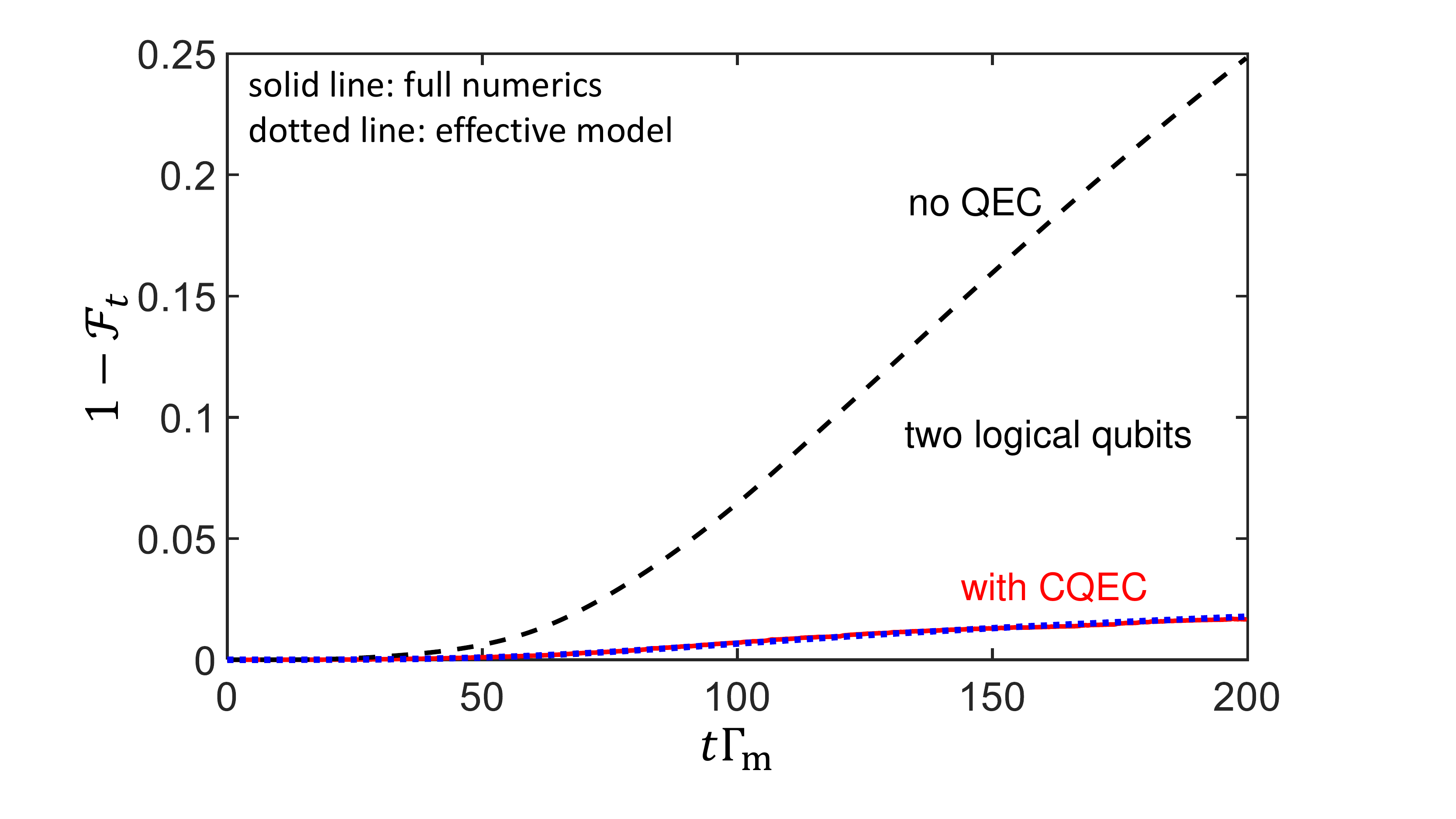}
\caption{Logical state infidelity for two logical qubits. Parameters: $\gamma=1.25\times10^{-3}\Gamma_{\rm m}$ (error rate of all qubits), $\Omega_0=0.1\Gamma_{\rm m}$ (Hamiltonian strength, see Eq.~\eqref{eq:h2L}), $\Theta_1=-0.54$ and $\Theta_2=0.8$, $\tau=2.1\Gamma_{\rm m}^{-1}$, $\eta=1$, and $\Gamma_{\rm m}$ denotes the strength of continuous measurement. {Blue dotted line: effective model of Eq.~\eqref{eq:rhoL-eom-2L}. Red solid line: full numerical calculations averaged over an ensemble of 5,000 realizations (see text)}.}
\label{fig:infidelity-2L}
\end{figure}

We now discuss how to generalize the effective open-system {model for one logical qubit, Eq.~\eqref{eq:rhoL-eom},} to the general case of {multiple} logical qubits. In this general case, we again have logical errors that come in two forms: logical $X$ errors, and logical errors {that are} characterized by a logical error operation $V_{q}^{(l)}$, where $q$ now labels the three physical qubits that encode the $l$th logical qubit. The logical error operations $V_q^{(l)}$ are again given by Eq.~\eqref{eq:Vq-def}, {where} $h_{\rm L}(t)$ (logical Hamitonian) and $h_{\rm spurious}(t)$ (spurious Hamiltonian) {are now specified respectively by} the code space diagonal submatrices of the system Hamiltonian $H(t)$ and  $X_qH(t)X_q$. Logical $X$ errors acting on the $l$th logical qubit occur at a rate $\Gamma_{\rm L}^{(l)}$ that is {also} given by Eq.~\eqref{eq:GL-v2}. Note that the set of parameters of the double threshold CQEC protocol ($\tau$, $\Theta_1$ and $\Theta_2$)  can differ for different logical qubits,  so $\Gamma_{\rm L}^{(l)}$  may not be the same for all logical qubits. The logical errors that are characterized by $V_q^{(l)}$ occur at the rate $\tilde{\gamma}_q=\gamma_q(1-p_{\rm misdiag}^{(X_q)})$  {(Eq.~\eqref{eq:rhoL-eom}).  This is} approximately equal to the bit-flip error rate $\gamma_q$ of the $q$th qubit, since the probability $p_{\rm misdiag}^{(X_q)}$ to misdiagnose the error $X_q$ is typically much smaller than one (see Fig.~\ref{fig:prob-misdiagnosis}). As an example,  we consider two logical qubits encoded by the physical qubits $q=1,2,3$ (logical qubit with label $l=1$) and $q=4,5,6$ (logical qubit with label $l=2$). {Consider the} two-qubit logical Hamiltonian 
\begin{align}
\label{eq:h2L}
\tilde{h}_{\textrm{L}}(t) =&\; -\Omega_{0}\,\Big[a(t)\left(\sigma_{x}^{(1)} + \sigma_{x}^{(2)}  \right) +b(t)\,\Big(\sigma_{z}^{(1)} + \sigma_{z}^{(2)} + \nonumber\\
&\;\sigma_{z}^{(1)} \sigma_{z}^{(2)}\Big)\Big], 
\end{align}
where $\sigma_x^{(l)}$ and $\sigma_z^{(l)}$are the Pauli $x$ and $z$ operators corresponding to the $l$th logical qubit ($l=1$, 2), and {the quantum annealing}  coefficients $a(t)$ and $b(t)$ are given in this example by Eq.~\eqref{eq:a-b}. We will assume that the initial condition for the target logical state evolution is  $|\psi_{\rm L}(0)\rangle = (|0_{\rm L}\rangle + |1_{\rm L}\rangle)\otimes(|0_{\rm L}\rangle + |1_{\rm L}\rangle)/2$. For this example of two logical qubits, the effective open-system model reads as 
\begin{align}
\label{eq:rhoL-eom-2L}
&\dot{\varrho}_{\rm L} =\, -i[\tilde h_{\rm L}(t),\varrho_{\rm L}] + \sum_{l=1,2}\Gamma^{(l)}_{\rm L}\,\big[\sigma^{(l)}_x\,\varrho_{\rm L}\, \sigma^{(l)}_x - \varrho_{\rm L}\big] + \nonumber \\
&\sum_{q=1,2,3}\!\!\gamma_q \big[V^{(1)}_{q}\varrho_{\rm L}{V^{(1)}_q}^{\dagger}\!  - \varrho_{\rm L}\big] +\!\! \sum_{q=4,5,6}\!\!\gamma_q \big[V^{(2)}_{q}\varrho_{\rm L}{V^{(2)}_q}^{\dagger} \! - \varrho_{\rm L}\big], 
\end{align} 
where $V_q^{(1)}$ (for $q=1$, 2, 3) is obtained from Eq.~\eqref{eq:Vq-def} with $h_{\rm L}(t)$ replaced by $\tilde h_{\rm L}(t)$ and $h_{\rm spurious}(t)$ now given by Eq.~\eqref{eq:h2L} with $\sigma_z^{(1)}$ replaced by $\sigma_z^{(1)}/3$. Similarly, $V_q^{(2)}$ (for $q=4$, 5, 6) is obtained from Eq.~\eqref{eq:Vq-def} with $h_{\rm L}(t)$ also replaced by $\tilde h_{\rm L}(t)$ and  $h_{\rm spurious}(t)$ given by Eq.~\eqref{eq:h2L} with $\sigma_z^{(2)}$ replaced by $\sigma_z^{(2)}/3$. Figure~\ref{fig:infidelity-2L} shows that the logical state infidelity obtained from the effective open-system model for two logical qubits [Eq.~\eqref{eq:rhoL-eom-2L}, ({dotted blue line})] agrees very well with the {corresponding infidelity} obtained from {the} full numerical calculations, {(solid red line)}. This indicates that the effective model can be used to {accurately} estimate and optimize the performance of our CQEC protocol in order to protect the coherent evolution of several logical qubits.  {Most importantly, both effective model and full numerical calculations show that the CQEC protocol provides a significant reduction in the final state infidelity by a factor of $\sim 14$ relative to the value obtained without error correction.}   For the two-logical qubit  Hamiltonian considered here, Eq.~\eqref{eq:h2L}, this reduction is similar to that obtained for the corresponding single logical qubit in Fig.~\ref{fig:accuracy-effective-model}.  However, in general, 
different values of $R(\gamma,\Omega_0)$ may be found, since the reduction also depends on the form of the coupling between the logical qubits. Thus if the coupling term is changed, the logical state infidelity can change since both $\tilde h_{\rm L}(t)$  and the $V_q^{(i)}$ terms in Eq.~\eqref{eq:rhoL-eom-2L} are dependent on this coupling (see Eq.~\eqref{eq:Vq-def}). For example, if the sign of the $\sigma_z^{(1)}\sigma_z^{(2)}$ term in Eq.~\eqref{eq:h2L} is flipped, then the final time reduction factor $R$ is close to 6 instead of 14.

\begin{figure}[t!]
\centering
\includegraphics[width=\linewidth,trim =1.8cm 0.35cm 2cm 1cm,clip=true]{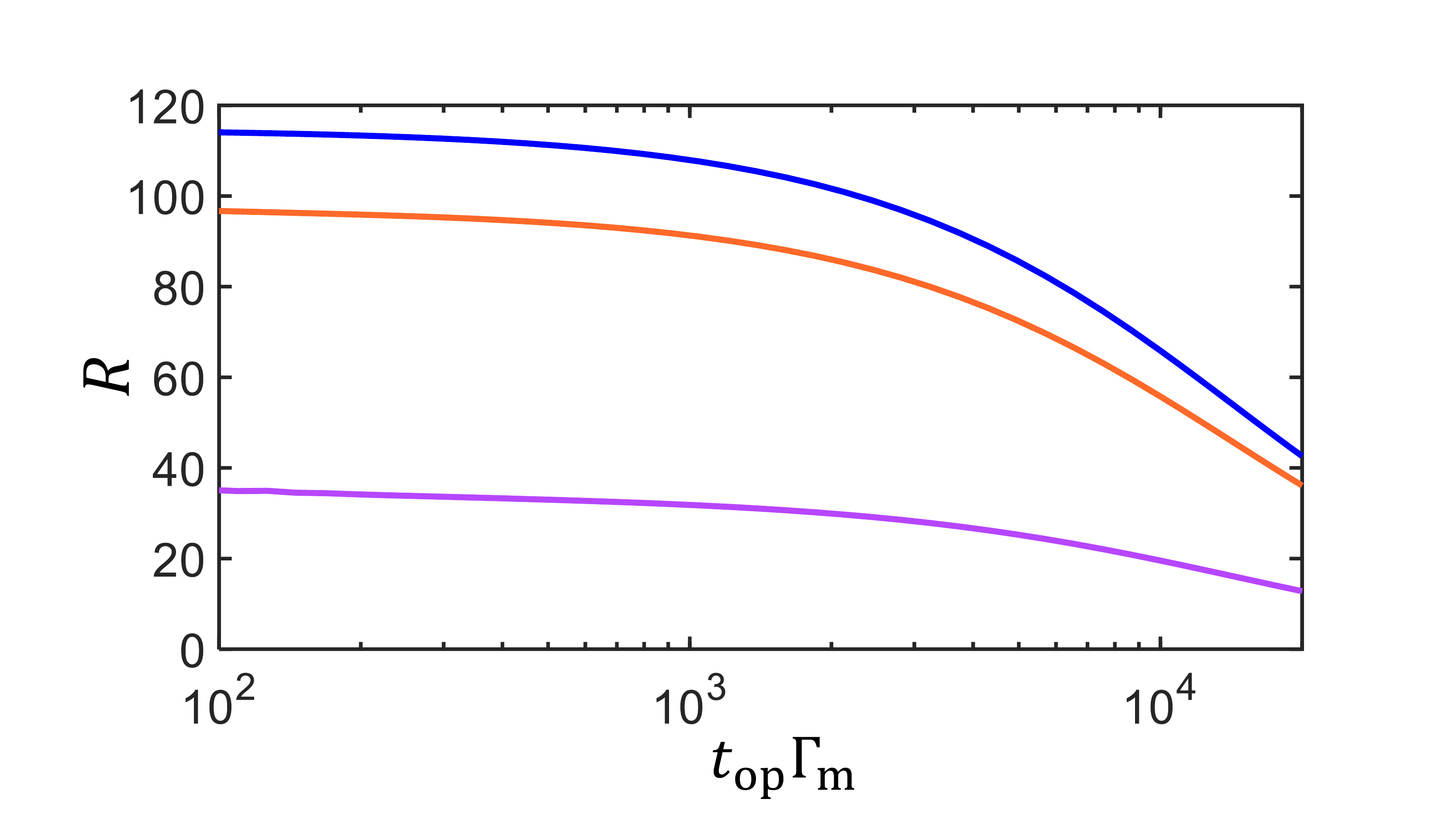}
\caption{Reduction factor $R$ of the final logical state infidelity as a function of the operation time $t_{\rm op}$ for single logical qubit annealing under CQEC. Blue, orange and purple lines depict the reduction factors  corresponding to the cases where the error correction operations are given by $\tilde C_{\rm op}(t)$ [Eq.~\eqref{eq:Cop-new}], $\tilde C_{\rm op}^{\rm approx}(t)$ [Eq.~\eqref{eq:Cop-appro}] and $C_{\rm op}=X_q$, respectively. Parameters: $\gamma = 1.25\times 10^{-4}\Gamma_{\rm m}$, $\Omega_0 = 0.1\Gamma_{\rm m}$, $\tau = 2.5\Gamma_{\rm m}^{-1}$, $\Theta_1 = -0.54$,  $\Theta_2 = 0.8$, and $\eta = 1$.}
\label{fig:reduction_factor_vs_top}      
\end{figure}

\subsection{Modified logical error operations}
\label{sec:correction_logical_errors}

Our analysis of {logical errors due to spurious coherent evolution in the error subspaces during error diagnosis (scenario 2 in Sec.~\ref{subsec:QME_logical})} employed 
the conventional error correction operators of the three-qubit bit flip code as the correction operators, i.e., $C_{\rm op}=X_q$ with $X_q=X_1$, $X_2$ or $X_3$ (see Eq.~\eqref{eq:V-def}). 
It is possible to further increase the reduction factors $R$ of the final logical state infidelity by using the following {modified} error correction operations:
\begin{align}
\tilde C_{\rm op}(t) =&\; \exp\left(-i t_{\rm det}^{(q)} H(t)\right)\exp\left(i t_{\rm det}^{(q)} X_q\,H(t)\,X_q\right) X_q, 
\label{eq:Cop-new}
\end{align}
where $t$ is the time moment when the error $X_q$ is detected, and $t_{\rm det}^{(q)}$ is the time that it takes to diagnose such error. To analyze the performance of the modified error correction operations $\tilde C_{\rm op}(t)$, we can use the effective evolution equation~\eqref{eq:rhoL-eom} for the logical state without the term describing the logical errors due to spurious evolution in the error subspaces; i.e., without the last term at the right-hand-side of Eq.~\eqref{eq:rhoL-eom}. This can be shown by carrying out the analysis of Sec.~\ref{subsec:QME_logical} with error correction operations given by Eq.~\eqref{eq:Cop-new}. We emphasize that spurious evolution is not fully compensated by using the modified error correction operations in an experimental setup because the error-detection times $t_{\rm det}^{(q)}$ are  random.

From the solution of Eq.~\eqref{eq:rhoL-eom} without the last term at its right-hand-side, we have estimated the reduction factor $R$ for the final logical state infidelity that corresponds to our CQEC operation with error correction operations given by Eq.~\eqref{eq:Cop-new}. The blue line in  Fig.~\ref{fig:reduction_factor_vs_top} shows our results for this situation and for a  bit-flip error rate $\gamma=1.25\times10^{-4}\Gamma_{\rm m}$. By comparing the blue and the purple lines in Fig.~\ref{fig:reduction_factor_vs_top}, we see that, by using the modified error correction operations, $R$ increases by a factor of $\simeq3.3$ with respect to the case of CQEC with conventional error correction operations, $C_{\rm op}=X_1$, $X_2$ or $X_3$. This improvement factor in $R$  depends on the  physical error rate $\gamma$; for instance, for $\gamma=1.25\times10^{-3}\Gamma_{\rm m}$ we obtain that $R$ is increased by a smaller factor $\simeq1.7$ by using the modified error correction operations. We have also calculated the reduction factor $R$ from (computationally expensive) full quantum trajectory simulations of Eq.~\eqref{eq:rho-eom-no-errors-v2} with the double-thresholding error detection protocol and error correction operations given by Eq.~\eqref{eq:Cop-new}, where   $t_{\rm det}^{(q)}$ is given by Eq.~\eqref{eq:t_det}. Our results show that the reduction factors indicated by the blue line in Fig.~\ref{fig:reduction_factor_vs_top} are somewhat larger than those from full trajectory simulations by $\simeq 15\%$, for operation times $t_{\rm op}$ in the range $100\,\Gamma_{\rm m}^{-1}$ to $1000\,\Gamma_{\rm m}^{-1}$, which is the region where the reduction factor is roughly independent of the operation time, see Fig.~\ref{fig:reduction_factor_vs_top}.

The modified error correction operations $\tilde C_{\rm op}(t)$,  Eq.~\eqref{eq:Cop-new}, may be challenging to implement, so we have also analyzed the performance of the following error correction operations that can be regarded as an approximate version of $\tilde C_{\rm op}(t)$:
\begin{equation}
    \tilde C_{\rm op}^{\rm approx}(t) = \exp\left(i t_{\rm det}^{(q)} \big(X_q\,H(t)\,X_q-H(t)\big)\right)X_q.
    \label{eq:Cop-appro}
\end{equation}
We note that the exponent in Eq.~\eqref{eq:Cop-appro} is $2\Omega_0b(t)Z_q/3$ for the Hamiltonian given in Eq.~\eqref{eq:HL}. If we carry out the analysis of Sec.~\ref{subsec:QME_logical} with error correction operations given by Eq.~\eqref{eq:Cop-appro}, we arrive to an effective evolution equation for the logical state that is similar to Eq.~\eqref{eq:rhoL-eom} with the operator $V_q(t)$ replaced by  
\begin{align}
\label{eq:V_q_lc}
&\tilde V_q^{\rm approx}(t) = \exp\left({i t_{\rm det}^{(q)}\,h_{\rm L}(t)}\right)\times\nonumber \\
&\exp\left(i t_{\rm det}^{(q)}(h_{\rm spurious}(t) - h_{\textrm L}(t))\right) \exp\left(-i t_{\rm det}^{(q)}h_{\rm spurious}(t)\right). 
\end{align}
From this effective evolution equation for the logical state we have estimated the performance of our CQEC operation with error correction operations given by Eq.~\eqref{eq:Cop-appro}. The reduction factors in this case are somewhat smaller than those corresponding to the case of using error correction operations given Eq.~\eqref{eq:Cop-new}, see orange and blue lines in Fig.~\ref{fig:reduction_factor_vs_top}, respectively. The main advantage of using $\tilde C_{\rm op}^{\rm approx}(t)$ as the error corrections operations is that they can be simpler to implement.    

\section{Discussion and conclusions}
\label{sec:discussion}

We have developed a continuous quantum error correction protocol for operation under both quantum memory and evolution under a time-dependent Hamiltonian.  We illustrated the approach with a detailed analysis of the continuous operation performance of the three-qubit bit flip code, which is designed to preserve the coherent evolution of the logical qubits against decoherence from  bit-flip errors. Error detection is carried out using a relatively simple and nearly optimal protocol that consists of filtering (time-averaging) the noisy bare readout signals and using a double thresholding scheme to diagnose the error syndrome in real time. In addition, immediately after diagnosing an error, discrete (i.e., instantaneous) error correction operations are applied to the physical qubits, as in the conventional code operation. We have shown that this combination of continuous detection of errors in real time with discrete correction of errors is very effective and yields, e.g., in the case of quantum memory operation, a logical $X$ error rate that exhibits a nearly quadratic scaling on the physical qubit error rate and has a magnitude that is slightly larger than the logical $X$ error rate of the linear variant of the optimal Wonham filter~\cite{Dressel2019}. The advantage of our double threshold CQEC protocol is that it can be simpler to implement.

Spurious coherent evolution of the system state in the error subspaces~\cite{KYoung2013}, due to a (time-dependent) encoded Hamiltonian, leads to a new type of logical errors, for which we have found the corresponding effective Kraus logical error operators, $V_q(t)$, that act on the instantaneous logical state. The Kraus logical error operator $V_q(t)$ is parametrized by the time $t_{\rm det}^{(q)}$ that the CQEC protocol takes to detect the error $X_q$, see Eq.~\eqref{eq:Vq-def}. The time $t_{\rm det}^{(q)}$   {should} be as small as possible in order to minimize the detrimental effect of logical errors due to spurious evolution on the performance of the double threshold CQEC protocol. For this protocol, $t_{\rm det}^{(q)}$ is estimated to be proportional to the averaging time parameter $\tau$ (Eq.~\eqref{eq:t_det}), which, however, cannot be arbitrarily small without degrading the performance of the CQEC protocol to correctly diagnose single bit-flip  errors $X_1$, $X_2$ or $X_3$. 

We have developed an effective open-system model for the logical qubit state [see Eq.~\eqref{eq:rhoL-eom}] that accounts for the two types of logical errors that are relevant for, e.g., quantum simulation and quantum annealing applications: logical errors due to spurious coherent evolution in the error subspaces and the usual logical $X$ errors of quantum memory operation. This effective model is very useful because it allows us to readily estimate and optimize the performance of the double threshold CQEC protocol without performing computationally expensive numerical calculations on the full encoding qubit system (full numerics). We have shown that the effective model accurately describes the actual logical state during the continuous operation, see Fig.~\ref{fig:accuracy-effective-model}. In addition, we have discussed how to generalize the effective model for multiple logical qubits, where we have again found excellent agreement with the {more cumbersome and computationally expensive} full numerics approach, see Fig.~\ref{fig:infidelity-2L}.  

Using the effective open-system model for one logical qubit, we have analyzed the  performance of the  double threshold CQEC protocol to preserve the coherent evolution of the logical qubit due to a quantum-annealing type Hamiltonian with a linear schedule. We have introduced the reduction factor $R$ of the final logical state infidelity, see Eq.~\eqref{eq:reduction-factor}, as a measure of the performance of the CQEC protocol. The performance depends on the relative magnitudes of three problem-specific parameters; namely, the bit-flip error rate, $\gamma$, the strength of the logical Hamiltonian, $\Omega_0$, and the strength of the continuous measurements of the code stabilizer generators, characterized by the measurement strength parameter $\Gamma_{\rm m}$.

For a given magnitude of the error rate $\gamma$, the reduction factor $R$ increases as the magnitude of $\Omega_0$ decreases, which our analysis shows is due to the fact that a reduction in $\Omega_0$ causes the contribution of spurious coherent evolution to logical errors to decrease.  
In contrast, for a given ratio $\Omega_0/\Gamma_{\rm m}$, the reduction factor $R$ increases as we decrease the physical qubit error rate $\gamma$, until $R$ reaches a plateau level $R_{\rm plateau}$.
This reduction factor is the analog of the reduction of error probability obtained with discrete operation of quantum error correcting codes, which becomes increasingly effective as the physical qubit error rate decreases~\cite{steane2007tutorial}. The plateau level depends on the relative magnitude of the logical Hamiltonian strength $\Omega_0$ and the strength of continuous measurements $\Gamma_{\rm m}$, as is evident in Fig.~\ref{fig:opt-results}. For instance, we obtain $R_{\rm plateau}\approx 37$, $184$ or $1002$ for measurement strengths $\Gamma_{\rm m} =  10\Omega_0,$ $30\Omega_0$ or $90\Omega_0$, respectively, assuming that continuous measurements are performed by ideal detectors ($\eta=1$). These reduction factors become $R_{\rm plateau}\approx 15$, $66$ or $340$, respectively, if the measurement efficiency is $\eta=0.5$ (nonideal detectors).

Overall, our results reveal that the region of optimal performance is defined by the following hierarchy of timescales:
\begin{equation*}
    \frac{10}{\Gamma_{\rm m}} \leq \Omega_0^{-1} \ll t_{\rm op} 
\end{equation*}
For example, for measurement rate $\Gamma_{\rm m} = (10\,{\rm ns})^{-1}\approx 2\pi\times 16\,$MHz, effective correction of bit flip errors can be achieved for annealing Hamiltonians of strength $\Omega_0 = 2\pi \times 1.6\,$MHz and operation time $t_{\rm op}\lesssim\gamma^{-1}$, with bit flip error rates  $\gamma=(10\,\mu{\rm s})^{-1}\approx2\pi\times 16\,$kHz. 
Such decoherence rates are the same order of magnitude as thermal relaxation rates of quantum annealers operating at temperatures of 8-20 mK~\cite{king2018observation,novikov2018exploring,weber2017coherent}.        
In this case, the reduction factors $R$ are approximately equal to 10 and 20 (see purple lines in Fig.~\ref{fig:opt-results}) for measurement quantum efficiencies $\eta=0.5$ and 1, respectively.

It is useful to consider here the feasibility of parity measurements on these timescales. Parity measurement with superconducting qubits~\cite{DiVincenzo2013} can be realized by dispersively coupling two qubits with a single resonator (cavity) such that their  interaction Hamiltonian is $H_{\rm int} = \chi(Z_1+Z_2)\hat{n}$, where $\chi$ is the dispersive coupling parameter and $\hat{n}$ is the intracavity photon number. In this setup, a superposition of odd-parity states $|01\rangle$ and $|10\rangle$ does not decohere due to measurement, because they do not shift the cavity resonance frequency. In contrast, a superposition of even-parity states $|00\rangle$ and $|11\rangle$ exhibits decoherence due to measurement at a rate that is proportional to $(\kappa/\chi)^2$ when $\chi\gg\kappa$. 
In this situation the odd parity states show an identical high response and the even parity states an identical low response to the cavity drive. 
To use this setup for parity measurement, it is required that $\chi\gg\kappa$ and the cavity is driven at a frequency equal to  the average of the cavity resonance frequencies corresponding to the four two-qubit states: $|c,d\rangle$ with $c,d=\{0,1\}$~\cite{DiCarlo2013}. The measurement rate is then $\Gamma_{\rm m}\approx\kappa\, \bar{n}_{01}/2$, where $\bar{n}_{01}$ is the average number of intracavity photons when the two-qubit state is $|01\rangle$. This measurement rate dictates the exponential decay of the off-diagonal matrix elements of the two-qubit density matrix connecting states of even and odd parities; i.e., $\langle\rho_{00,01}(t)\rangle \sim e^{-\Gamma_{\rm m}t}$.
Now, assuming realistic parameter values of  $\kappa/2\pi=3\,$MHz and $\bar{n}_{01}=10$, 
we find that $\Gamma_{\rm m}\approx (10.6\,{\rm ns})^{-1}$.

The CQEC approach developed in this work is applicable to a wide range of analog quantum simulations, allowing the encoded logical quantum states to be efficiently corrected as they evolve under a general time-varying Hamiltonian and in the presence of environmental decoherence. The reduction factor $R_t$ measures the overall effectiveness and success of the CQEC strategy under a given Hamiltonian evolution, while the quadratic scaling of the logical error rate $\Gamma_{\rm L}$ with physical qubit error rate $\gamma$ guarantees that the strategy is correcting single physical qubit errors. As in discrete quantum error correction, this active correction of errors for greater fidelity of quantum simulation comes with an overhead of additional qubits for the logical encoding. However, unlike discrete protocols, we do not require additional ancilla qubits and entangling operations to transfer the logical state information for measurement.

The reduction factor expresses the decrease in error probability due to quantum error correction.  This is the metric of success that is used for analysis of performance of error correcting codes in quantum computation, and can be also be used for quantum simulations. The cost of CQEC will depend on the specific application.
Since quantum annealing is a heuristic analog approach that is not guaranteed to find the global minimum of the final Hamiltonian in a single run and instead finds it with a less than unit success probability $p_s$, it is conventional to measure the cost of quantum annealing on a specific device by measuring the time to achieve a desired success probability at least once during multiple runs, defined as the time-to-solution (TTS), often also taking the parallelization possible on a given device into account \cite{ronnow2014defining}. This TTS metric, which is applied to both algorithmic and error mitigation protocols, reflects a trade-off between a high single-run success probability with a long operation time, and a low single-run success probability with a short operation time \cite{Tameem2018}. It is generally defined for quantum annealing as
\begin{align}
        \text{TTS}(t_{\rm op}) &= t_{\rm  op} \mathcal{R}_s(t_{\rm  op})\frac{N}{N_{\rm max}}  \label{eq:TTS}
\end{align}
with
\begin{align}
        \mathcal{R}_s(t_{\rm  op}) &= \frac{\log(1-p_d)}{\log(1-p_s(t_{\rm  op}))}. \label{eq:TTS_R}
\end{align} 
This refers just to the pure annealing time and not to the wall-clock time that also contains the setup, cooling and readout times on a real device. 
In Eq.~\eqref{eq:TTS_R}, $p_d$ is the desired success probability to get the solution at least once, $p_s(t_{\rm op})$ is the success probability of a single run, $N$ is the required number of qubits in a single run, and $N_{\textrm max}$ is the total available number of qubits in the device. $\mathcal{R}_s$ is the number of total runs needed to get the solution at least once with desired probability $p_d$ and the factor ${N}/N_{\rm  max}$ takes into account the possible reduction by parallelization of runs. A separate analysis applies to the time for readout and state preparation. Comparing the TTS for our CQEC protocol with a classical protection strategy utilizing parallelization alone shows that 
CQEC provides a smaller TTS for the quantum annealing dynamics than does the classical strategy.  Detailed results and analysis are presented in  Appendix~\ref{sec: appendix_TTS}. This improvement over the classical strategy is noteworthy since at small encodings such as the three-qubit code employed here, proposals for suppression of errors during quantum annealing using stabilizer encodings with penalty Hamiltonians have shown that these perform worse than the classical parallelization scheme~\cite{pudenz2014error}.

Overall, the high level of reduction of infidelity achieved by this CQEC for a small three-qubit code requiring overhead of just two additional qubits in both quantum memory and quantum simulation under a time-evolving Hamiltonian is excellent. The reduction in time-to-solution relative to classical parallelization for quantum annealing shows the benefits of this in application to analog quantum simulations.  In principle, one can expect even better performance with larger encodings.  Indeed,
the continuous time quantum error correction protocol presented in this work can be readily applied to any {subspace} stabilizer QEC code, such as the three-qubit repetition code studied here, and can also be extended to subsystem {stabilizer} codes \cite{Atalaya2020}. In practice, the limitations will be achieving continuous measurement of multi-qubit stabilizer operators. For larger systems, subsystem codes~\cite{Poulin2005} such as the Bacon-Shor codes~\cite{Bacon2006,Cross2007} or generalizations of this~\cite{bravyi2011subsystem,marvian2019robust} would be preferred since these require measurement of only two-body operators.

An important direction for further work is to apply the CQEC protocol to other error models. Clearly arbitrary single qubit errors can be corrected using this approach with larger stabilizer codes.  Of particular interest for quantum annealing is correction of thermal errors.  
The bit-flip code alone is not effective here, since it cannot correct $X$ and  $Y$ errors at the same time. However, active continuous correction of thermal errors can be achieved by implementing the present CQEC protocol in a adiabatic frame and combining this with error suppression techniques in which an energy penalty consisting of the negative of the bit flip code stabilizer operators is added to the time-dependent Hamiltonian \cite{Atalaya2020b}.

More generally, one would like to develop CQEC protocols for architecture-specific errors, such as biased noise. In the future, developing error correction diagnostics for physical errors encountered in realistic devices may be assisted by the use of machine learning techniques \cite{Foesel2018} {or filters for non-Markovian noise \cite{tsang2015volterra}}. In addition, experimental implementations frequently see drift of the key parameters such as the measurement rate $\Gamma_{\rm m}$ and efficiency $\eta$ as well as slow temporal variations of the offset of the measurement signals. Exploring the use of machine learning techniques to track these parameters and adjust the CQEC protocol accordingly in real time during an experiment would be a useful direction for further work.
Another important direction for further study based on CQEC is the extension of this approach to fault tolerant error correction~\cite{Cross2007,linke2017fault,li2018direct}.

The favorable performance of the CQEC protocol seen for the quantum annealing application presented here, in particular the lack of {any} significant {decrease} in performance going from one to two logical qubits, indicates the potential viability of modular approaches to quantum error correction for quantum simulation and for quantum annealing in particular.  For quantum computation and simulation on near term quantum machines, it is advantageous to use encodings that generate only low weight logical operators, while also requiring only low weight measurement operators. Since the {weight of the} logical operators of stabilizer codes, whether subspace or subsystem, always grow with the number of {encoding} qubits, small codes are therefore highly attractive from this perspective. {Indeed, quantum annealing Hamiltonians of the Ising spin glass form, i.e., containing only terms of the form $H_Z = \sum_i h_i^z \sigma_i^z,  H_X = \sum_i h_i^x \sigma_i^x, H_{ZZ}= \sum_{ij} J_{ij} \sigma_i^z \sigma_j^z$, that are encoded with the three-qubit stabilizer code {result in logical operator terms of only weight two and three}.} 
{The three-qubit code} thus presents an attractive {modular} option for implementing error correction of quantum annealing with large numbers of logical qubits.

\begin{acknowledgements}
This work was supported by the Office of
the Director of National Intelligence (ODNI), Intelligence Advanced
Research Projects Activity (IARPA), via the U.S. Army Research Office
contract W911NF-17-C-0050. The views and conclusions contained herein are
those of the authors and should not be interpreted as necessarily
representing the official policies or endorsements, either expressed or
implied, of the ODNI, IARPA, or the U.S. Government. The U.S. Government
is authorized to reproduce and distribute reprints for Governmental
purposes notwithstanding any copyright annotation thereon. 
This work was also partially supported by the National Aeronautics and Space Administration under Grant/Contract/Agreement No.80NSSC19K1123 issued through the Aeronautics Research Mission Directorate.

 *J.A. and S.Z. contributed equally to this work.
\end{acknowledgements}

\begin{appendix}
\section{Numerical method to generate discretized readout signals and density matrix evolution} \label{sec: appendix_numerics}

We {describe here} the numerical approach {used} to generate discretized realizations of the readout signals $I_{1}(t)$ and $I_2(t)$ [see Eq.~\eqref{eq:Ik}], the filtered readout signals $\mathcal{I}_1(t)$ and $\mathcal{I}_2(t)$ [see Eq.~\eqref{eq:filter-eq}], and the system density matrix $\rho(t)$, {with a timestep $dt$.  $\rho(t)$ evolves according to the combined action of Eq.~\eqref{eq:rho-eom-no-errors-v2} and the error correction operations} $C_{\rm op}$ [Eq.~\eqref{eq:Cop}] that are applied on the physical qubits whenever an error is diagnosed by the double threshold CQEC protocol. 

{We use the Bayesian update method of Ref.~\cite{Korotkov2001} to obtain the discretized}  readout signals $\bar I_k(t+dt)$ that correspond to the averages of $I_k(t)$ during the time interval $(t,t+dt)$ and {hence} to measurement of the stabilizer generators $S_k$ [see Eq.~\eqref{eq:stabilizers}]. {$\bar I_k(t+dt)$} is obtained {from}
\begin{align}
\label{eq:Ik-numerics}
    \bar I_k(t+dt) =&\; s_k + \sqrt{\tau_{\rm m}/dt}\, \zeta_k, 
\end{align}
{with} $s_k=\pm1$ is a binary random number that has the value of $+1$ with probability equal to $\rho_{000,000}(t)$ + $\rho_{001,001}(t)$ + $\rho_{110,110}(t)$ + $\rho_{111,111}(t)$ {for $k=1$ (i.e., $S_1=Z_1Z_2$)} {and} with probability equal to $\rho_{000,000}(t)$ + $\rho_{011,011}(t)$ + $\rho_{100,100}(t)$ + $\rho_{111,111}(t)$ {for $k=2$ (i.e., $S_2=Z_2Z_3)$}, and $\zeta_k$ is a Gaussian random number with zero mean and variance 1. {We employed a timestep $dt=5\times10^{-3}\Gamma_{\rm m}^{-1}$ in all our numerical calculations.}  
\newline
The {quantum state of the system} is then updated according to the information, $\bar I_k(t+dt)$,  {obtained  from this} measurement of $S_k$, {according to}:
\begin{align}
\label{eq:Bayesian-update}
    \rho_{ij}(t+dt) =&\; \frac{\sqrt{p_i\big(\bar I_k(t+dt)\big)\,p_j\big(\bar I_k(t+dt)\big)}}{p\big(\bar I_k(t+dt)\big)}\,e^{-\gamma_{ij}dt}\times\nonumber\\
    &\;\rho_{ij}(t).
\end{align}
{Here} $p_i(I)=\exp{[-(I - \langle i|S_k|i\rangle)^2/2\mathcal{D}]}/\sqrt{2\pi \mathcal{D}}$ is the conditional probability density for the ouput signal $I$ given that the system is in the state $|i\rangle$, where $|i\rangle$ indicates one of the  three-qubit computational states, {i.e.,} $\langle i|S_k|i\rangle=\pm1$, and $\mathcal{D}=\tau_{\rm m}/dt$. In addition, $\gamma_{ij}=\Gamma_{\rm m}(1-\eta)\left(\langle i|S_k|i\rangle - \langle j|S_k|j\rangle\right)^2/4$. Note that for  ideal measurements ($\eta=1$), we have  $\gamma_{ij}=0$. The denominator of Eq.~\eqref{eq:Bayesian-update} is the probability  distribution of the continuous random variable $I$, {defined by} $p(I) = \sum_{i=0,1,..7}\rho_{ii}(t)\,p_i(I)$, where the sum is over all three-qubit computational states $|i\rangle$. Equations~\eqref{eq:Ik-numerics}--\eqref{eq:Bayesian-update} 
{provide Bayesian updates} for the discretized readout signals $\bar I_k(t)$ as well as the corresponding conditional state $\rho(t)$, which is conditioned on the recorded readout signal $\bar I_k(t)$, at {all} times $t=n\,dt$, {$n = 0,1,\dots n_{\rm op}$ where $n_{\rm op}dt = t_{\rm op}$}. 

The discretized filtered readout signals {$\mathcal{I}_k(t)$} are {then readily} obtained  from the discretized readout signals $\bar I_k(t)$ 
{using Eq.~\eqref{eq:filter-eq}}:
\begin{align}
    \mathcal{I}_k(t+dt) = \left(1-\frac{dt}{\tau}\right)\mathcal{I}_k(t) + \frac{dt}{\tau}\,\bar I_k(t+dt).  
\end{align}
The system quantum state $\rho(t)$ also evolves due to {the} Hamiltonian $H(t)$ and to decoherence ({bit-flip errors in this work}). {The state update due solely to Hamiltonian-induced  evolution during the time step $dt$} is obtained as
\begin{align}
\label{eq:state-update-Ht}
    \rho(t+dt) = \mathcal{U}(t,t+dt)\,\rho(t)\,\mathcal{U}^\dagger(t,t+dt),
\end{align}
where the unitary evolution operator  $\mathcal{U}(t,t+dt)$ is approximated using the first-order Magnus expansion~\cite{Magnus},  
\begin{align}
    \mathcal{U}(t,t+dt)\approx \mathcal{U}_{\rm M}(t,t+dt) =  \exp[-iH(t+dt/2)\,dt].
\end{align}
The state update due only to decoherence is evaluated as 
\begin{align}
\label{eq:state-update-decoh}
    \rho(t+dt) = \rho(t) + \dot{\rho}_{\rm decoh}(t)\,dt,
\end{align}
where $\dot{\rho}_{\rm decoh}(t)$ is given in Eq.~\eqref{eq:rho-decoh-eom}. 

To account for {all three processes of} measurement, coherent evolution and decoherence at each timestep, we apply the quantum Bayesian update twice ({once} for measurement of $S_1=Z_1Z_2$ and {once} for measurement of $S_2=Z_2Z_3$), followed by state update due to Hamiltonian-induced evolution  [Eq.~\eqref{eq:state-update-Ht}], and then state update due to decoherence [Eq.~\eqref{eq:state-update-decoh}]. {After this we} use the double threshold CQEC protocol to determine whether or not we need to apply an error correction operation $C_{\rm op}$ to the system state at the moment $t+dt$: $\rho(t+dt)\to C_{\rm op}\rho(t+dt)C_{\rm op}$. For example, if $\mathcal{I}_1(t+dt)<\Theta_1$ and $\mathcal{I}_2(t+dt)>\Theta_2$, then the diagnosed error syndrome is $(Z_{12}=-1,Z_{23}=+1)$, the diagnosed error is $X_1$ and so we have to apply the error correction operation  $C_{\rm op}=X_1$. After error correction, we also reset the filtered readout signals: $\mathcal{I}_k(t+dt)\to +1$ {for $k=1,2$}. If there is no error correction {operation in this timestep}, the filtered readout signals are not reset. 

\section{Probability of misdiagnosing bit-flip error $X_2$}
\label{sec:appendix_prob}
{We derive here} the result Eq.~\eqref{eq:p-mis-2} for the probability $p_{\rm misdiag}^{(X_2)}$ to misdiagnose the bit-flip error $X_2$.  In contrast to the conventional implementation {of the bit-flip QEC}, in the continuous operation misdiagnosis of single bit-flip errors occurs when relatively large fluctuations affect one or both filtered readout signals $\mathcal{I}_k(t)$. It is however more likely that only one of the filtered readout signals exhibits a large fluctuation, so we consider this situation to {obtain an estimate for} the probability  $p_{\rm misdiag}^{(X_2)}$. The bit-flip error $X_2$ is misidentified as $X_1$ if, at the moment when the filtered readout signal $\mathcal{I}_1(t)$ exits the "syndrome uncertainty region"  by crossing the lower error threshold $\Theta_1$ {(see Fig.~\ref{fig:filtered-signals-X2-err})}, the filtered readout signal $\mathcal{I}_2(t)$ is above the upper error threshold  $\Theta_2$ due to a {unusually} large positive fluctuation of size larger than $\Theta_2-\Theta_1$. The probability that this situation occurs is given by the probability that $\Delta \mathcal{I}(t)\equiv\mathcal{I}_{2}(t)-\mathcal{I}_1(t)\geq \Theta_2-\Theta_1$. From Eq.~\eqref{eq:filter-eq}, we {have}
\begin{align}
\label{eq:DeltaI-eom}
    \frac{d\,\Delta \mathcal{I}(t)}{dt} = -\frac{\Delta \mathcal{I}(t)}{\tau} + \frac{\sqrt{\tau_{\rm m}}}{\tau}\left(\xi_2(t)-\xi_1(t)\right), 
\end{align}
where the noises $\xi_1(t)$ and $\xi_2(t)$ are the uncorrelated noises of the bare readout signals $I_k(t)$ (see Eqs.~\eqref{eq:Ik}--\eqref{eq:noises-corr-func}). Note that Eq.~\eqref{eq:DeltaI-eom} is valid {both} before and after the occurrence of the bit-flip error $X_2$ because the "signal parts" of the  readout signals $I_1(t)$ and $I_2(t)$ cancel each other in $\Delta \mathcal{I}(t)$. {Specifically,} before (after) occurrence of the error $X_2$, the "signals parts" of $I_1(t)$ and $I_2(t)$ are both {equal to} $+1$ ($-1$). This {implies} that the probability that $\Delta \mathcal{I}(t)\geq\Theta_2-\Theta_1$ can be obtained from the stationary probability distribution, $p_{\rm st}(\Delta \mathcal{I})$, of $\Delta \mathcal{I}(t)$. From Eq.~\eqref{eq:DeltaI-eom}, we obtain
\begin{align}
    p_{\rm st}(\Delta \mathcal{I}) = \left[\frac{\tau}{2\pi\tau_{\rm m}}\right]^{1/2} e^{-(\Delta\mathcal{I})^2\tau/2\tau_{\rm m}}.
    \label{eq:prob-st-DeltaI}
\end{align}
The probability that $\Delta \mathcal{I}(t)$ is larger than $\Theta_2-\Theta_1$ is then equal to %
\begin{align}
    p(\Delta\mathcal{I}\geq\Theta_2-\Theta_1) =&\; \frac{1}{2}\left[1 - {\rm erf}\left(\sqrt{\frac{\tau}{2\tau_{\rm m}}}\, (\Theta_2-\Theta_1)\right) \right], \nonumber\\
    \approx&\;\frac{1}{\sqrt{2\pi}}\, \frac{e^{-(\Theta_2-\Theta_1)^2\tau/2\tau_{\rm m}}}{(\Theta_2-\Theta_1)\sqrt{{\tau}/{\tau_{\rm m}}}},
    \label{eq:prob-deriv1}
\end{align}
where ${\rm erf}(\cdot)$ is the error function and the approximation applies in the limit of large averaging time parameters $\tau$. The result~\eqref{eq:prob-deriv1} is our estimation for the probability that the bit-flip error $X_2$ is misdiagnosed as the error $X_1$. The same result is also  obtained for the probability that the error $X_2$ is misdiagnosed as the error $X_3$. Therefore, the probability that the  error $X_2$ is misdiagnosed is given by 

\begin{align}
    p_{\rm misdiag}^{(X_2)} = &\; c\, \frac{e^{-(\Theta_2-\Theta_1)^2\tau/2\tau_{\rm m}}}{(\Theta_2-\Theta_1)\sqrt{{\tau}/{\tau_{\rm m}}}}
    \label{eq:Prob-misdiag-X2}
\end{align}
The numerical coefficient $c$ that follows from the above {analysis} is $\sqrt{2/\pi}\approx 0.7979$. {By fitting our numerical results to} Eq.~\eqref{eq:Prob-misdiag-X2}, we obtain that the coefficient $c$ is larger, specifically, {fitting to the data in Fig.~\ref{fig:prob-misdiagnosis} yields $c\approx 1.607$}. Equation~\eqref{eq:Prob-misdiag-X2} with $c\approx 1.607$ has been successfully tested against numerical results for various values of the error threshold parameters $\Theta_1$ and $\Theta_2$
in addition to the values indicated in Fig.~\ref{fig:prob-misdiagnosis}.
\begin{figure}[t!]
\centering
\includegraphics[width=0.9\linewidth]{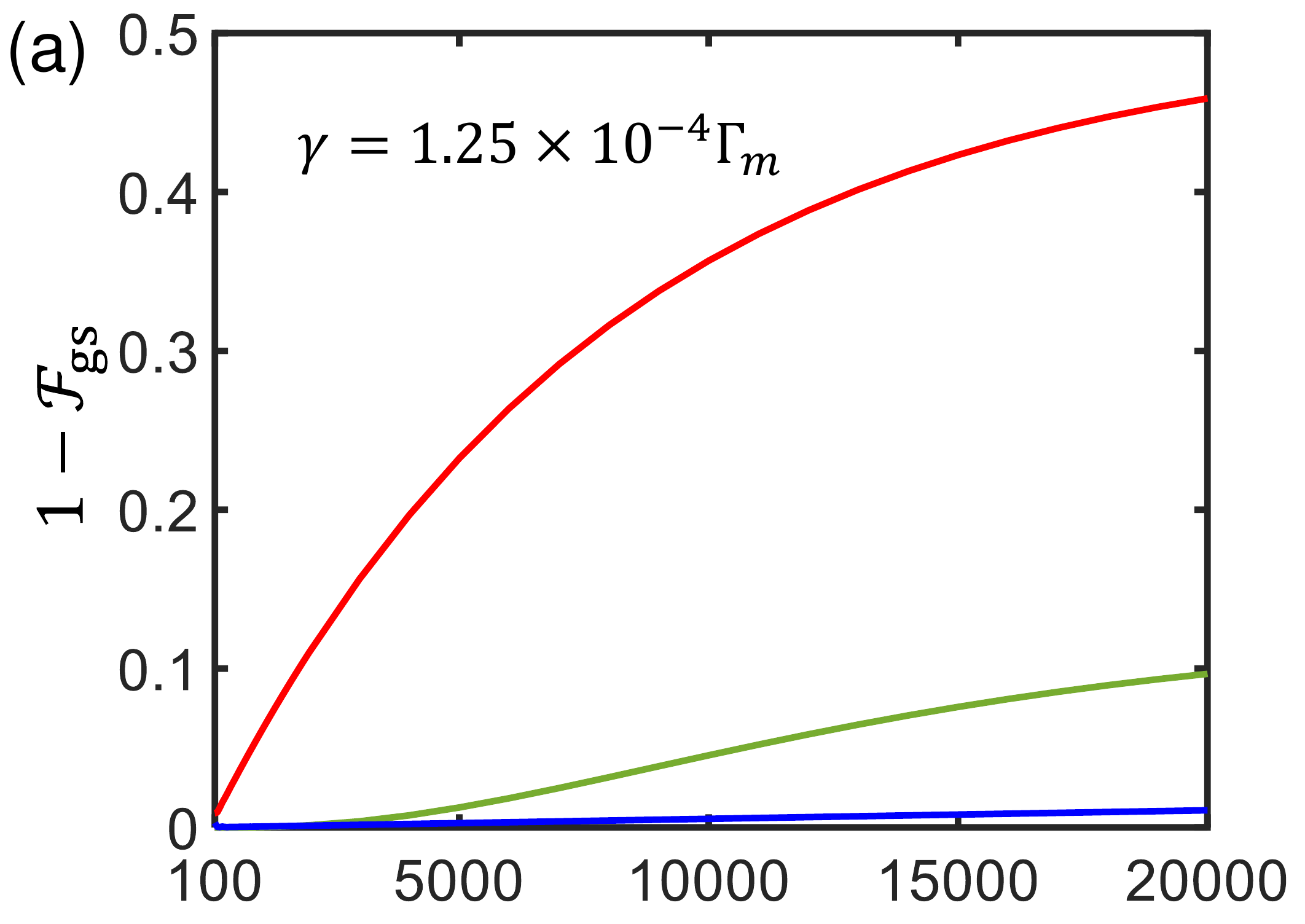}
\includegraphics[width=0.9\linewidth]{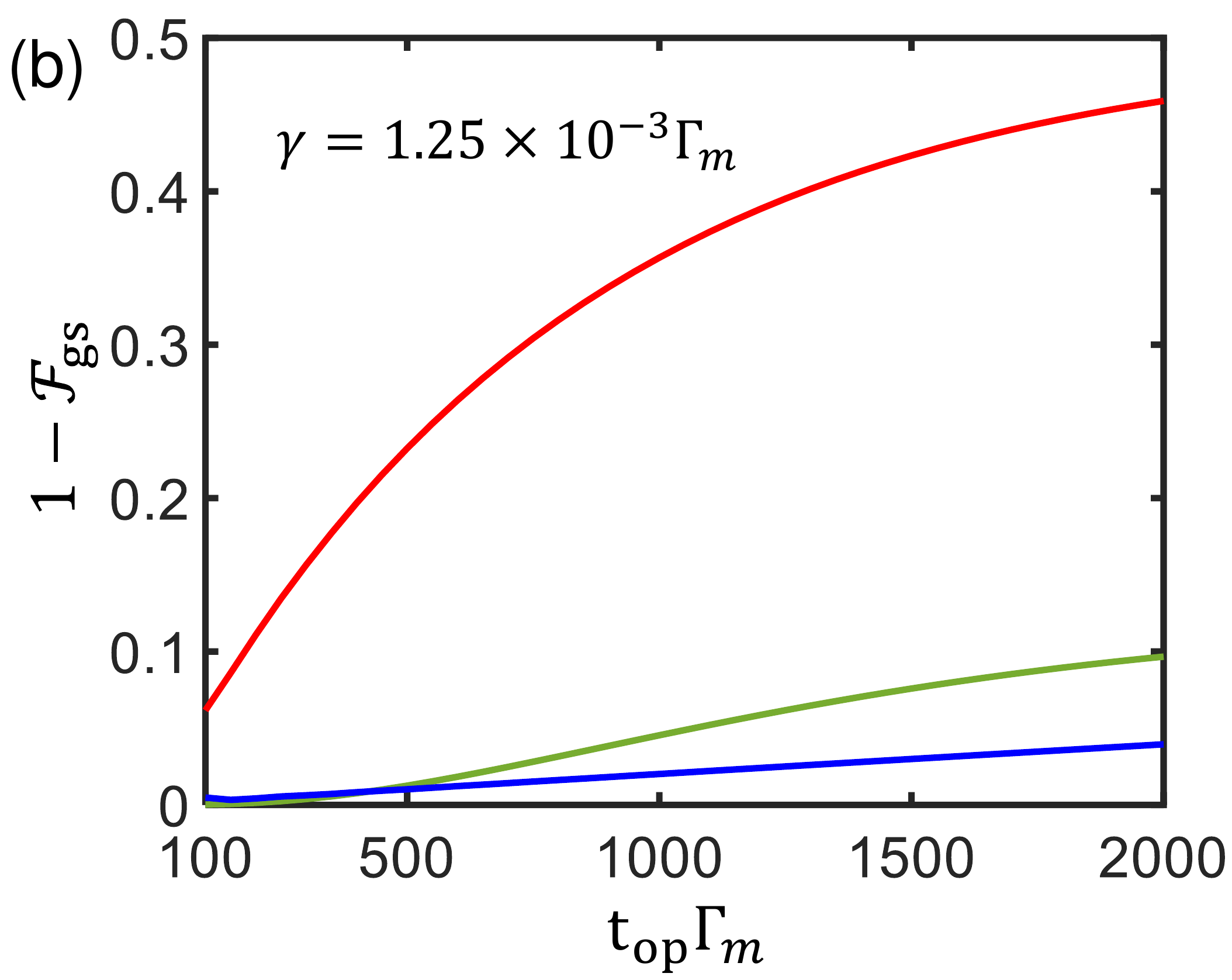}
\caption{Final infidelity with respect to the ground state of the problem Hamiltonian for one qubit annealing as a function of operation time $t_{\rm op}$. Panel (a): $\gamma = 1.25\times 10^{-4}\Gamma_{\rm m}$. Panel (b): $\gamma = 1.25\times 10^{-3}\Gamma_{\rm m}$.  Red line: one unencoded physical qubit, no error correction. Green line: classic protection scheme of Ref.~\cite{pudenz2014error} 
(see text). Blue line: one logical qubit with CQEC and modified error correction, Eq.~\eqref{eq:Cop-new}.  Parameter values: $\eta = 1$, $\Omega = 0.1\Gamma_{\rm m}$, $\theta_1 = -0.54$, $\theta_2 = 0.8$, and $\tau = 2.5\Gamma_{\rm m}^{-1}$. 
}
\label{fig:final_infidelity_vs_top}
\end{figure}

\begin{figure}[t]
\centering
\includegraphics[width=0.9\linewidth]{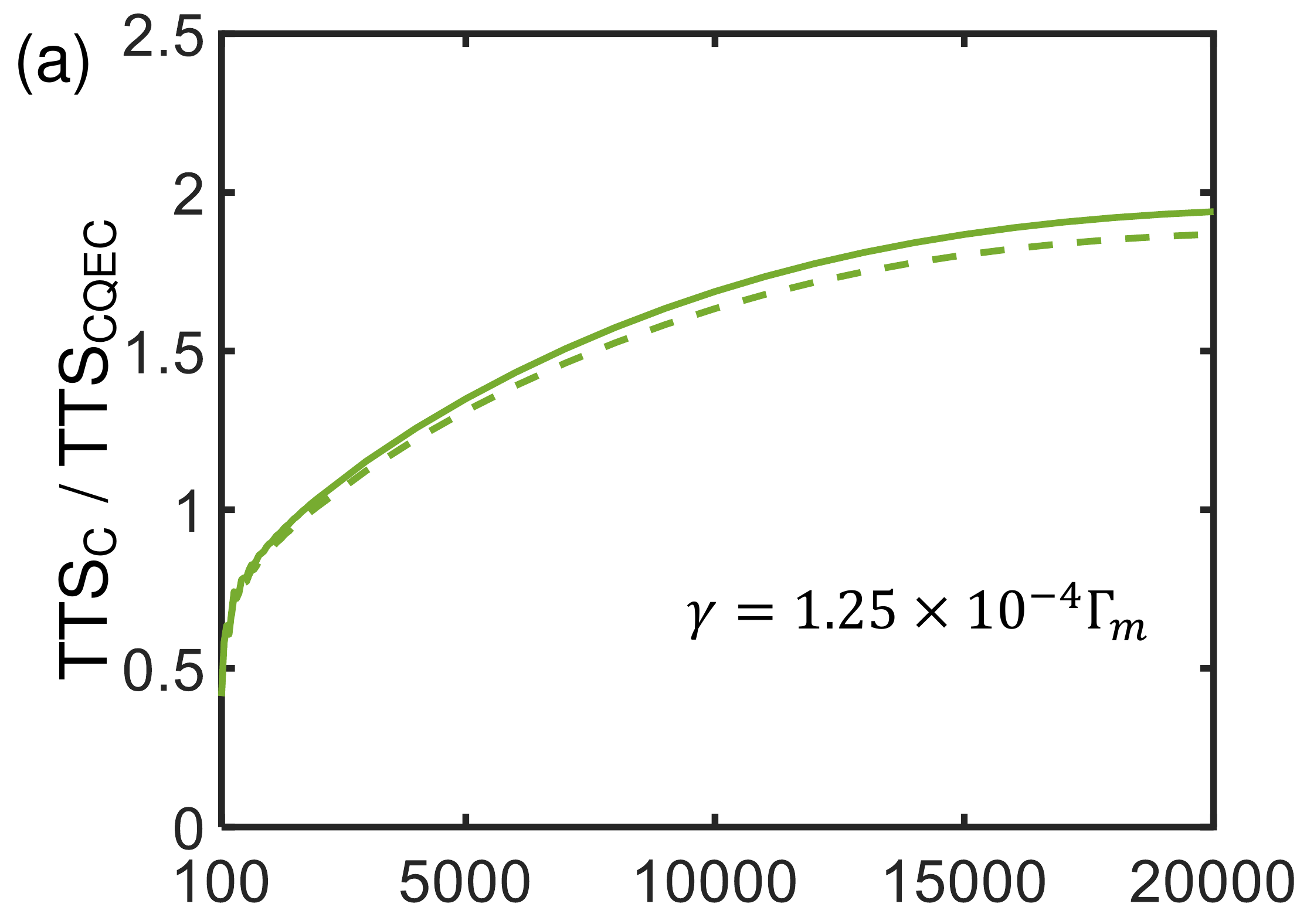}
\includegraphics[width=0.9\linewidth]{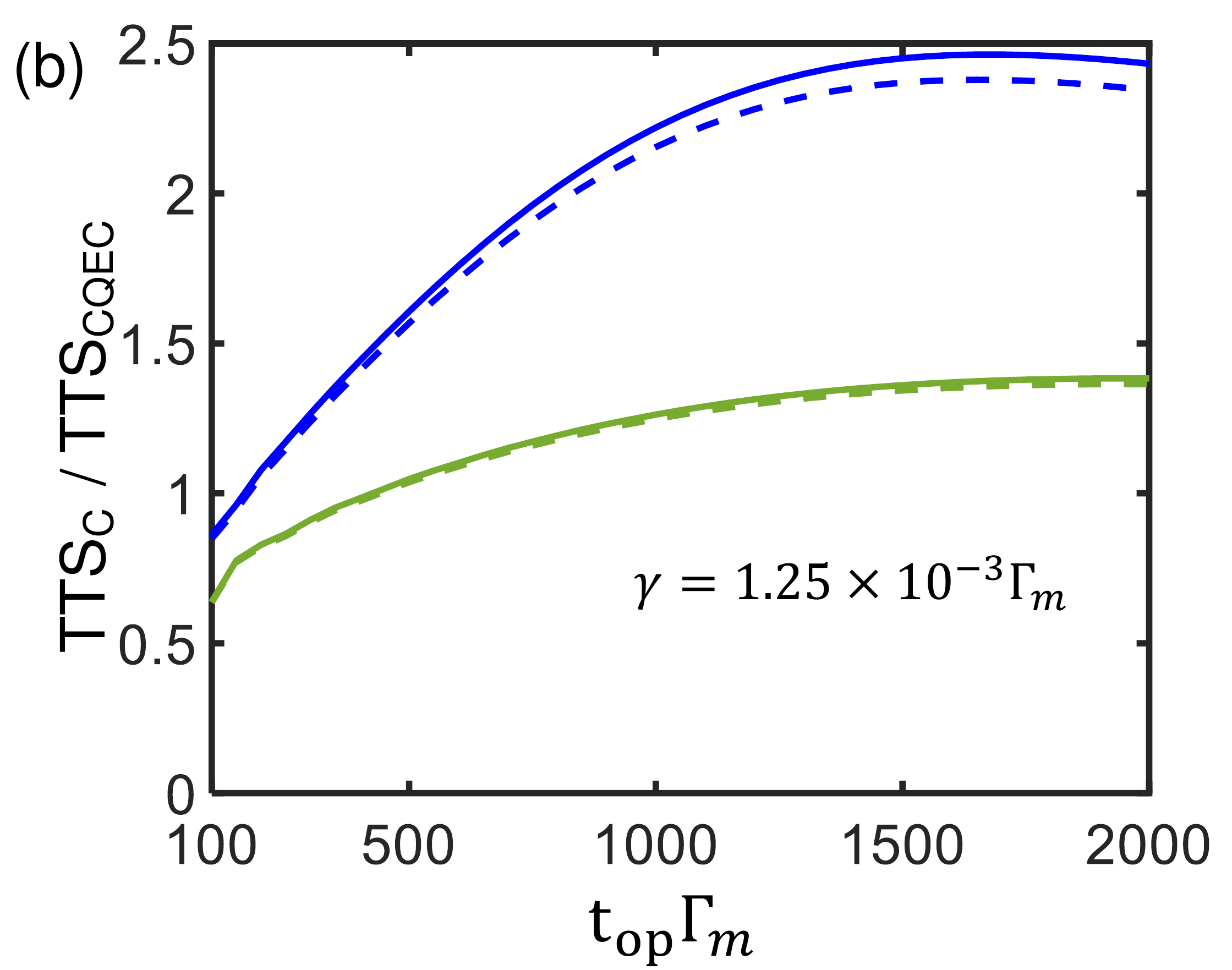}
\caption{Ratio of the TTS value for the classical (C) strategy to that for CQEC, as a function of $t_{\rm op}$. Panel (a): one logical qubit, $\gamma = 1.25\times 10^{-4}\Gamma_{\rm m}$. Panel (b): one and two logical qubits, $\gamma = 1.25\times 10^{-3}\Gamma_{\rm m}$. Green lines are for one logical qubit, with error correction operations in the CQEC calculations given by Eq.~\eqref{eq:Cop-new} (solid lines) and Eq.~\eqref{eq:Cop-appro} (dashed lines). 
Blue lines are for two logical qubits, with error correction operations in the CQEC calculations given by Eq.~\eqref{eq:Cop-new} (solid line) and Eq.~\eqref{eq:Cop-appro} (dashed line). Parameters: one logical qubit calculations use the same values as in Fig.~\ref{fig:final_infidelity_vs_top}, two logical qubit calculations employ $\tau = 2.1\Gamma_{\rm m}^{-1}$, with all other parameter values the same as for one logical qubit.  
}
\label{fig:TTS_ratio_vs_top}
\end{figure}

\section{Time-to-solution metric}\label{sec: appendix_TTS}

We present here analysis of the application of our CQEC protocol to quantum annealing with the time-to-solution (TTS) metric. In particular, we show how the value of this metric depends on the annealing operation time $t_{\rm op}$. We compare the performance of our CQEC protocol with regard to TTS with that of the so-called classical (C) strategy of Ref.~\cite{pudenz2014error}, in which annealing is carried out on $m$ unencoded copies in parallel and the infidelity estimated from the binomial distribution function $(1-p^{\rm unenc}_{\rm s}(t_{\rm op}))^m$, with $p^{\rm unenc}_{\rm s}(t_{\rm op})$ the probability of successfully reaching the 
ground state of the problem Hamiltonian, i.e., the final Hamiltonian at $t=t_{\rm op}$,
for a single unencoded qubit (see Ref.~\cite{pudenz2014error}). 
We use the example of single qubit quantum annealing from Section~\ref{sec:Optimization} for this detailed analysis, 
before summarizing the performance of both single qubit and two-qubit quantum annealing problem at a particular value of $t_{\rm op}$.

Figure \ref{fig:final_infidelity_vs_top} shows how the final CQEC infidelity (blue lines, with modified logical error correction operations, see Eq.~\eqref{eq:Cop-new}) depends on the operation time $t_{\rm op}$ for single qubit quantum annealing, and compares this with the corresponding infidelities obtained for the unencoded dynamics (red lines) and the classical (C) strategy (green lines). Note that here we are plotting the final fidelity with respect to the ground state of the problem Hamiltonian.  The two panels show results for two different physical error rates $\gamma$, with the value of the product $\gamma t_{\rm op}$ for the maximal value of $t_{\rm op}$ held constant at 2.5. This corresponds to the same overall average number of errors at the largest shown value of $t_{\rm op}$ in both panels. The calculations related to the blue lines in Fig.~\ref{fig:final_infidelity_vs_top} were carried out using the effective evolution equation Eq.~\eqref{eq:rhoL-eom} without the last term at its right-hand-side. The accuracy of this was confirmed by full quantum trajectory simulation of  Eq.~\eqref{eq:rho-eom-no-errors-v2} together with the double-thresholding error detection protocol and error correction operations given by Eq.~\eqref{eq:Cop-new}. For example, for $\gamma=1.25\times 10^{-4}\Gamma_{\rm m}$ and $t_{\rm op} = 10^{4}\Gamma_{\rm m}^{-1}$, we obtain infidelities of $6.6\times 10^{-3}$ from the quantum trajectory simulations (averaging over 1000 trajectories) and $5.14\times10^{-3}$ from the effective evolution equation~\eqref{eq:rhoL-eom}, respectively. 
The unencoded calculations (red lines) are simulated directly from Eq.~\eqref{eq:rho-eom-no-errors-v2} without the measurement terms, and the classical strategy values (green lines) are obtained as the third power of the unencoded values.  

It is evident that CQEC performs better than both the unencoded annealing and the classical strategy at longer operation times.  The final infidelity of unencoded annealing of the single qubit tends towards {that of} the completely mixed state on the time scales shown here, consistent with the lack of any error correction,  {while that of the classical strategy tends to the corresponding binomial function of this.}   In contrast, once the operation time is long enough that non-adiabatic errors are small, the CQEC strategy is successful in correcting the bit-flip errors due to coupling to the environment {and can achieve significantly lower infidelities.}

The data in Fig. \ref{fig:final_infidelity_vs_top} are then used to evaluate the TTS for {single qubit annealing}, according to Eqs.~\eqref{eq:TTS}--\eqref{eq:TTS_R} in the main text. We have chosen {the target success probability} $p_d = 0.99$.  Figure~\ref{fig:TTS_ratio_vs_top} shows how the ratio of the TTS value for the classical strategy to that for the CQEC strategy varies with operation time $t_{\rm op}$, for a given value  of physical error rate $\gamma$. At the longer operation times in the time scale range shown here, this TTS ratio is larger than $1$, indicating that the classical  strategy will require a longer time to reach the desired ground state solution for the annealing problem.
 In terms of this TTS metric, the CQEC strategy thus clearly  outperforms the classical {strategy} at the longer operation times $t_{\rm op}$ 
 that suppress non-adiabatic errors and are preferred for quantum annealing.

To illustrate the performance on larger systems, we have also calculated the corresponding ratio of TTS values for quantum annealing of two logical qubits, using the effective two qubit model Eq.~\eqref{eq:rhoL-eom-2L}. The blue lines in Fig.~\ref{fig:TTS_ratio_vs_top} panel (b) show the resulting ratio TTS$_{\rm C}/$TTS$_{\rm CQEC}$  as a function of $t_{\rm op}$. Comparing with the green lines for one logical qubit, it is evident that in terms of the TTS metric, the CQEC protocol performs increasingly better than the classic strategy when scaled up to two qubits. This is encouraging for the use of a modular form of CQEC for larger systems.

\end{appendix}


\newpage
\begin{thebibliography}{99}
\bibitem{lidar1998decoherence} D. A. Lidar, I. L. Chuang, and K. B. Whaley, \textit{Decoherence-Free Subspaces for Quantum Computation}, Phys. Rev. Lett. {\bf 81}, 2594 (1998). 

\bibitem{lidar2003decoherence}  D. A. Lidar and K. B. Whaley, {\it Decoherence-free Subspaces and Subsystems}, in ”Irreversible Quantum Dynamics”, F. Benatti and R. Floreanini (Eds.), pp. 83-120 (Springer Lecture Notes in Physics vol. 622, Berlin, 2003).

\bibitem{jordan2006error} S. P. Jordan, E. Farhi, and P. W. Shor, \textit{Error correcting codes for adiabatic quantum computation}, Phys. Rev. A {\bf 74}, 052322 (2006). 

\bibitem{bookatz2015error} A. D. Bookatz, E. Farhi, and L. Zhou, \textit{Error suppression in Hamiltonian-based quantum computation using energy penalties}, Phys. Rev. A {\bf 92}, 022317 (2015).

\bibitem{marvian2017error} M. Marvian and D. A. Lidar, \textit{Error suppression in Hamiltonian-based quantum computation using energy penalties}, Phys. Rev. A {\bf 95},  032302 (2017). 

\bibitem{viola1999dynamical} L.  Viola, E.  Knill, and S. Lloyd, \textit{Dynamical Decoupling of Open Quantum Systems}, Phys. Rev. Lett. {\bf 82}, 2417 (1999). 

\bibitem{pasini2010optimized} S. Pasini and G. S. Uhrig, \textit{Optimized dynamical decoupling for power-law noise spectra}, Phys. Rev. A {\bf 81}, 012309 (2010). 

\bibitem{lidar2014review} D. A. Lidar, \textit{Review of Decoherence‐Free Subspaces, Noiseless Subsystems, and Dynamical Decoupling}, Adv. Chem. Phys. {\bf 154}, 295 (2014).

\bibitem{Gorman2012pulses} D. J. Gorman, K. C. Young, and K. B. Whaley, \textit{Overcoming dephasing noise with robust optimal control}, Phys. Rev. A {\bf 86}, 012317 (2012).

\bibitem{gottesman2002introduction} D. Gottesman, \textit{An introduction to to quantum error correction}, Proc. Sympos. Appl. Math. {\bf 58}, 221 (2002).  

\bibitem{devitt2013quantum} S. J Devitt, W. J. Munro, and K. Nemoto, \textit{Quantum error correction for beginners}, Rep. Prog. Phys.  {\bf 76}, 076001 (2013). 

\bibitem{lidar2013quantum} D. A. Lidar and T. A. Brun, {\it Quantum Error Correction} (Cambridge University Press, 2013).

\bibitem{terhal2015quantum} B. M. Terhal, \textit{Quantum error correction for quantum memories}, Rev. Mod. Phys. {\bf 87}, 307 (2015). 

\bibitem{shor1996fault} P. W. Shor, \textit{Fault-Tolerant Quantum Computation} in {\it Proc. 37th Symposium on the Foundations of Computer Science}, (IEEE Computer Society Press), 56 (1996).

%
\bibitem{Shor1995} P. W. Shor, \textit{Scheme for reducing decoherence in quantum computer memory}, Phys. Rev. A {\bf 52}, R2493 (1995).

\bibitem{Steane1996} A. M. Steane, \textit{Error Correcting Codes in Quantum Theory}, Phys. Rev. Lett. {\bf 77}, 793 (1996).

\bibitem{Gottesman1996}  D. Gottesman, \textit{Class of quantum error-correcting codes saturating the quantum Hamming bound}, Phys. Rev. A {\bf 54}, 1862 (1996).

\bibitem{ChuangBook}  M. A. Nielsen and I. L. Chuang, {\it Quantum Computation
and Quantum Information} (Cambridge University Press, Cambridge, 2000). 

\bibitem{KrausBook} K. Kraus, {\it States, Effects, and Operations: Fundamental Notions of Quantum Theory} (Springer, Berlin, 1983).

\bibitem{Carmichael1993} H. J. Carmichael, {\it An Open System Approach to Quantum Optics, Lecture Notes in Physics} (Springer, Berlin, 1993). 

\bibitem{Wiseman2010} H. M. Wiseman and G. J. Milburn, {\it Quantum Measurement and Control} (Cambridge University Press, Cambridge, UK, 2010). 

\bibitem{Diosi1988} L. Di\'osi, \textit{Continuous quantum measurement and itô formalism}, Phys. Lett. A {\bf 129}, 419 (1988). 

\bibitem{Molmer1992} J. Dalibard, Y. Castin, and K. M\o lmer, \textit{Wave-function approach to dissipative processes in quantum optics}, Phys. Rev. Lett.
68, 580 (1992).

\bibitem{Korotkov2016} A. N. Korotkov, \textit{Quantum Bayesian approach to circuit QED measurement with moderate bandwidth}, Phys. Rev. A {\bf 94}, 042326 (2016). 

\bibitem{Katz2006}  N. Katz et al.,  \textit{Coherent state evolution in a superconducting qubit from partial-collapse measurement}, Science {\bf 312}, 1498 (2006). 

\bibitem{PalaciosLaloy2010} A. Palacios-Laloy, F. Mallet, F. Nguyen, P. Bertet, D. Vion, D. Esteve, and A. N. Korotkov, \textit{Experimental violation of a Bell’s inequality in time with weak measurement}, Nat. Phys. {\bf 6}, 442 (2010).

\bibitem{Devoret2013} M. Hatridge et al., \textit{Quantum Back-Action of an Individual Variable-Strength Measurement}, Science {\bf 339}, 178 (2013). 

\bibitem{KMurch2013} K. W. Murch, S. J. Weber, C. Macklin, and I. Siddiqi, \textit{Observing single quantum trajectories of a superconducting quantum bit}, Nature (London) {\bf 502}, 211 (2013). 

\bibitem{DiCarlo2013}D. Rist\'e, M. Dukalski, C. A. Watson, G. de Lange, M. J. Tiggelman, Ya. M. Blanter, K. W. Lehnert, R. N. Schouten, and L. DiCarlo, \textit{Deterministic entanglement of superconducting qubits by parity measurement and feedback}, Nature (London) {\bf 502}, 350 (2013). 

\bibitem{Hacohen-Gourgy2016} S. Hacohen-Gourgy, L. S. Martin, E. Flurin, V. V. Ramasesh, K. B. Whaley, and I. Siddiqi, \textit{Quantum dynamics of simultaneously measured non-commuting observables}, Nature (London) {\bf 538}, 491 (2016).

\bibitem{Huard2018} Q. Ficheux, S. Jezouin, Z. Leghtas, and B. Huard, \textit{Dynamics of a qubit while simultaneously monitoring its relaxation and dephasing}, Nat. Commun. {\bf 9}, 1926 (2018). 

\bibitem{Zurek1998} J. P. Paz and W. H. Zurek, \textit{Continuous error correction}, Proc. R. Soc. Lond. A {\bf 454}, 355 (1998).

\bibitem{Ahn2002} C. Ahn, A. C. Doherty, and A. J. Landahl, \textit{Continuous quantum error correction via quantum feedback control}, Phys. Rev. A {\bf 65}, 042301 (2002).

\bibitem{Ahn2003} C. Ahn, H. M. Wiseman, and G. J. Milburn, \textit{Quantum error correction for continuously detected errors}, Phys. Rev. A {\bf 67}, 052310 (2003).
 
\bibitem{Ahn2004} C. Ahn, H. M. Wiseman, and K. Jacobs, \textit{Quantum error correction for continuously detected errors with any number of error channels per qubit}, Phys. Rev. A {\bf 70}, 024302 (2004).

\bibitem{Sarovar2004} M. Sarovar, C. Ahn, K. Jacobs, and G. J. Milburn, \textit{Practical scheme for error control using feedback}, Phys. Rev. A {\bf 69}, 052324 (2004).

\bibitem{Sarovar2005} M. Sarovar and G. J. Milburn, \textit{Continuous quantum error correction by cooling}, Phys. Rev. A {\bf 72}, 012306 (2005).

\bibitem{Brun2007} O. Oreshkov and T. A. Brun, \textit{Continuous quantum error correction for non-Markovian decoherence}, Phys. Rev. A {\bf 76}, 022318 (2007).

\bibitem{Geremia2008} B. A. Chase, A. J. Landahl, and J. M. Geremia, \textit{Efficient feedback controllers for continuous-time quantum error correction}, Phys. Rev. A {\bf 77}, 032304 (2008).

\bibitem{Mabuchi2009} R. van Handel and H. Mabuchi, \textit{Optimal error tracking via quantum coding and continuous syndrome measurement}, arXiv:0511221; 
\bibitem{Mabuchi_new_j_phys2009} H. Mabuchi, \textit{Continuous quantum error correction as classical hybrid control}, New J. Phys. {\bf 11}, 105044 (2009).

\bibitem{Brun2016} K. C. Hsu and T. A. Brun, \textit{Method for quantum-jump continuous-time quantum error correction}, Phys. Rev. A {\bf 93}, 022321 (2016).

\bibitem{Dressel2019}  R. Mohseninia, J. Yang, I. Siddiqi, A. N. Jordan, and J. Dressel, \textit{Always-On Quantum Error Tracking with Continuous Parity Measurements}, arXiv:1907.08882 (2019).

\bibitem{Cardona2019} G. Cardona, A. Sarlette, and P. Rouchon, \textit{Continuous-time quantum error correction with noise-assisted quantum feedback}, IFAC-PapersOnLine, {\bf 52}, 198 (2019). 

\bibitem{Atalaya2017} J. Atalaya, M. Bahrami, L.P. Pryadko, and A.N. Korotkov, \textit{Bacon-Shor code with continuous measurement of noncommuting operators}, Phys. Rev. A {\bf 95}, 032317 (2017). 

\bibitem{Atalaya2020} J. Atalaya, A. N. Korotkov, and K. B. Whaley, \textit{Error correcting Bacon-Shor code with continuous measurement of noncommuting operators}, arXiv:1910.08272 (2019). 

\bibitem{Lidar2018} T. Albash and D. A. Lidar, \textit{Adiabatic quantum computation}, Rev. Mod. Phys. {\bf 90}, 015002 (2018). 

\bibitem{Nori2014} I. M. Georgescu, S. Ashhab, and Franco Nori, \textit{Quantum simulation}, Rev. Mod. Phys. {\bf 86}, 153 (2014). 

\bibitem{KYoung2013} K. C. Young, M. Sarovar, and R. Blume-Kohout, \textit{Error suppression and error correction in adiabatic quantum computation: Techniques and challenges}, Phys. Rev. X {\bf 3}, 041013 (2013). 

\bibitem{Korotkov2001} A.N. Korotkov, \textit{Selective quantum evolution of a qubit state due to continuous measurement}, Phys. Rev. B {\bf 63}, 115403 (2001).

\bibitem{Slichter2015} D. H. Slichter, C. M\"uller, R. Vijay, S. J. Weber, A. Blais, and I. Siddiqi, \textit{Quantum Zeno effect in the strong measurement regime of circuit quantum electrodynamics}, New J. Phys. {\bf 18}, 053031 (2016). 

\bibitem{HorowitzBook} P. Horowitz and W. Hill, The Art of Electronics (Cambridge University Press, Cambridge, 2015) 

\bibitem{steane2007tutorial} A. M. Steane, \textit{A tutorial on quantum error correction},
PROCEEDINGS-INTERNATIONAL SCHOOL OF PHYSICS ENRICO FERMI, IOS Press, {\bf 162}, 1 (2007).

\bibitem{king2018observation} A. D. King et al., \textit{Observation of topological phenomena in a programmable lattice of 1,800 qubits}, Nature {\bf 560} 456 (2018).

\bibitem{novikov2018exploring} S. Novikov, R. Hinkley, S. Disseler, J. I. Basham, T. Albash, A. Risinger, D. Ferguson, D. A. Lidar, and K. M. Zick,  \textit{Exploring More-Coherent Quantum Annealing}, IEEE International Conference on Rebooting Computing (ICRC), 1 (2018).

\bibitem{weber2017coherent} S.J. Weber et al., \textit{Coherent coupled qubits for quantum annealing}, Phys. Rev. Applied {\bf 8}, 014004 (2017).

\bibitem{DiVincenzo2013} D. P. DiVincenzo and F. Solgun, \textit{Multi-qubit parity measurement in circuit quantum electrodynamics}, New J. Phys. {\bf 15} 075001 (2013). 

\bibitem{ronnow2014defining} T. F. R{\o}nnow, Z. Wang, J. Job, S. Boixo, S. V. Isakov, D. Wecker, J. M. Martinis, D. A. Lidar, and M. Troyer, \textit{Defining and detecting quantum speedup}, Science {\bf 345}, 420 (2014).

\bibitem{Tameem2018} T. Albash and D. A. Lidar \textit{Demonstration of a scaling advantage for a quantum annealer over simulated annealing}, Phys. Rev. X {\bf 8}, 031016 (2018)

\bibitem{pudenz2014error} K. L. Pudenz, T. Albash, and D. A. Lidar, \textit{Error-corrected quantum annealing with hundreds of qubits}, Nature communications {\bf 5}, 1 (2014).

\bibitem{Poulin2005} D. Poulin, \textit{Stabilizer Formalism for Operator Quantum Error Correction}, Phys. Rev. Lett. {\bf 95}, 230504 (2005).

\bibitem{Bacon2006} D. Bacon, \textit{Operator quantum error-correcting subsystems for self-correcting quantum memories}, Phys. Rev. A {\bf 73}, 012340 (2006).

\bibitem{Cross2007} P. Aliferis and A. W. Cross, \textit{Subsystem fault tolerance with the Bacon-Shor code}, Phys. Rev. Lett. {\bf 98}, 220502 (2007).

\bibitem{bravyi2011subsystem}S. Bravyi, \textit{Subsystem codes with spatially local generators}, Phys. Rev. A {\bf 83}, 012320 (2011). 

\bibitem{marvian2019robust} M. Marvian and S. Lloyd, \textit{Robust universal Hamiltonian quantum computing using two-body interactions}, arXiv:1911.01354 (2019).

\bibitem{Atalaya2020b} J. Atalaya and K. B. Whaley, in preparation (2020). 

\bibitem{Foesel2018} T. F{\"o}sel, P. Tighineanu, T. Weiss, and F. Marquardt, \textit{Reinforcement learning with neural networks for quantum feedback}, Phys. Rev. X {\bf 8}, 031084 (2018).

\bibitem{tsang2015volterra} M. Tsang, \textit{Volterra filters for quantum estimation and detection}, Phys. Rev. A {\bf 92}, 062119 (2015).

\bibitem{linke2017fault} N. M. Linke, M. Gutierrez, K. A. Landsman, C. Figgatt, S, Debnath, K. R.  Brown, and C. Monroe, \textit{Fault-tolerant quantum error detection}, Science advances, {\bf 3}, e1701074 (2017).

\bibitem{li2018direct} M. Li, D. Miller, and K. R. Brown, \textit{Direct measurement of Bacon-Shor code stabilizers}, Phys. Rev. A {\bf 98}, 050301(R), (2018).

\bibitem{Magnus} S. Blanes, F. Casas, J.A. Oteo, and J. Ros, \textit{The Magnus expansion and some of its applications}, Phys. Reports {\bf 470}, 151 (2009). 


\end{thebibliography}
\end{document}